\documentclass[final,3p,times]{elsarticle}

\biboptions{sort&compress}
\usepackage{amssymb}
\usepackage{amsmath}
\usepackage{float}
\usepackage{subcaption}
\usepackage{xurl}
\usepackage{hyperref}

\begin{document}

\begin{frontmatter}



\title{Hybrid electrolyzer systems: Smart strategy or economic fallacy?} 

\author[label1,label2]{Marie Arnold$^{*,}$} 
\author[label2]{Jonathan Brandt$^{*,}$} 
\author[label1]{Geert Tjarks} 
\author[label2]{Richard Hanke-Rauschenbach} 

\affiliation[label1]{organization={EWE GASSPEICHER GmbH},
            city={Oldenburg},
            postcode={26122}, 
            state={Lower Saxony},
            country={Germany}}

\affiliation[label2]{organization={Leibniz Universität Hannover, Institute for Electric Power Systems},
            city={Hanover},
            postcode={30167}, 
            state={Lower Saxony},
            country={Germany}}

\cortext[cor1]{Corresponding authors. Emails: marie.arnold@ewe.de, brandt@ifes.uni-hannover.de}

\begin{abstract}
Hybrid electrolyzer systems combining alkaline water electrolysis and proton exchange membrane water electrolysis have been investigated in the literature motivated by the expectation that their contrary techno-economic characteristics compensate for the individual technical and economic restrictions of each technology, thereby improving the profitability of green hydrogen production. To reassess the economic potential of hybrid electrolyzer systems beyond these technology-specific assumptions, we independently vary two key characteristics, electrolyzer efficiency and investment cost, in a large-scale sensitivity analysis. For each generated parameter configuration, we performed a techno-economic optimization of a green hydrogen supply chain, including two electrolyzers. The resulting system design, cost objective, and dispatch behavior are subsequently analyzed. Consequently, hybrid electrolyzer systems are identified as optimal if they provide a cost benefit over single electrolyzer systems. The analysis reveals that hybrid electrolyzer systems represent the optimal solution in at most 5.0\% of the investigated cases. Furthermore, the maximum cost benefit is 0.057 €/kgH2, which corresponds to only about 1\% of the total green hydrogen production cost. Additional analyses considering variations in energy purchase prices, storage fees, availability of renewable energy, and baseline electrolyzer efficiency yield negligible changes to these results. Hence, considering that hybrid electrolyzer systems offer marginal cost benefits and prove economically optimal in very few cases, they seem more likely to represent an economic fallacy than a smart strategy.
\end{abstract}


\end{frontmatter}

{\footnotesize

\noindent Supplementary material for this preprint is available as the ancillary file Supplementary\_material.pdf on arXiv.
\par}

\section{Introduction}
\subsection*{Motivation}
\noindent Global green hydrogen production is facing an implementation gap due to early market risks as well as technical and economic challenges \cite{Odenweller2025, Knoebl2026}. The latter, particularly in Europe, result from high investment cost for electrolyzer technologies as well as high electricity purchase prices for renewable energy sources (RES), which are required for green hydrogen production \cite{Omar2025,Brandt2024}.
The technical challenges arise from uncertainties in efficiency and degradation, with the latter becoming increasingly relevant under the fluctuating electrolyzer operation characteristic of RES-based power supply \cite{Grigoriev2020,Wei2019}. Both uncertainties are associated with additional cost, since efficiency directly affects electricity consumption, while degradation reduces efficiency and shortens stack lifetime \cite{Arnold2025}. Thus, these technical uncertainties lead to increased costs and represent additional economic challenges for a proceeding hydrogen market ramp-up \cite{Ffe2025}. The lowest technical uncertainties are currently associated with alkaline water electrolysis (AWE) and proton exchange membrane water electrolysis (PEMWE), which have the highest levels of technological readiness \cite{IEA2025}. Both technologies are attributed with different technical and economic properties. For instance, AWE is generally assigned with lower investment cost and lower efficiency in the literature, whereas PEMWE is considered to be more expensive yet more efficient \cite{Hu2025}. Therefore, an advantage could result from the combination of both electrolyzer technologies to benefit from the mentioned techno-economic differences, thus addressing technical and economic challenges \cite{Tan2025}.

\subsection*{Literature review}
\noindent The techno-economic differences of AWE and PEMWE mentioned above form the basis of various studies that compare both technologies. While Guo et al. \cite{Guo2019} perform a review based comparison, Wang et al. \cite{Wang2023} conducted an experimental comparison and Munther et al. \cite{Munther2025} simulated both technologies on a large scale. In all three studies, investment cost and efficiency are assumed to be contrary for AWE and PEMWE as indicated above. This contrary difference motivated several studies to examine both technologies within a hybrid electrolyzer system (HES), enabling the aforementioned potential techno-economic advantages to be investigated. The objective of several of these studies is to optimize dispatch, while assuming a fixed, predefined design of the respective HES. As a result, Tan et al. \cite{Tan2025}, Zhang et al. \cite{Zhang2024}, Wang et al. \cite{Wang2026} and Yang et al. \cite{Yang2025} achieved a higher efficiency in hydrogen production. Additionally, Tan et al. \cite{Tan2025} reached higher revenues, which Yu et al. \cite{Yu2024} did as well. Moreover, Yu et al. \cite{Yu2024} found a longer lifetime of the HES compared to the single AWE, respectively, PEMWE system. These techno-economic advantages have also been identified in other studies that specifically focus on configuring the design of HES. Using various optimization approaches, Hu et al. \cite{Hu2025}, Xu et al. \cite{Xu2024}, Shi et al. \cite{Shi2026} and Tang et al. \cite{Tang2025} come to the same conclusion, in which the HES configuration contains a smaller proportion of PEMWE than AWE. Similar results are found in Liang et al. \cite{Liang2024} and Mingxuan et al. \cite{Mingxuan2024}, which analyze various AWE and PEMWE ratio configurations based on scenario analyses. A comparable approach is used in Shin et al. \cite{Shin2026}; however, the most cost-effective configuration identified in this study consists solely of PEMWE. This is also consistent with one finding in Ma et al. \cite{Ma2025}, where different optimization methods were compared. \newline

\noindent This clarifies that the results of studies on HES vary significantly depending on the underlying model assumptions and that the selection of a HES does not necessarily correspond to the most techno-economic efficient solution. Nevertheless, although the topic of HES has been widely studied, only Hu et al. \cite{Hu2025} and Shin et al. \cite{Shin2026} conducted a compact sensitivity analysis of investment cost to verify their findings. Furthermore, current manufacturer data sheets and literature references indicate that the contrary ratio between efficiency and investment cost cannot be clearly attributed to the technologies AWE and PEMWE \cite{GreenHydrogenSystems,QuestOne,HydrogenPro,NelAWE,NelPEMWE,Accelera,ShivaKumar2022,Sayed-Ahmed2024,Sezer2025}. Thus, this study applies a comprehensive sensitivity analysis varying both investment cost and efficiency to prove the profitability of HES independent of uncertain contrary techno-economic properties of AWE and PEMWE. In addition, this study applies a free design optimization approach that allows zero capacities and does not enforce the implementation of a HES. This technology-independent and design-open approach enables a direct evaluation of whether HES represent a smart strategy or rather an economic fallacy with negligible techno-economic benefits.

\subsection*{Aim and contribution}
\noindent To answer this research question, we proceed as follows:
Design and dispatch of an integrated hydrogen supply chain including Power Purchase Agreement (PPA) contracting, two electrolyzers, hydrogen cavern storage and hydrogen demand are optimized by a cost minimization. The methodology is presented in Section \ref{Sec: System_under_consideration}. An extensive sensitivity analysis is conducted to comprehensively examine the model including the techno-economic feasibility of HES. The underlying approach is described in the study design in Section \ref{Sec:Study_design}, along with a guide through the result sections. Section \ref{Sec:Exemplary HES result} initially presents a specific exemplary HES result to support the understanding of the model behavior. Section \ref{Sec: System design and cost} contains the key results for the entire sensitivity analysis space comprising design optima as well as cost objective for the comprehensive techno-economic investigation of HES. Also based on the sensitivity analysis methodology conducted in this study, this section contains a quantitative deep dive into the trade-off between investment cost and efficiency of a single electrolyzer system to provide a cost-based ratio of both parameters. Section \ref{Sec: Dispatch results} completes the results by presenting the dispatch of single electrolyzer systems and HES, highlighting their differing operational behavior and indicating how electrolyzers are optimally utilized under varying parameter conditions. \newpage

\section{System under consideration}\label{Sec: System_under_consideration}

\begin{figure}[H]
    \centering \includegraphics[width=0.65\textwidth]
    {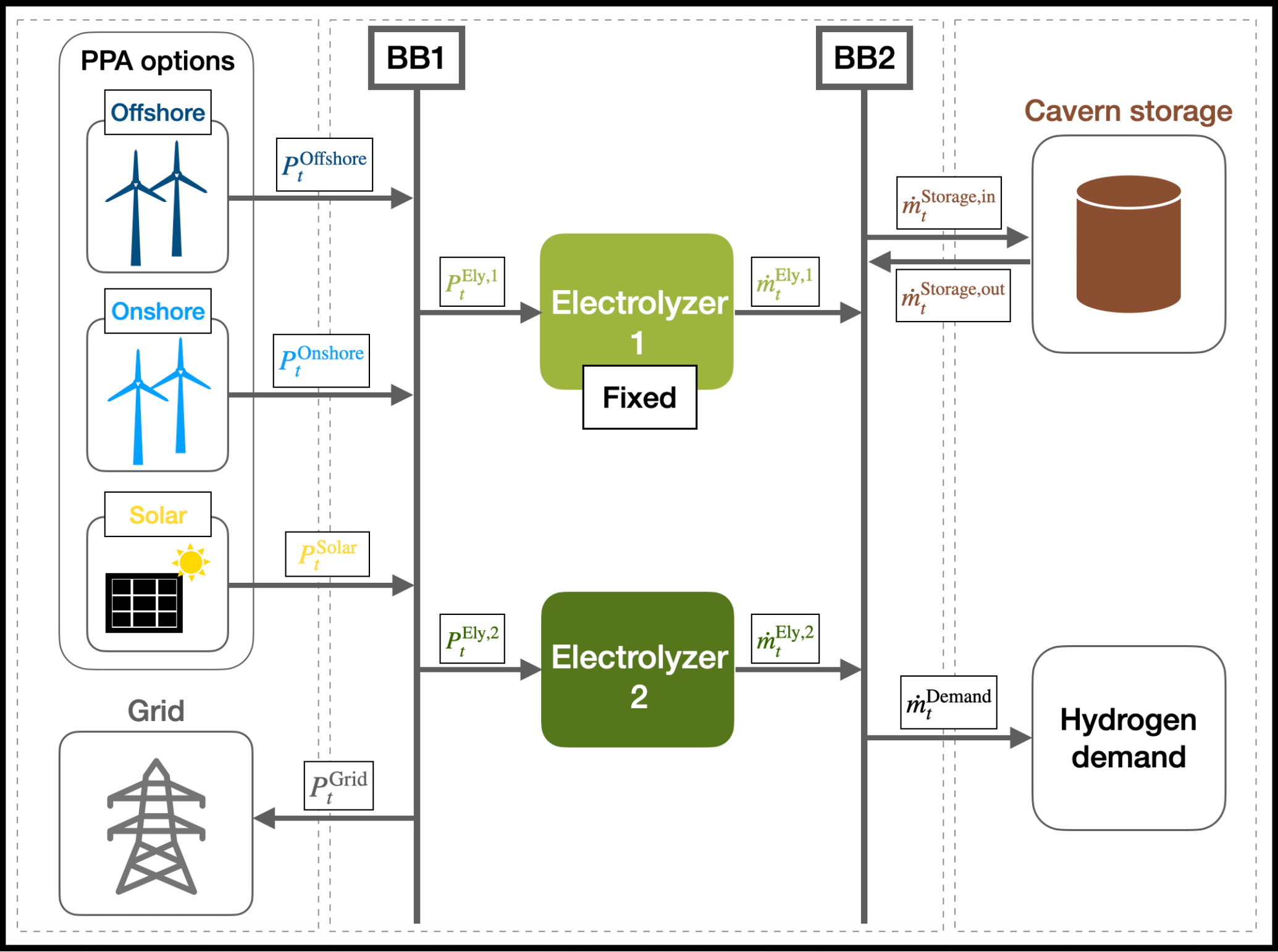}
    \caption{Set up of the system under consideration. The dashed box on the left shows the available power purchase options including onshore wind, offshore wind and solar energy as well as the electricity grid. The middle dashed box presents two electrolyzer options. The right dashed box shows the hydrogen cavern storage as well as a consumer characterized by a certain hydrogen demand. BB1 marks the electricity bus bar, BB2 marks the hydrogen bus bar.}
    \label{Fig: System under consideration}
\end{figure}

\noindent This study takes the perspective of a hydrogen provider that operates two electrolyzers integrated into a hydrogen supply chain, which is shown in Figure \ref{Fig: System under consideration}. On the left, the renewable PPA options including onshore wind, offshore wind and solar power are shown. In addition, the grid as an option for the surplus electricity sales is shown on this side. In the middle, the electrolyzers are presented, which produce green hydrogen using the booked PPA capacities. The produced hydrogen can then be injected into the hydrogen cavern storage or delivered directly to the customer having a predefined hydrogen demand. Both last-mentioned components of the hydrogen supply chain are presented on the right. In the following analysis, design and dispatch of this infrastructure are optimized according to Equation \eqref{Eq.: objective function}. The corresponding optimization variables, grouped in vector $X$, include the design of PPA options, electrolyzers and cavern storage, as well as the dispatch of surplus sales, electrolyzer power and cavern storage utilization. These variables are used to mathematically minimize total expenditures, which comprises PPA contracting costs ($C^{\text{PPA}}$), cavern storage booking and operating costs ($C^{\text{Storage}}$) and the costs of $N\in\{1,2\}$ electrolyzer systems ($\sum_{i=1}^{N}{C_i}^{\text{Electrolyzer}}$), minus surplus sales revenues ($R^\text{Surplus}$).

\begin{align}\label{Eq.: objective function}
    \min_X \quad C^{\text{PPA}} + C^{\text{Storage}} + \sum_{i=1}^N C_i^{\text{Electrolyzer}} - R^{\text{Surplus}}
\end{align}

\noindent The solved optimization problem provides various variables of interest such as PPA and cavern storage sizing as well as the sizing of one or two electrolyzer systems. These variables of interest are used to calculate the levelized cost of hydrogen ($LCOH$), which serves as an indicator for the economic performance of the analyzed infrastructure. For this purpose, the minimized annual costs are divided by the annual sum of the predefined hydrogen demand. Equation \eqref{Eq: LCOH} shows this calculation.

\begin{align}\label{Eq: LCOH}
   LCOH = \frac{C^\text{PPA} + C^\text{Storage}+ \sum_{i=1}^N C_i^{\text{Electrolyzer}}-R^\text{Surplus}}{\sum_{t=1}^T \dot{{m}}^\text{Demand}_{t} \cdot \Delta t}
\end{align}

\noindent The complete optimization problem formulation including all constraints is linear. The corresponding set of constraints is listed in \ref{app1}. The optimization time frame is one year with an hourly resolution, ensuring compliance with European regulations for green hydrogen production \cite{EuropeanCommission2023}. The PPA options are pay-as-produced. To ensure comparability within Europe, this study, using Germany as an example, assumes that no grid fees or taxes are charged for the electrolyzer operation \cite{BMJ_EnWG, BMJ_StromStG}. Regarding the surplus sales, a constant pricing assumption is used. For both electrolyzers, annualized installation cost occur. For simplicity, the electrolyzers are modeled as single stacks, omitting any modular system architecture. The storage fees consist of bundled capacity booking plus operational costs. The hydrogen demand is set to 5500 kg/h, which leads to a total electrolyzer capacity of approximately 500 MW with 5000 full load hours (FLH), depending on the respective parameter assumptions. The constant hydrogen demand assumption is representative of large-scale industrial applications, such as the chemical sector, which is one of the first major consumers of domestically produced green hydrogen in Europe \cite{IRENA2020,RWEAG2025}. Further information on the cost and technical assumptions as well as the underlying capacity factor time series for the PPAs are provided in \ref{app2}. All economic input parameters are set to present cost values derived from recent publications and converted into 2024 euros (€$_{2024}$) using the Chemical Engineering Plant Cost Index (CEPCI) \cite{CEPCI}. 

\section{Study design}\label{Sec:Study_design}

\begin{figure}[H]
    \centering \includegraphics[width=0.55\textwidth]
    {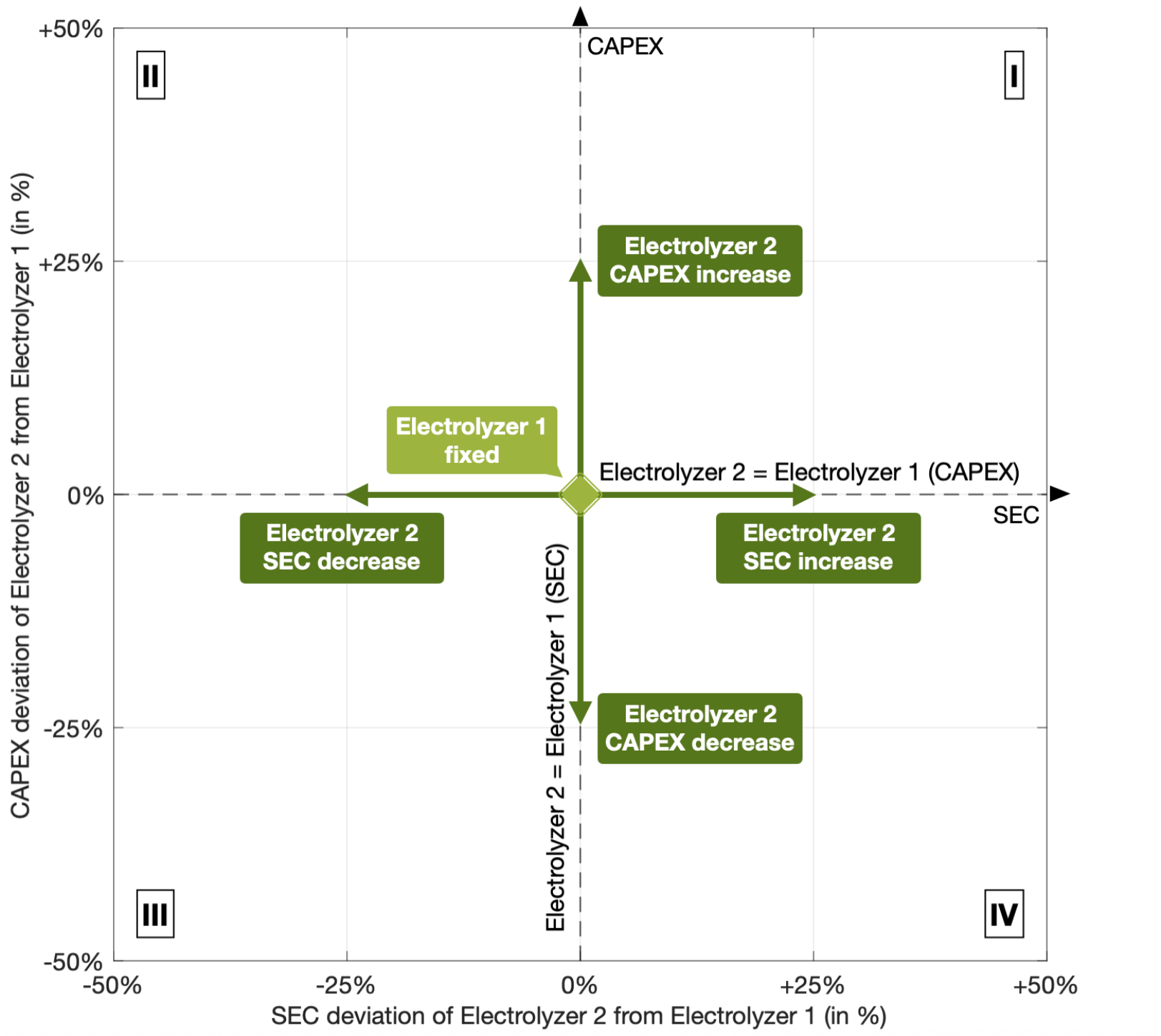}
    \caption{Illustration of CAPEX/SEC parameter variation space. CAPEX deviation of electrolyzer 2 from electrolyzer 1 in $\%$ is shown on the y-axis. SEC deviation of electrolyzer 2 from electrolyzer 1 in $\%$ is shown on the x-axis. The dashed lines mark where electrolyzer 2 has the same CAPEX value respectively SEC value as electrolyzer 1. In the corners, the quadrants of the coordinate system are numbered. The light green square marks the fixed baseline value of electrolyzer 1, the dark green arrows visualize the variation of electrolyzer 2.}
    \label{Fig: Study design}
\end{figure}

\noindent In this study a comprehensive sensitivity analysis is performed that varies both investment cost and efficiency of the two electrolyzer options shown in Figure \ref{Fig: System under consideration} to investigate the research question of HES as a smart strategy or an economic fallacy. The investment cost correspond to the capital expenditures (CAPEX) and the efficiency corresponds to the specific energy consumption (SEC) in the following. CAPEX have a direct impact on the total cost and ar  e measured in €/kW. The SEC's cost influence is indirect, as it defines the energy consumption of the electrolyzer in kWh per kg produced hydrogen. Thus, the higher the SEC, the more energy is needed for hydrogen production and must be procured accordingly via PPAs. Therefore, the SEC's cost impact is reflected in the PPA cost. For the analysis of this study, both CAPEX and SEC are fixed for electrolyzer 1 to baseline values of 1450 €/kW and 52 kWh/kg, respectively. Starting from this baseline, both parameters are varied between a 50\% decrease and a 50\% increase for electrolyzer 2. The resulting parameter variation space is illustrated in Figure \ref{Fig: Study design}. CAPEX deviation of electrolyzer 2 from electrolyzer 1 in \% is shown on the y-axis. SEC deviation of electrolyzer 2 from electrolyzer 1 in \% is shown on the x-axis. The dashed lines indicate where electrolyzer 2 has the same baseline CAPEX and SEC value as electrolyzer 1. In the corners, the quadrants of the coordinate system are numbered. The light green square marks the fixed baseline value of electrolyzer 1, the dark green arrows visualize the variation of electrolyzer 2. For each parameter combination assigned to electrolyzer 1 and electrolyzer 2, the optimization problem described in Section \ref{Sec: System_under_consideration} is solved. \newline

\noindent The following guide explains the structure of the subsequent analysis of the results from the parameter variation space. First, a specific, exemplary HES result is shown and discussed regarding cost, design and dispatch in Section \ref{Sec:Exemplary HES result}. This ensures a detailed insight and understanding of the modeled hydrogen supply chain. Based on this knowledge, the entire parameter variation space is examined in the next step. In Section \ref{Sec: System design and cost}, the key results regarding design and cost are investigated first. This involves determining the resulting designs of all components of the considered hydrogen supply chain as well as the cost objective $LCOH$. This analysis is complemented by an examination of the results regarding the sensitivities energy price, storage fee, availability of RES and baseline SEC value. Independent of HES, this section provides a deeper analysis of the cost ratio between CAPEX and SEC. The same methodology and parameter variation space are considered for this purpose, but with focus on a single electrolyzer system. The result is further evaluated with respect to the sensitivities energy price, storage fee, availability of RES and baseline SEC value as well. Finally, beyond evaluating system design and cost, a detailed analysis of the electrolyzer operation is conducted in Section \ref{Sec: Dispatch results} to understand the optimal dispatch behavior of HES and single electrolyzer systems. The identified operational behaviors provide the basis for deriving fundamental techno-economic efficient operating mechanisms.

\section{Results and discussion}\label{Sec:Results_and_discussion}

\subsection{Exemplary hybrid electrolyzer system}\label{Sec:Exemplary HES result}
\begin{figure}[H]
  \centering
  \begin{minipage}[t]{0.46\textwidth}
    \centering
    \begin{subfigure}[t]{\textwidth}
    \includegraphics[width=\linewidth]{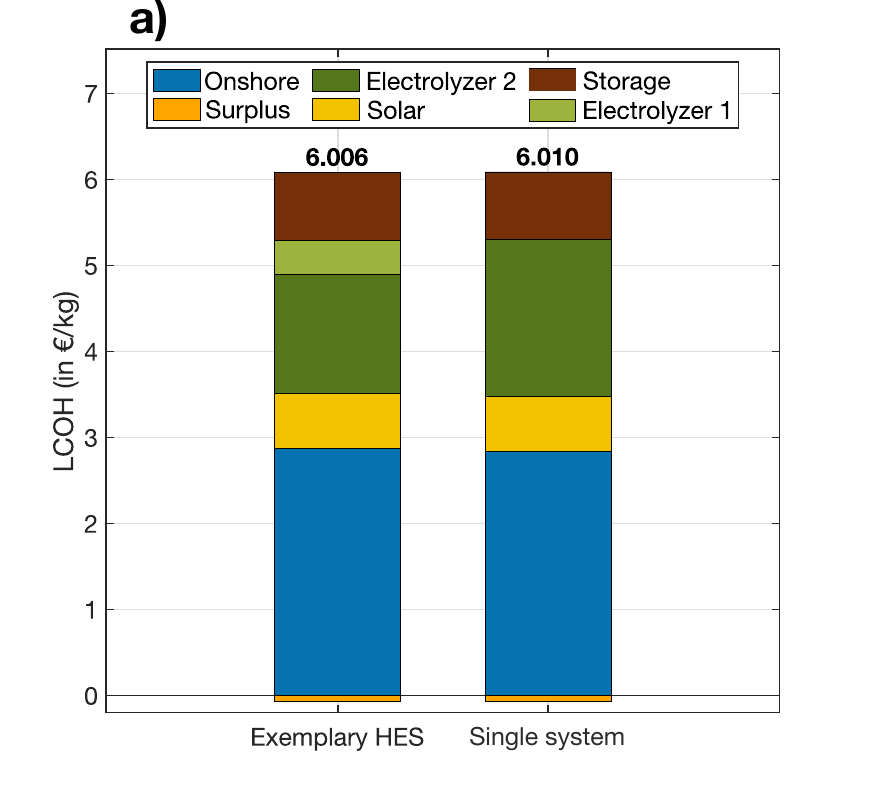}
    \end{subfigure}
    \begin{subfigure}[t]{\textwidth}
      \includegraphics[width=\linewidth]{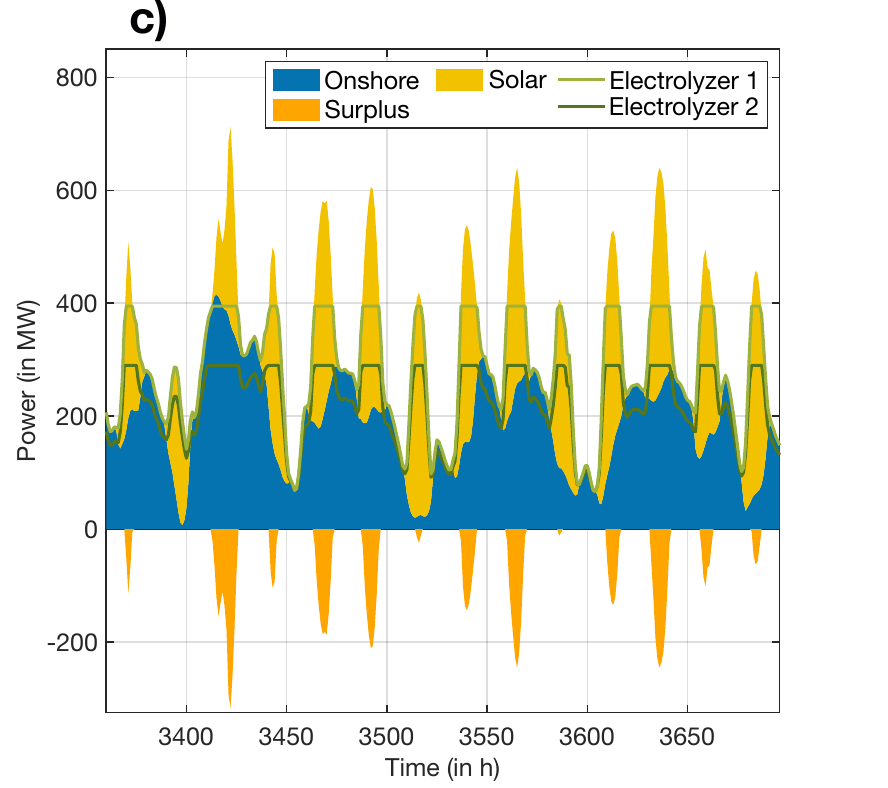}
    \end{subfigure}
    \begin{subfigure}[t]{\textwidth}
      \includegraphics[width=\linewidth]{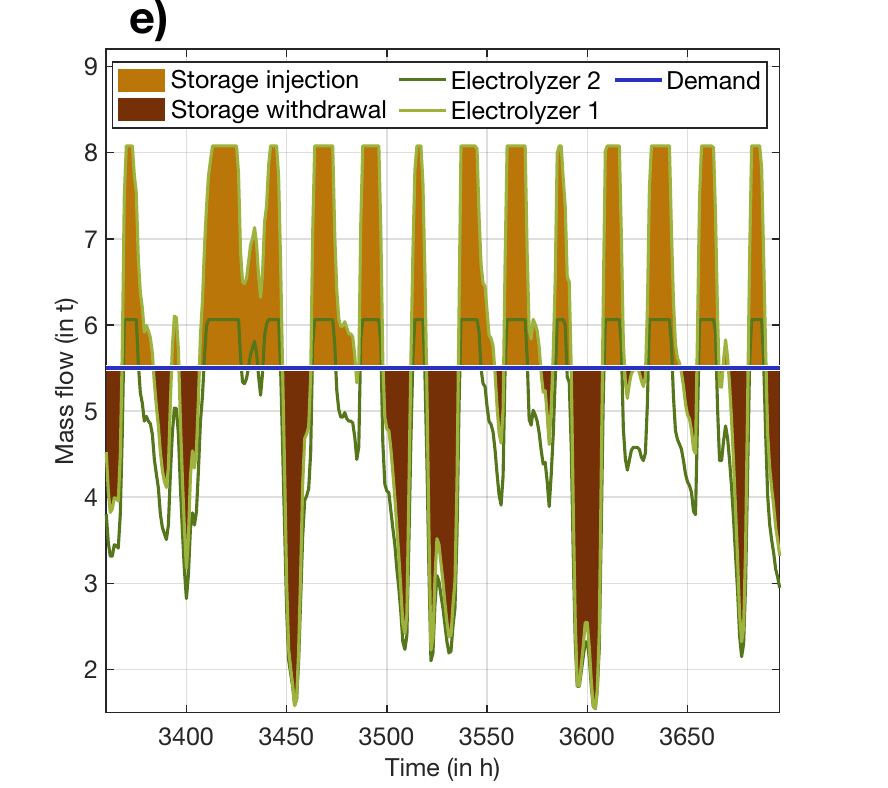}
    \end{subfigure}
  \end{minipage}
  \hfill
  \begin{minipage}[t]{0.46\textwidth}
    \centering
    \begin{subfigure}[t]{\textwidth}
      \includegraphics[width=\linewidth]{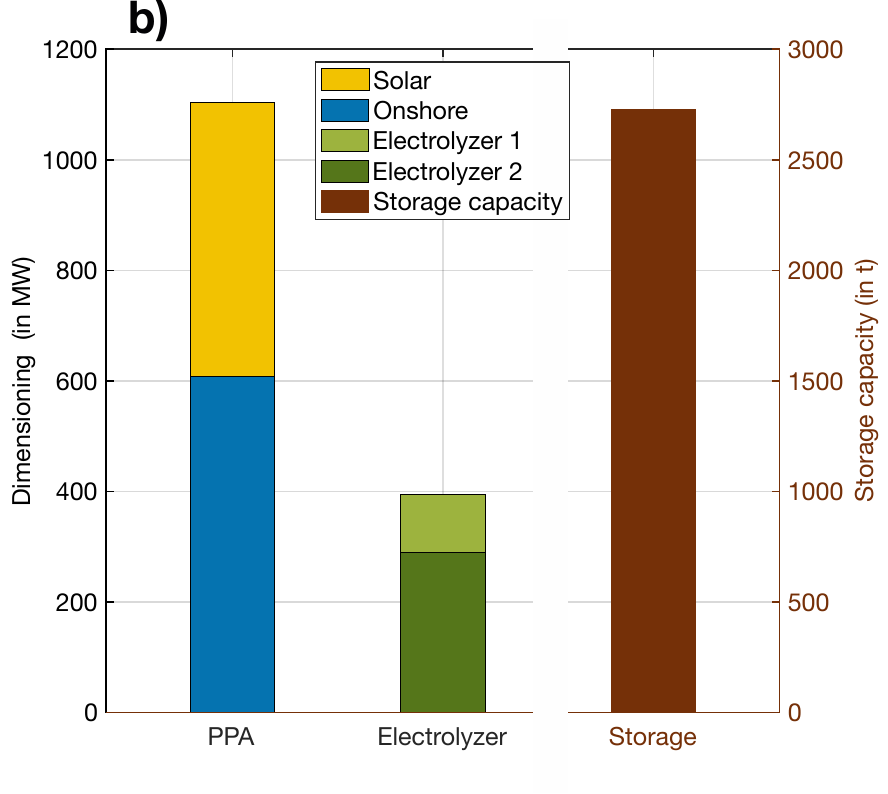}
    \end{subfigure}
    \begin{subfigure}[t]{\textwidth}
      \includegraphics[width=\linewidth]{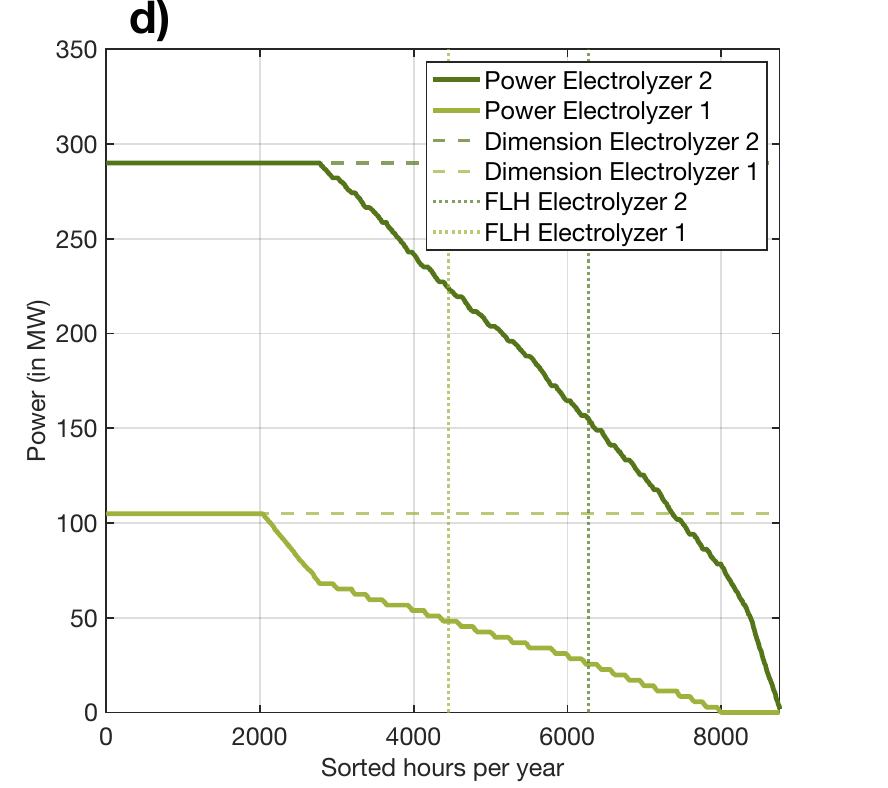}
    \end{subfigure}
    \begin{subfigure}[t]{\textwidth}
      \includegraphics[width=\linewidth]{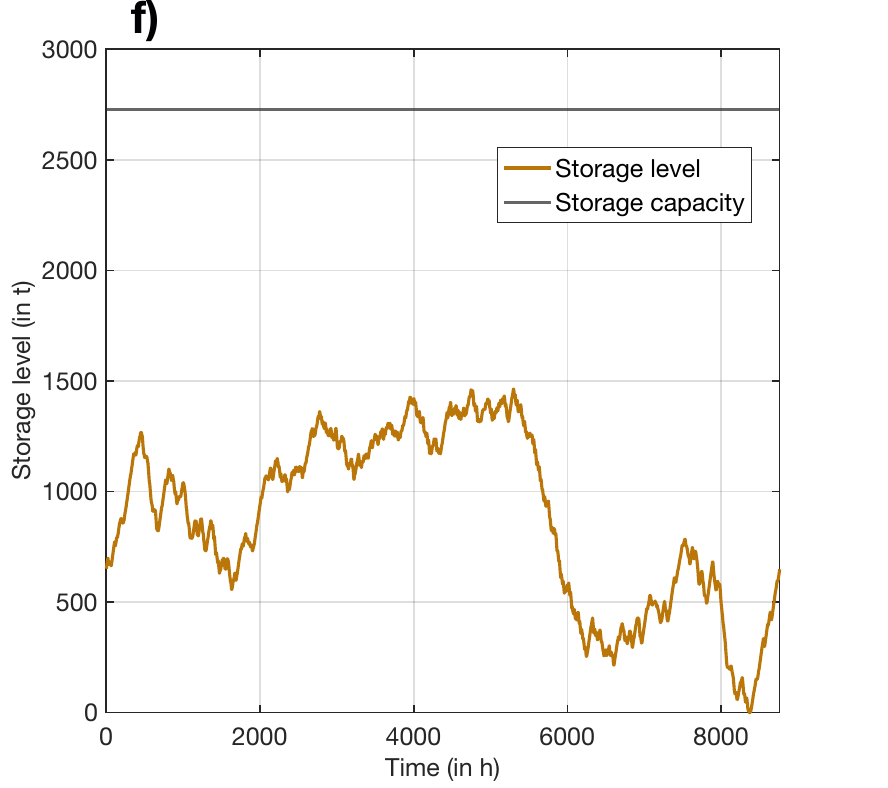}
    \end{subfigure}
  \end{minipage}
  \caption{Results of exemplary HES. a) $LCOH$ of exemplary HES and single electrolyzer system. b) Dimensioning of the components of the system under consideration: PPA and electrolyzers in MW and storage in tonnes (t). c) Hourly  power flow in MW for a two week summer period. d) Annual power duration curves for electrolyzer 1 and electrolyzer 2. e) Hourly hydrogen mass flow for a two week summer period. f) Storage level and storage capacity in tonnes of hydrogen depending on the hours of one year.}
  \label{Fig: Exemplary HES}
\end{figure}

\vspace{0.5em}
\noindent This section presents and discusses an exemplary HES case to provide a detailed understanding of the modeled hydrogen supply chain described in Section \ref{Sec:Study_design}. A representative data point located in quadrant II of the CAPEX/SEC parameter variation space shown in Figure \ref{Fig: Study design} is selected, resulting in a HES. Figure \ref{Fig: Exemplary HES} presents the optimized cost, design and dispatch of the exemplary HES. Figure \ref{Fig: Exemplary HES} a) shows $LCOH$ for the exemplary HES and the next best single electrolyzer system with respect to cost in comparison, Figure \ref{Fig: Exemplary HES} b) shows the designs of the exemplary HES infrastructure, Figure \ref{Fig: Exemplary HES} c) shows the power flows, Figure \ref{Fig: Exemplary HES} d) shows the electrolyzer operation, Figure \ref{Fig: Exemplary HES} e) shows the hydrogen mass flows and Figure \ref{Fig: Exemplary HES} f) shows the storage operation and dimension. In every subfigure, onshore wind is presented in blue, solar power is presented in yellow, surplus power is presented in orange, electrolyzer 1 is presented in light green, electrolyzer 2 is presented in dark green and storage is presented in brown.\newline

\noindent The $LCOH$ in €/kgH2 are shown in Figure \ref{Fig: Exemplary HES} a). The different cost components are stacked and visualized in different colors for the exemplary HES on the left and for the next best single electrolyzer system on the right. The total $LCOH$ resulting for the exemplary HES are 6.006 €/kgH2. The biggest $LCOH$ share is given by the PPA cost of 3.5 €/kgH2 followed by the total electrolyzer cost of about 1.7 €/kgH2. The storage cost are approximately 0.8 €/kgH2 and the surplus revenues make up 0.07 €/kgH2. In comparison, the total cost of the single electrolyzer system amounts to 6.010 €/kgH2. Thus, the cost benefit of the exemplary HES over the single electrolyzer system is 0.004 €/kgH2. The higher cost of the single electrolyzer system result from higher total electrolyzer cost. The design of the exemplary HES and the corresponding hydrogen infrastructure are shown in Figure \ref{Fig: Exemplary HES} b). The left y-axis shows the dimension in megawatts (MW) and the right y-axis shows the dimension in tonnes of hydrogen (t). The different component dimensions are type-dependently stacked and visualized in different colors. On the left, the PPA dimensions are presented consisting of onshore wind and solar power with a total dimension of about 1100 MW. The onshore wind dimension is about 600 MW, which is slightly larger than the dimension of 500 MW solar power. In comparison, the resulting total electrolyzer dimension presented in the middle is clearly smaller with in total 400 MW. This indicates a substantial oversizing of booked PPAs compared to electrolyzer dimension, which is typical given the volatile nature of RES. The dimension of electrolyzer 2 is approximately three times larger at 290 MW than the dimension of electrolyzer 1, which is sized at 105 MW. On the right, the resulting storage capacity of about 2750 t is shown. The power flow is shown in Figure \ref{Fig: Exemplary HES} c), with the power in MW as a function of time in hours over a two-week period in summer. The PPA power and the electrolyzer power are stacked independent of each other. The results show that electrolyzer 1 primarily operates during periods of peak solar power, while electrolyzer 2 operates more continuously. In addition, solar power peaks often exceed the total electrolyzer capacity, resulting in surplus power. Therefore, surplus sales do not appear to affect the electrolyzer dispatch. The electrolyzer operation is shown in Figure \ref{Fig: Exemplary HES} d) in the form of annual power duration curves (APDCs) of both electrolyzers, which present power as a function of sorted hours per year. The operation of electrolyzer 2 is characterized by a continuous, base load behavior, reflected by 6272 FLH. Electrolyzer 1, by contrast, achieves approximately 1817 fewer FLH and predominantly operates during peak-load periods. This underlines the operating behavior of the electrolyzers described in Figure \ref{Fig: Exemplary HES} c). The mass flow is shown in Figure \ref{Fig: Exemplary HES} e) by tonnes of hydrogen as a function of the same two-week period in summer as in Figure \ref{Fig: Exemplary HES} c). The electrolyzer mass flow is stacked. The mass flow shows that the storage injection periods correlate with the solar power periods in Figure \ref{Fig: Exemplary HES} c). This demonstrates the need for storage to balance short-term fluctuations in hydrogen production by RES in order to meet the predefined constant hydrogen demand. 
However, storage is also required to compensate for longer-term fluctuations in RES. For example, solar power generation is typically higher during summer months than during winter. The impact of these long-term fluctuations is illustrated in Figure \ref{Fig: Exemplary HES} f), which presents the storage level and storage capacity in tonnes as a function of the hours of the year. Generally, the storage level increases during summer period and decreases during winter period. This confirms the storage injection correlation with solar power periods observed in Figure \ref{Fig: Exemplary HES} e), as these periods are mostly given in summer. Moreover, Figure \ref{Fig: Exemplary HES} f) demonstrates, that the storage level never reaches the available storage capacity. This is due to the bundled marketing of storage capacity and operation, as mentioned in Section \ref{Sec: System_under_consideration}. Further details on the optimization constraints are provided in \ref{app1}. \newline

\noindent Overall, the detailed analysis of the exemplary HES case shows that the resulting system configuration and operation are consistent with the underlying techno-economic characteristics of the modeled supply chain. With respect to $LCOH$, the difference between the HES and the single electrolyzer system is rather small. Furthermore, regarding the operation of the HES, electrolyzer 2 tends to operate in base load behavior, whereas electrolyzer 1 operates rather in peak load behavior. The subsequent analyses investigate these results across the entire parameter variation space for verification and a more comprehensive understanding of these findings. \newline 

\subsection{System design and cost}\label{Sec: System design and cost}

\begin{figure}[H]
  \centering
  \begin{minipage}[t]{0.45\textwidth}
    \centering
    \begin{subfigure}[t]{\textwidth}
    \includegraphics[width=\linewidth]{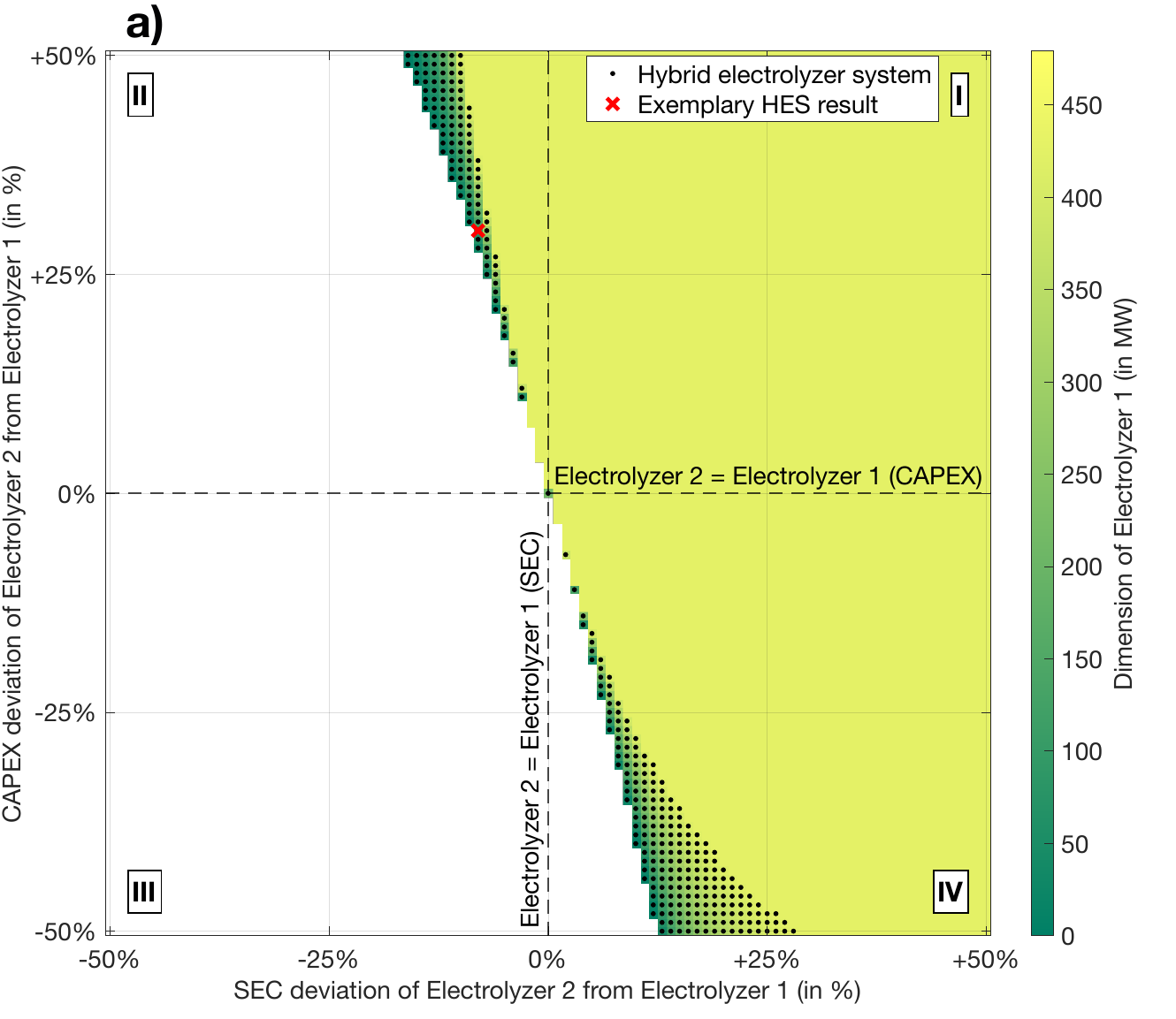}
    \end{subfigure}
    \begin{subfigure}[t]{\textwidth}
      \includegraphics[width=\linewidth]{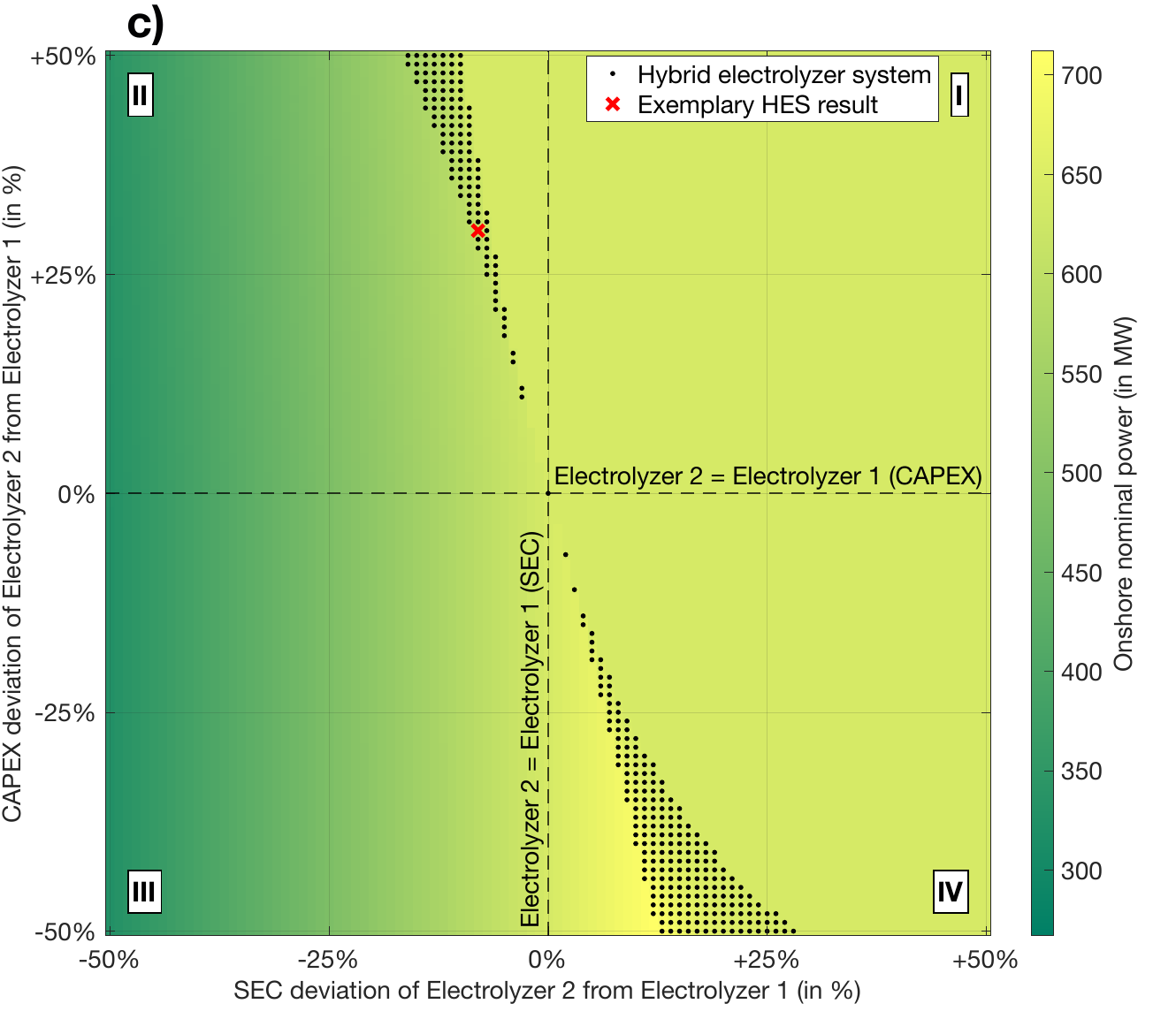}
    \end{subfigure}
    \begin{subfigure}[t]{\textwidth}
      \includegraphics[width=\linewidth]{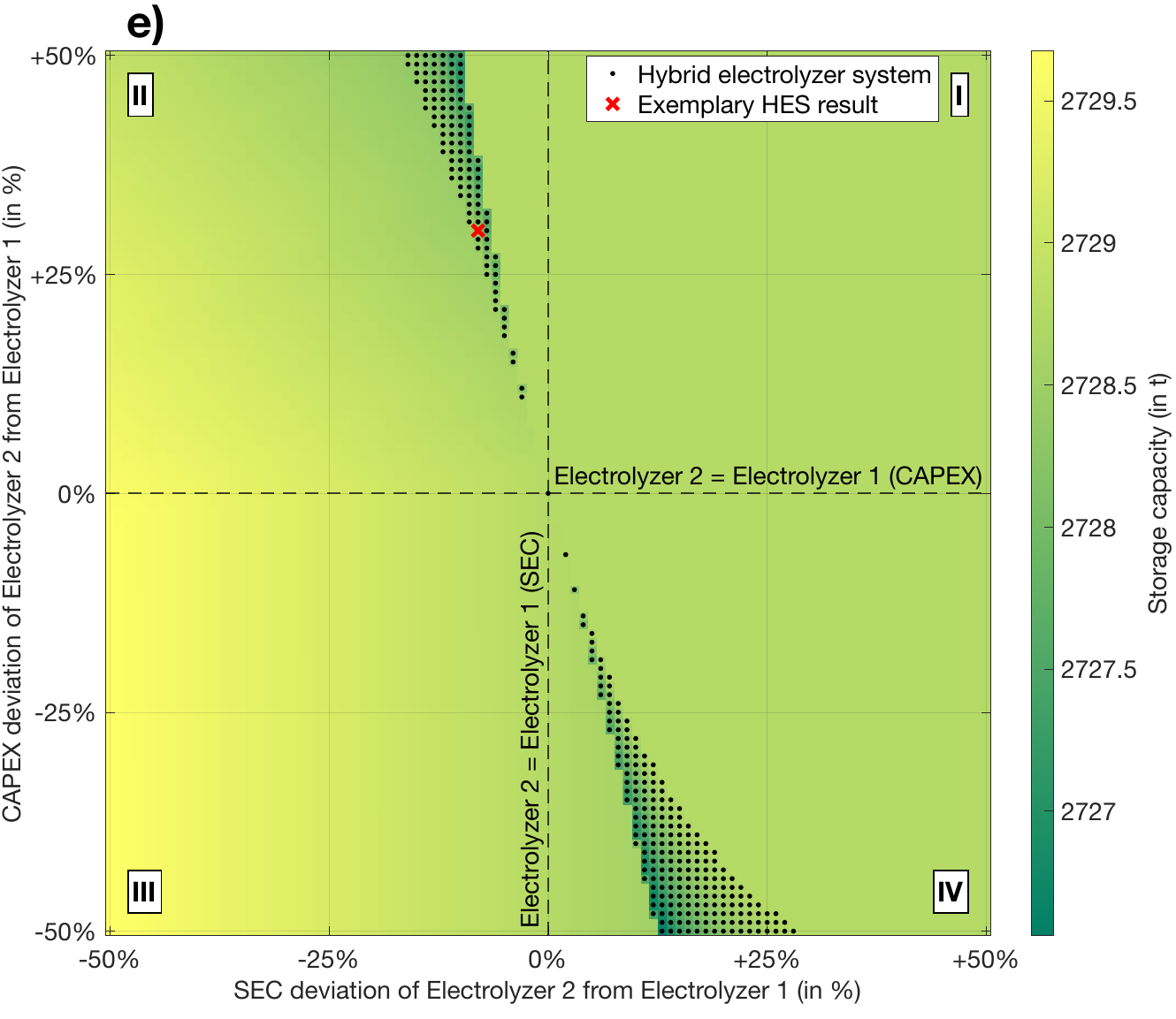}
    \end{subfigure}
  \end{minipage}
  \hfill
  \begin{minipage}[t]{0.45\textwidth}
    \centering
    \begin{subfigure}[t]{\textwidth}
      \includegraphics[width=\linewidth]{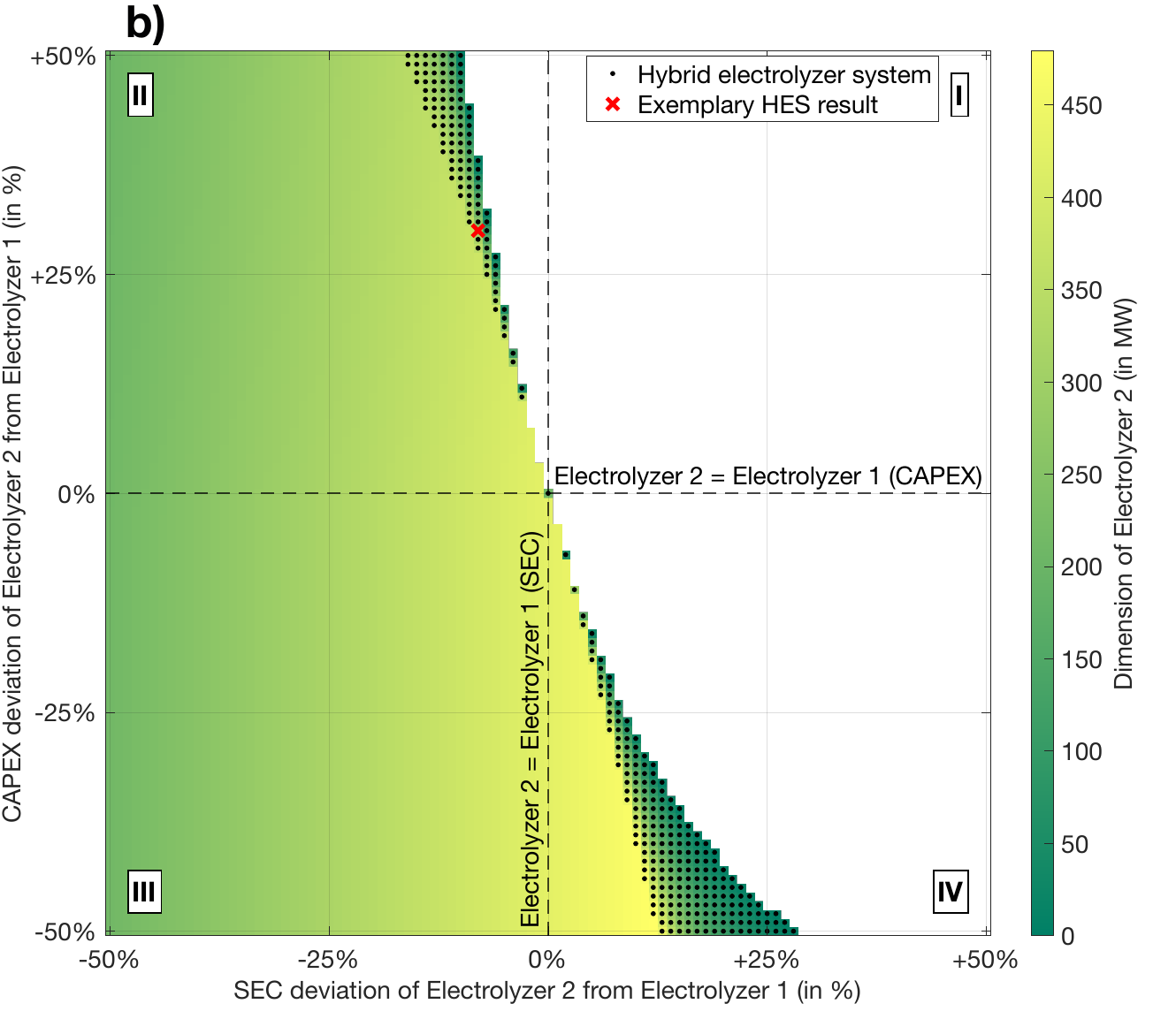}
    \end{subfigure}
    \begin{subfigure}[t]{\textwidth}
      \includegraphics[width=\linewidth]{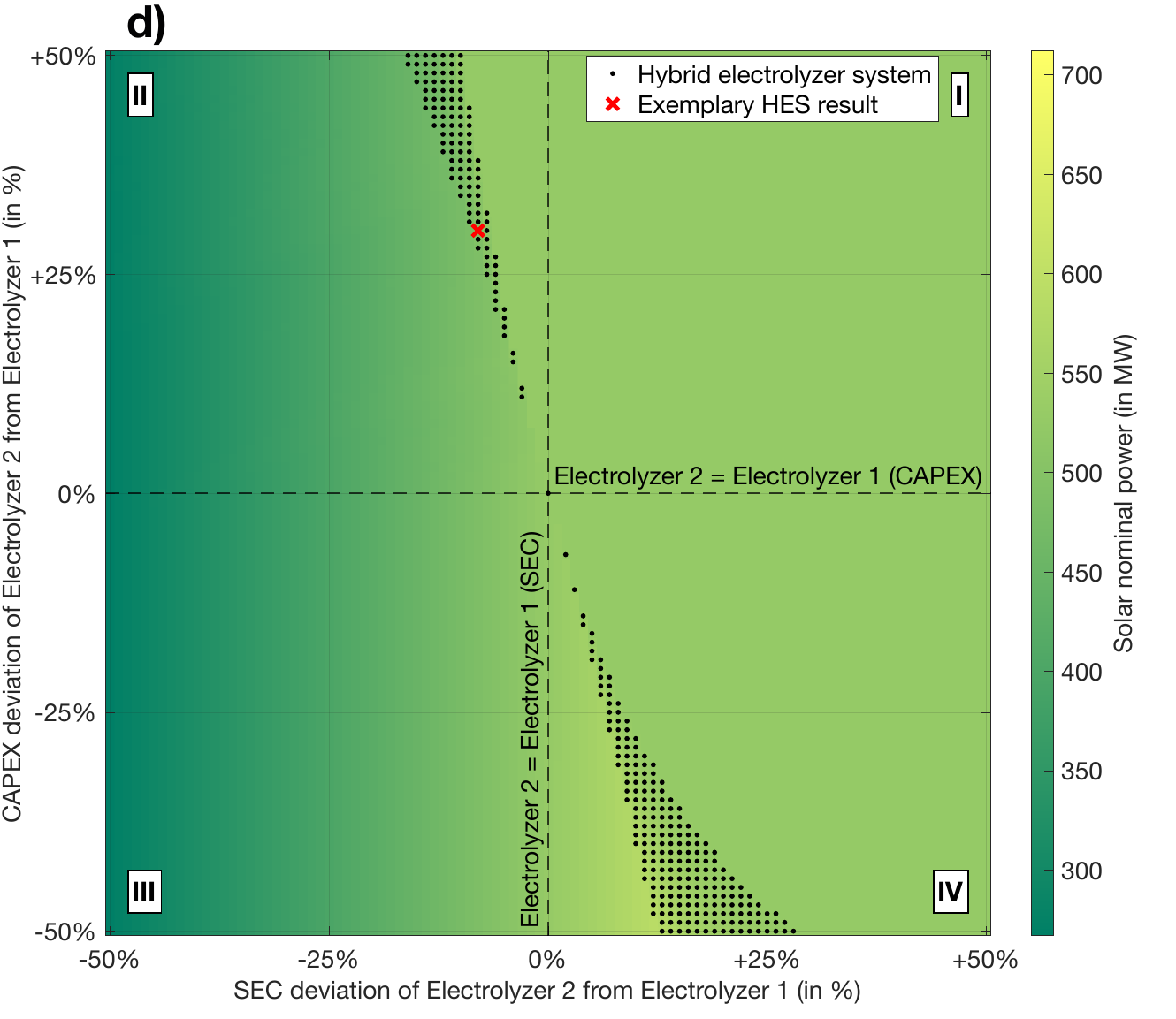}
    \end{subfigure}
  \end{minipage}
  \caption{Heat maps of hydrogen infrastructure component designs for the entire parameter variation space. CAPEX deviation of electrolyzer 2 from electrolyzer 1 in \% is shown on the y-axis. SEC deviation of electrolyzer 2 from electrolyzer 1 in \% is shown on the x-axis. The dashed lines mark where electrolyzer 2 has the same CAPEX, respectively SEC value as electrolyzer 1. CAPEX/SEC parameter combinations leading to HES are marked by black dots. The exemplary HES result is marked by a red cross.  (a) Dimension heat map of electrolyzer 1 in MW. (b) Dimension heat map of electrolyzer 2 in MW. (c) Dimension heat map of onshore wind power in MW. (d) Dimension heat map of solar power in MW. (e) Dimension heat map of storage capacity in t.}
  \label{Fig: Design Heat maps}
\end{figure}

\noindent This section investigates system design and cost for the entire parameter variation space based on the knowledge gained from the exemplary HES analysis in Section \ref{Sec:Exemplary HES result}. Figure \ref{Fig: Design Heat maps} presents system design heat maps illustrated in Figure \ref{Fig: Study design} and described in Section \ref{Sec:Study_design}. Every subfigure \ref{Fig: Design Heat maps} a)-e) shows the CAPEX deviation of electrolyzer 2 from electrolyzer 1 in \% on the y-axis and the SEC deviation of electrolyzer 2 from electrolyzer 1 in \% on the x-axis. The dashed lines mark where electrolyzer 2 has the same CAPEX value respectively SEC value as electrolyzer 1. In the corners, the quadrants of the coordinate system are numbered. Parameter combinations of CAPEX and SEC leading to HES are marked by black dots. The exemplary HES result is marked by a red cross for orientation. The marked HES regions are the same in every subfigure of Figure \ref{Fig: Design Heat maps}. Generally, HES result in quadrant II as well as in quadrant IV. In quadrant II, SEC is decreased and CAPEX is increased for electrolyzer 2. Thus, electrolyzer 2 is more expensive and more efficient in this area than electrolyzer 1. In quadrant IV, SEC is increased and CAPEX is decreased for electrolyzer 2. Thus, electrolyzer 2 is less expensive and less efficient in this area than electrolyzer 1. Consequently, HES result if the parameters of electrolyzer 1 and electrolyzer 2 are opposite. In total, the HES share is 3.3\% of the investigated cases. Electrolyzer 1 as a single system accounts for 47.4\% of the parameter variation space. Electrolyzer 2 as a single system accounts for 49.3\% of the parameter variation space. Thus, in total, shares of electrolyzer 2 are higher than shares of electrolyzer 1. In comparison to single electrolyzer system shares, HES share is minor. \newline

\noindent Figure \ref{Fig: Design Heat maps} a) shows the dimensioning heat map of electrolyzer 1. Parameter combinations resulting in an electrolyzer 1 dimension of zero are visualized in white. The heat map shows a broad plateau in quadrant I, where only electrolyzer 1 is installed, with a constant dimension of about 430 MW. In this quadrant, electrolyzer 1 is more efficient and less expensive than electrolyzer 2, which is why it is cost optimal as a single system. Entering the HES regions in quadrants II and IV, the size of electrolyzer 1 decreases. The decline in electrolyzer 1 size is steeper in quadrant II than in quadrant IV. This indicates that the more efficient electrolyzer 2 dominates the sizing decision in the HES region of quadrant II. In contrast, electrolyzer 1 dominates the sizing decision in the HES region of quadrant IV due to its higher efficiency. Consequently, in a HES configuration, the more efficient electrolyzer is assigned a larger dimension than the less efficient one. This indicates a higher SEC impact compared to CAPEX. Figure \ref{Fig: Design Heat maps} b) shows the dimensioning heat map of electrolyzer 2. Parameter combinations resulting in an electrolyzer 2 dimension of zero are visualized in white. In quadrant III, electrolyzer 2 is installed as a single electrolyzer system because it is both less expensive and more efficient. The size of electrolyzer 2 increases slightly with increasing SEC. The largest size of electrolyzer 2, and the largest electrolyzer size overall, occurs in quadrant IV left next to the HES region, reaching approximately 470 MW. When entering the HES regions, the size of electrolyzer 2 decreases. This trend in quadrant II and quadrant IV is consistent with the behavior observed for electrolyzer 1. This observation further confirms the dominant impact of SEC on electrolyzer sizing. Figures \ref{Fig: Design Heat maps} c) and d) present dimensioning heat maps of the PPA options onshore wind and solar power. In general, solar power dimension is about 100 MW smaller than onshore wind power dimension, which is similar to the finding in Section \ref{Sec:Exemplary HES result}. Both heat maps demonstrate a similar sizing trend, with the largest sizes occurring at low SEC values and further increasing as SEC increases. Thereby, the onshore power size varies more than the solar power size across the entire parameter variation space. Notably, both heat maps exhibit a peak in the same region in quadrant IV where electrolyzer 2 reaches its maximum size, as shown in Figure \ref{Fig: Design Heat maps} b). In this region, the maximum electrolyzer size leads to the highest PPA demand. Figure \ref{Fig: Design Heat maps} e) shows the storage dimensioning in tonnes of hydrogen. Generally, the variation in the storage dimension resulting from variations in CAPEX and SEC is relatively small, with a maximum difference of 3 t, corresponding to only 0.1\% of the maximum storage size of 2729,6 t. This maximum occurs at lowest SEC and CAPEX values, where PPA size is minimal. For increasing SEC and CAPEX values, the storage size decreases within the regions of the single electrolyzer system. The smallest storage dimensions are given in both HES regions. This suggests, that HES configurations can slightly reduce storage requirements, potentially due to the increased operational flexibility provided by the combination of both electrolyzer systems. 

\begin{figure}[H]
\centering

\begin{subfigure}[t]{0.75\textwidth}
    \centering
    \includegraphics[width=\linewidth]{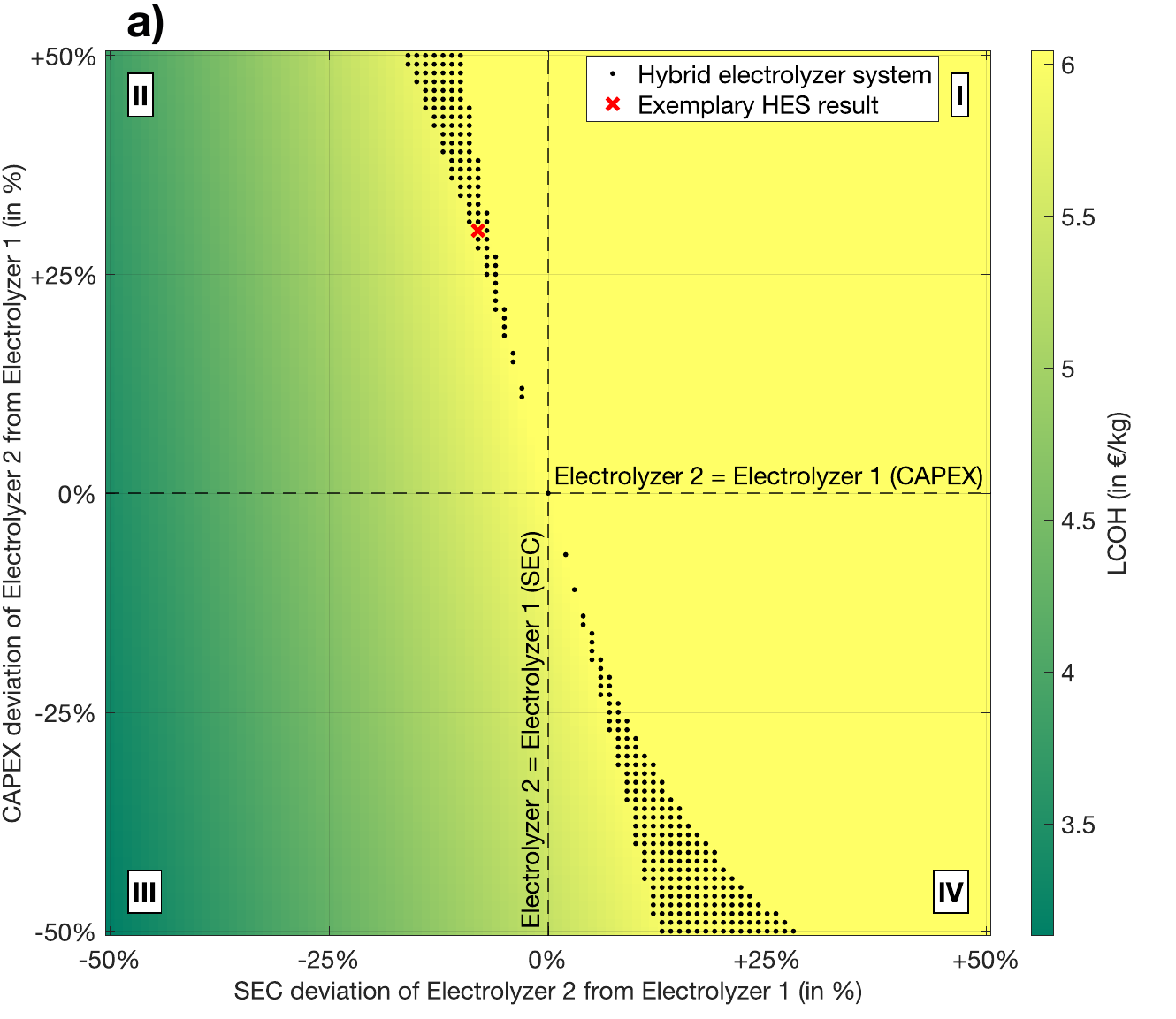}
\end{subfigure}

\vspace{0.8em}

\begin{minipage}{0.75\textwidth}
\centering
\begin{subfigure}[t]{0.45\textwidth}
    \centering
    \includegraphics[width=\linewidth]{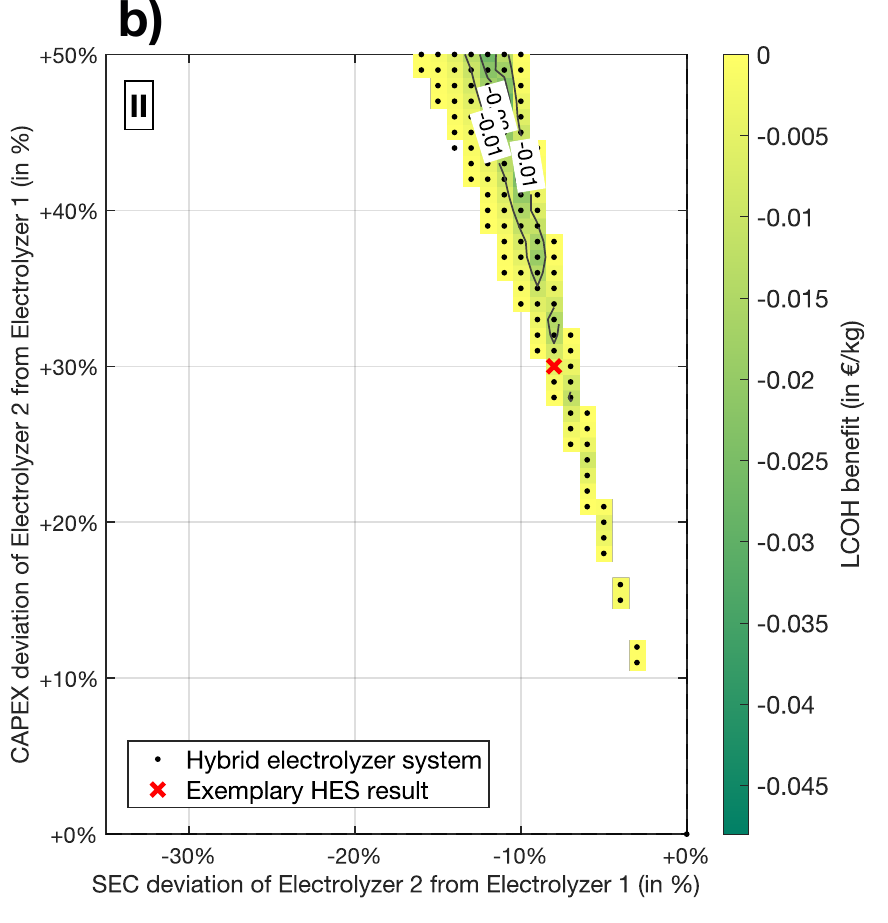}
\end{subfigure}
\begin{subfigure}[t]{0.45\textwidth}
    \centering
    \includegraphics[width=\linewidth]{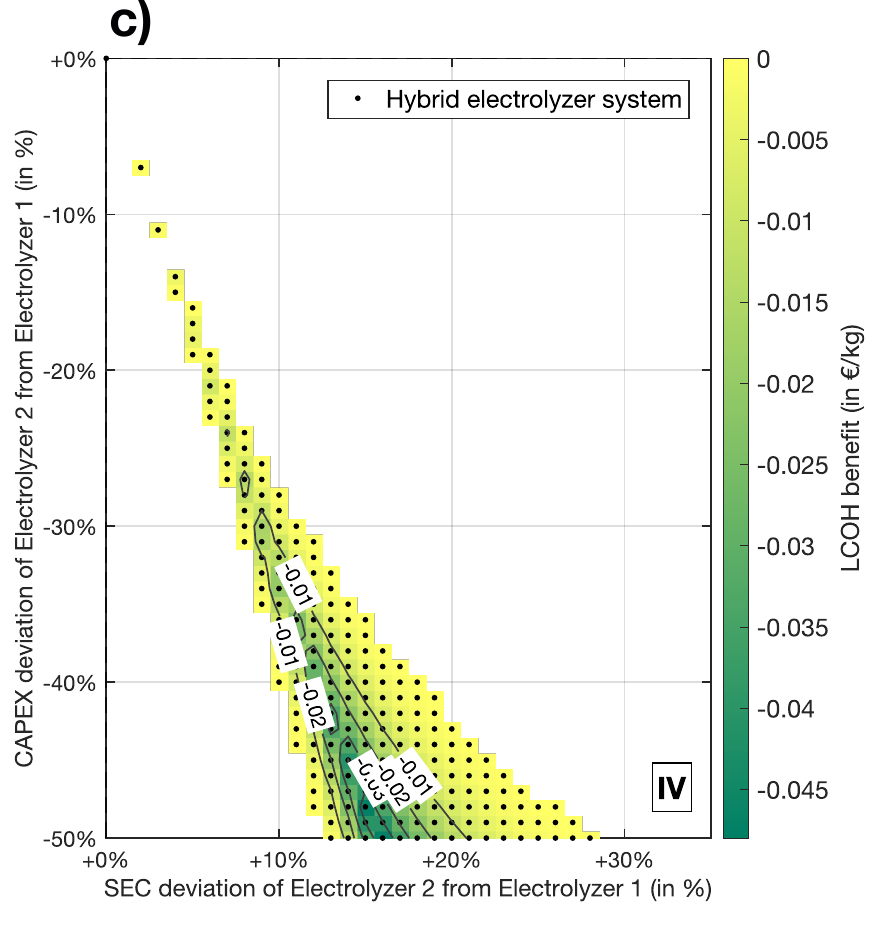}
\end{subfigure}
\end{minipage}
\caption{$LCOH$ heat map and $LCOH$ benefit. CAPEX deviation of electrolyzer 2 from electrolyzer 1 in \% is shown on the y-axis. SEC deviation of electrolyzer 2 from electrolyzer 1 in \% is shown on the x-axis. CAPEX and SEC parameter combinations leading to HES are marked by black dots. The exemplary HES result is marked by a red cross. a) $LCOH$ heat map in €/kgH2. The dashed lines mark where electrolyzer 2 has the same CAPEX value respectively SEC value as electrolyzer 1. b) $LCOH$ benefit of quadrant II in €/kgH2. c) $LCOH$ benefit of quadrant IV in €/kgH2.}
\label{Fig: LCOH}
\end{figure}

\noindent Having discussed the system design, the corresponding cost results of the HES are now analyzed across the entire parameter variation space. Figure \ref{Fig: LCOH}  presents the resulting $LCOH$ introduced in Equation \ref{Eq: LCOH} and the cost benefit of HES. Figure \ref{Fig: LCOH} a) shows the $LCOH$ heat map, equivalent to the presentation of the design heat maps in Figure \ref{Fig: Design Heat maps}. A plateau of the highest $LCOH$ of about 6 €/kgH2 are observed in yellow in the same region in quadrant I as the plateau of the electrolyzer 1 dimension shown in Figure \ref{Fig: Design Heat maps} a). The lowest $LCOH$ of approximately 3.5 €/kgH2 occur for the lowest SEC and CAPEX values in quadrant III. Although increases in both SEC and CAPEX lead to higher $LCOH$, the effect of SEC is considerably more dominant than that of CAPEX. This indicates a generally higher impact of SEC on the economics of the system under consideration. Independent of the specific characteristics observed in the subfigures of Figure \ref{Fig: Design Heat maps} with respect to the dimensioning of the system components, the increase in $LCOH$ remains continuous across the parameter variation space. For example, the region in quadrant IV with the highest overall electrolyzer and PPA dimensions does not correspond to the region with the highest $LCOH$ values. This suggests that no single design characteristic dominates the overall cost performance. The indicates that the combined effects of the different design characteristics tend to offset each other, resulting in the continuous $LCOH$ gradient observed across the parameter space. Figure \ref{Fig: LCOH} b) and c) show the $LCOH$ benefit of the HES relative to the cheapest single electrolyzer system. The $LCOH$ benefit is calculated by comparing the $LCOH$ of each HES configuration with those of both single electrolyzer systems and selecting the minimum difference. Negative values indicate a cost benefit of the HES, meaning that the hybrid configuration achieves lower hydrogen production costs than the cheapest single electrolyzer system. Figure \ref{Fig: LCOH} b) presents the HES region in quadrant II. The maximum $LCOH$ benefit of 0.02 €/kgH2 occurs for CAPEX values of +45\% and above. Figure \ref{Fig: LCOH} c) presents the HES region in quadrant IV. The maximum $LCOH$ benefit of 0.048 €/kgH2 occurs for CAPEX values of -47\% and below. Both CAPEX values represent outliers with regard to recent literature references.  \newline

\begin{figure}[H]
\centering
\begin{subfigure}[t]{0.19\textwidth}
    \includegraphics[width=\linewidth]{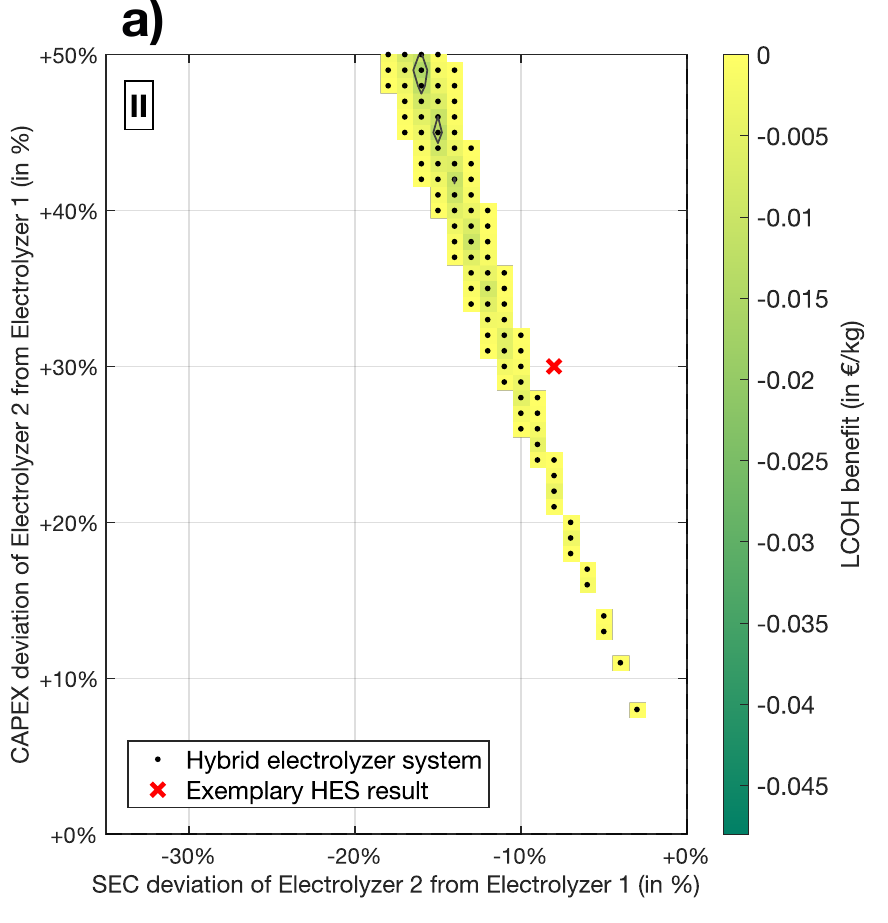}
\end{subfigure}
\hfill
\begin{subfigure}[t]{0.19\textwidth}
    \includegraphics[width=\linewidth]{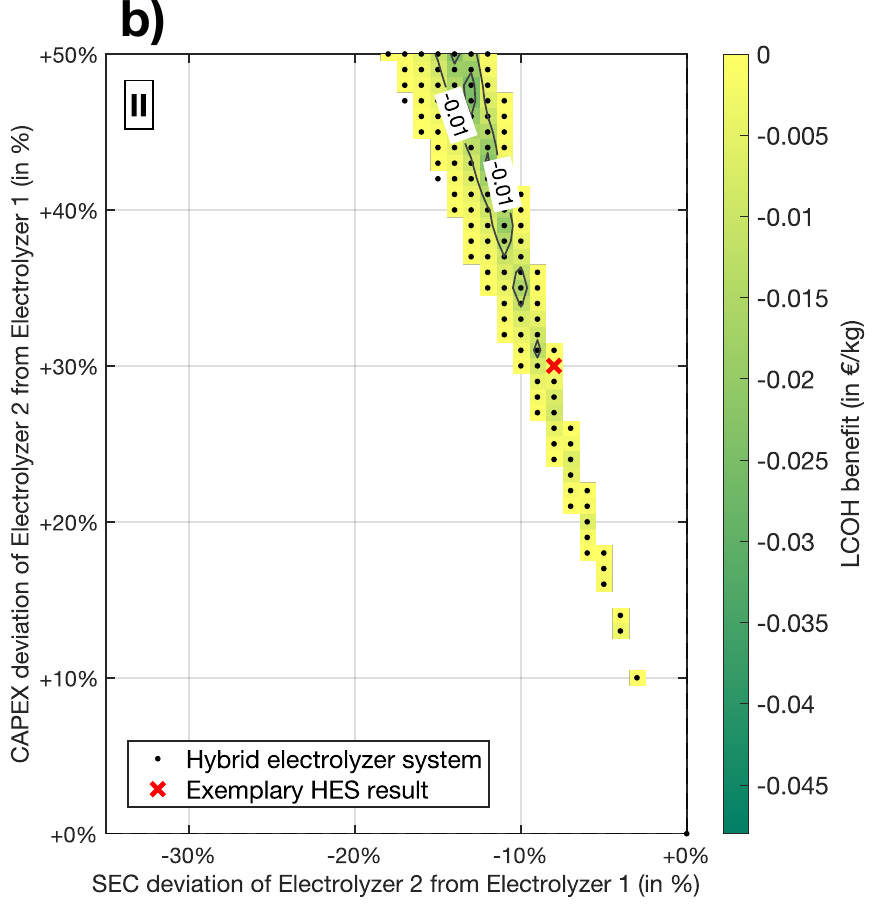}
\end{subfigure}
\hfill
\begin{subfigure}[t]{0.19\textwidth}
    \includegraphics[width=\linewidth]{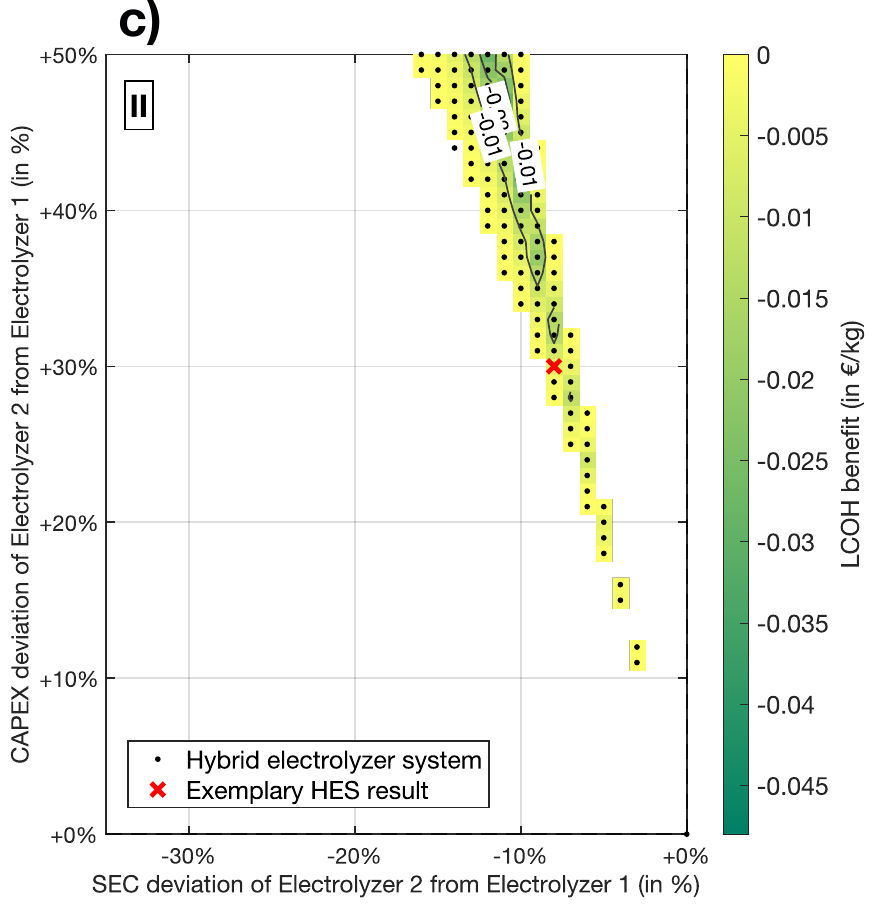}
\end{subfigure}
\hfill
\begin{subfigure}[t]{0.19\textwidth}
    \includegraphics[width=\linewidth]{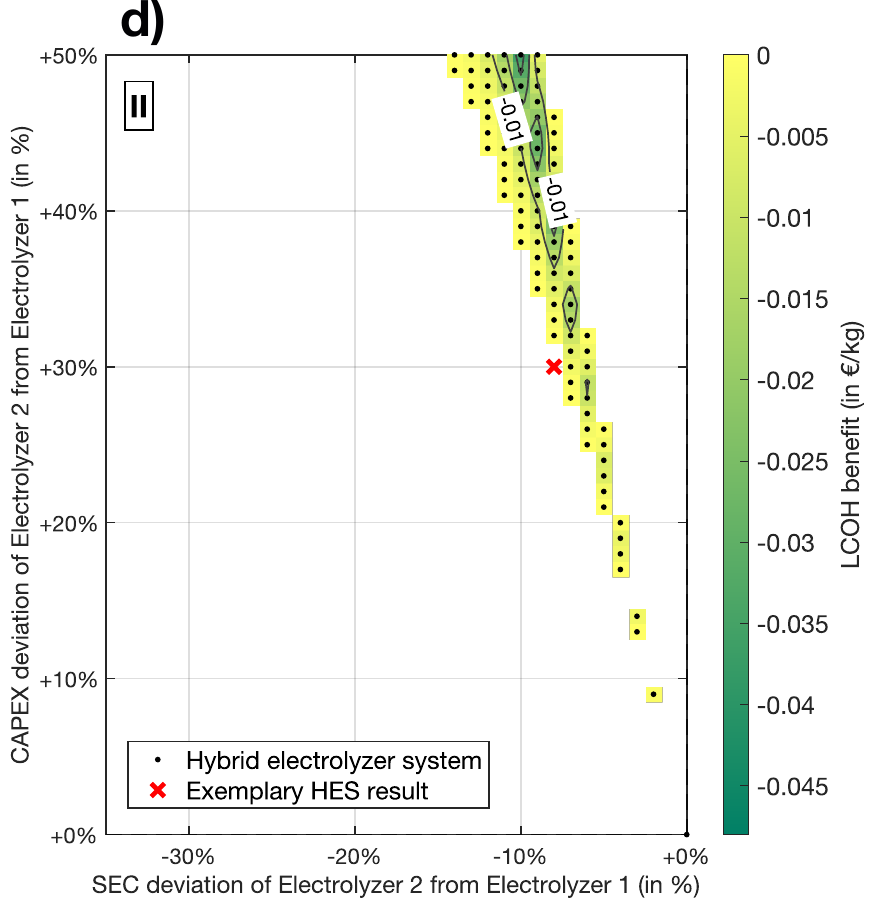}
\end{subfigure}
\hfill
\begin{subfigure}[t]{0.19\textwidth}
    \includegraphics[width=\linewidth]{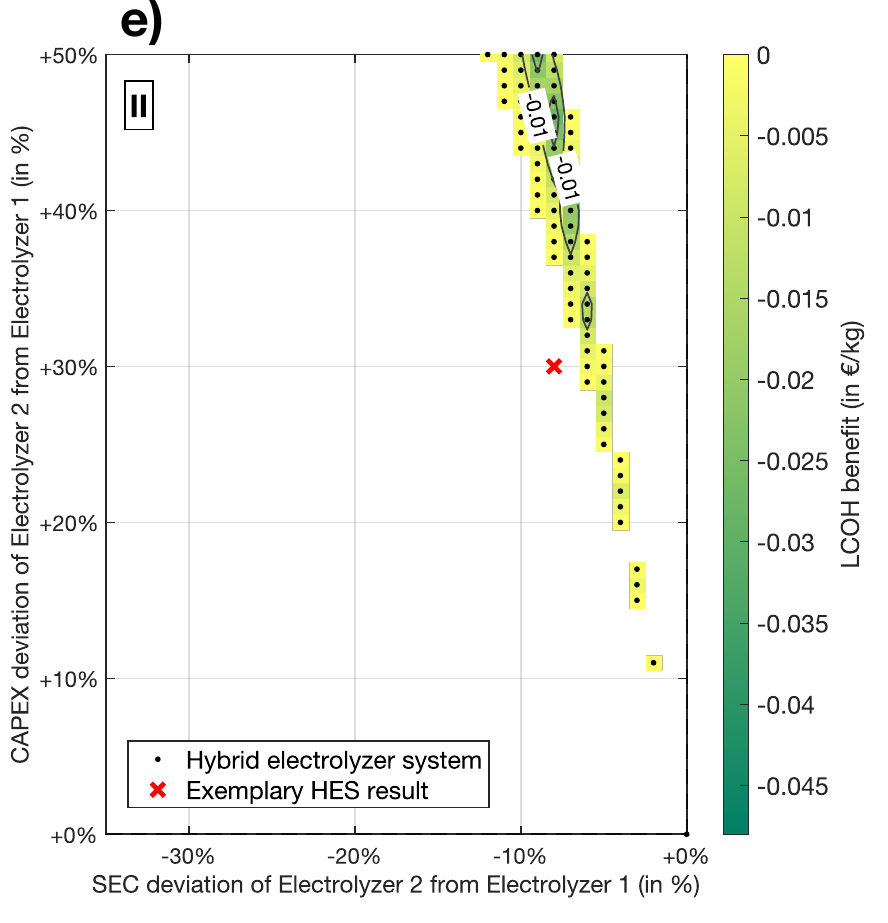}
\end{subfigure}
\vspace{0.5em}
\begin{subfigure}[t]{0.19\textwidth}
    \includegraphics[width=\linewidth]{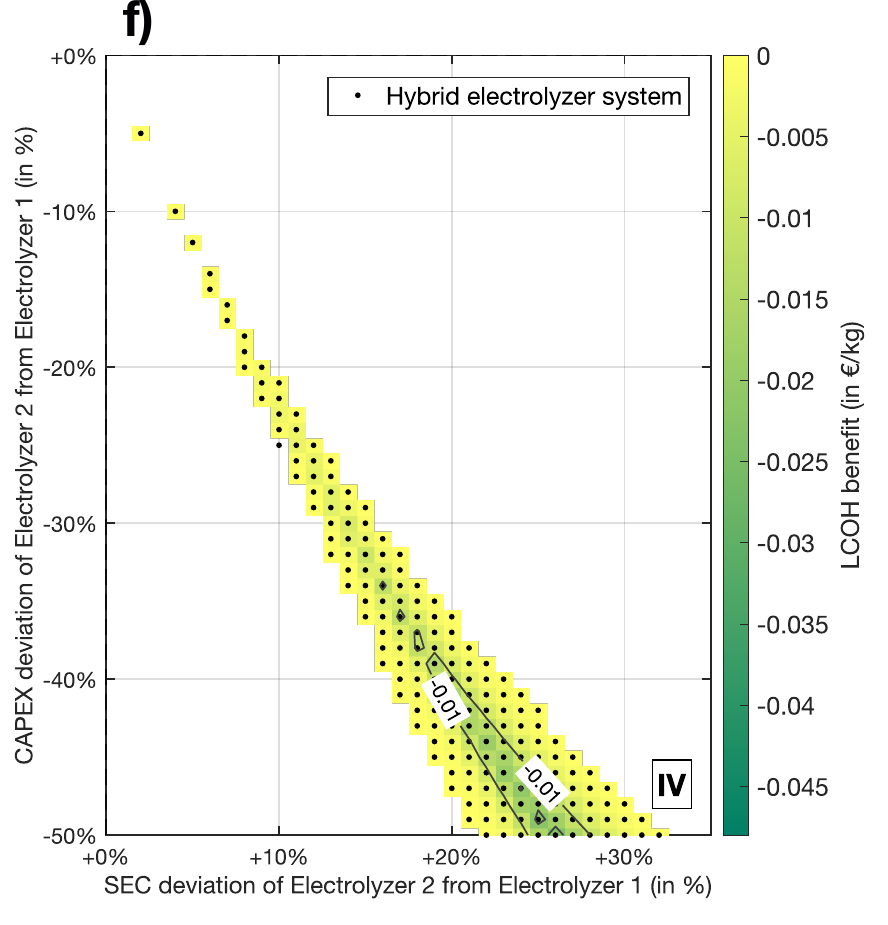}
\end{subfigure}
\hfill
\begin{subfigure}[t]{0.19\textwidth}
    \includegraphics[width=\linewidth]{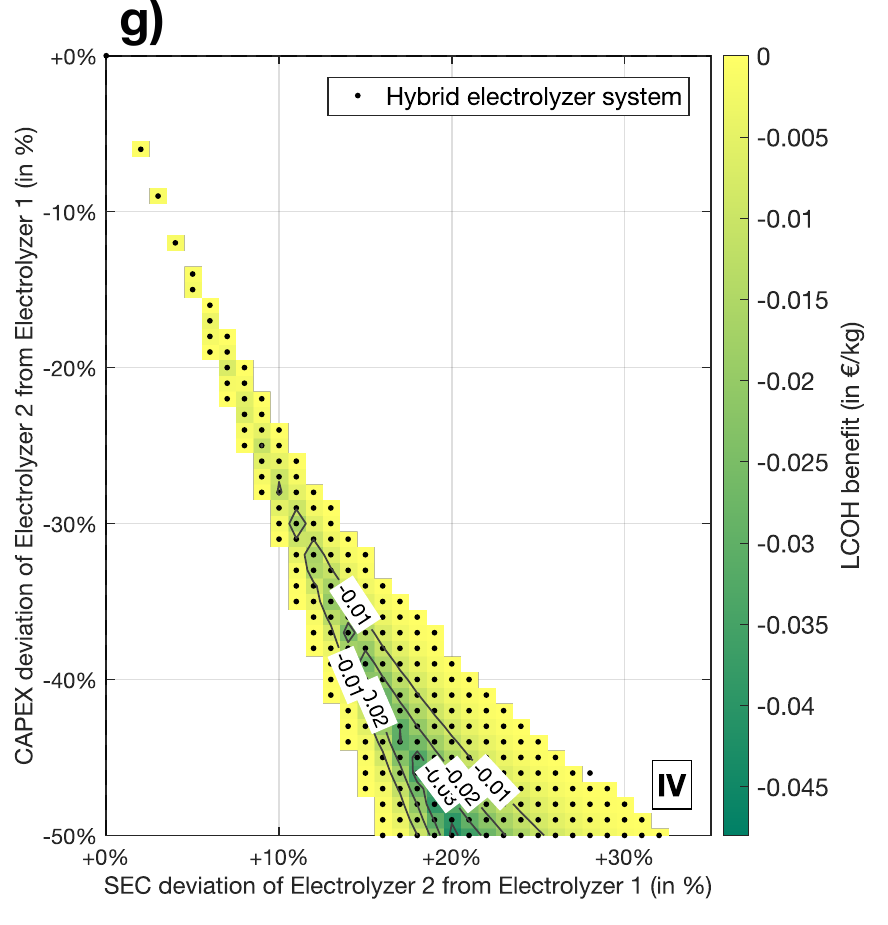}
\end{subfigure}
\hfill
\begin{subfigure}[t]{0.19\textwidth}
    \includegraphics[width=\linewidth]{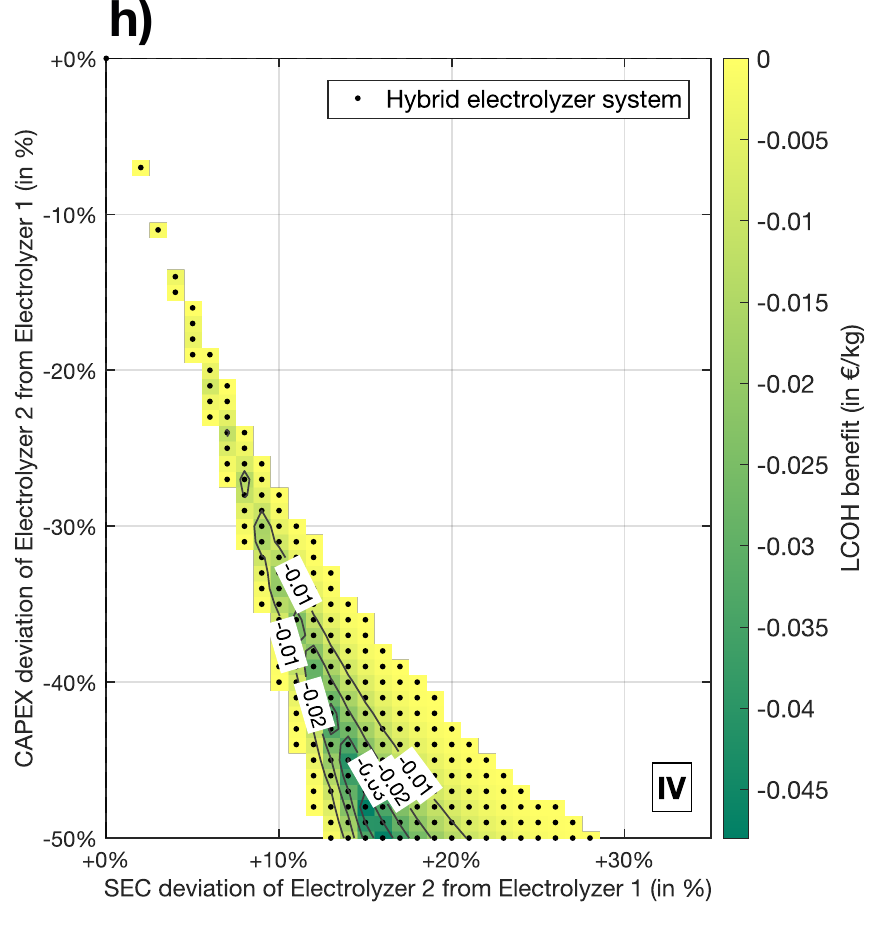}
\end{subfigure}
\hfill
\begin{subfigure}[t]{0.19\textwidth}
    \includegraphics[width=\linewidth]{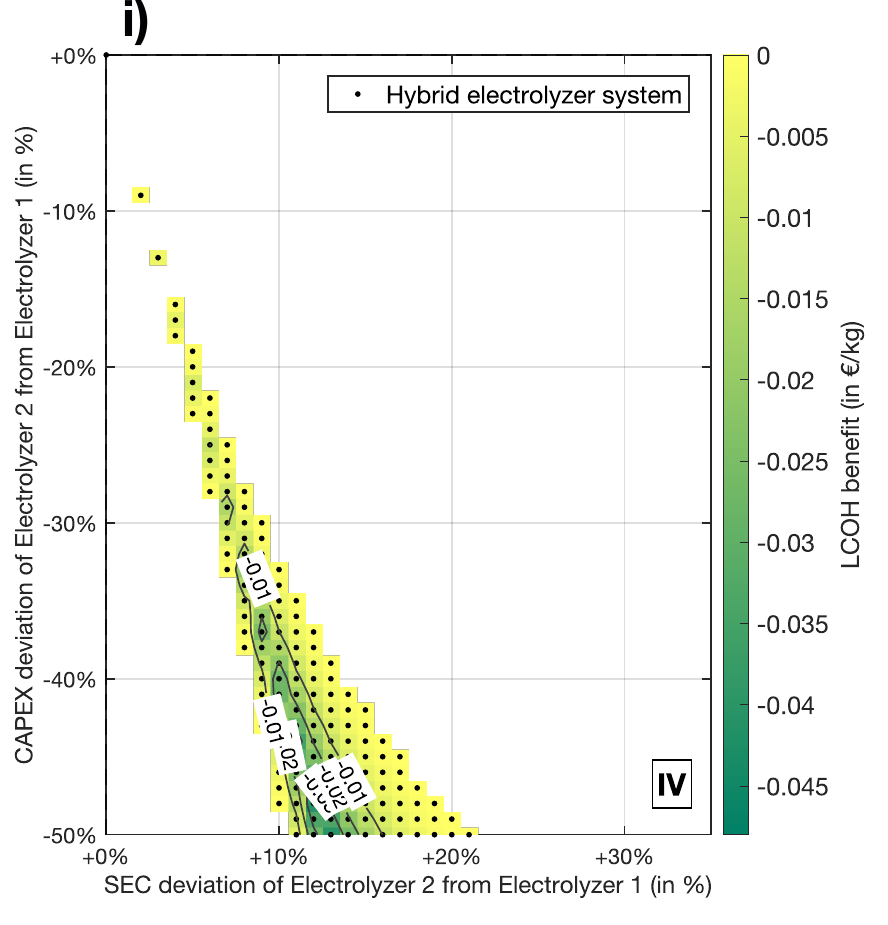}
\end{subfigure}
\hfill
\begin{subfigure}[t]{0.19\textwidth}
    \includegraphics[width=\linewidth]{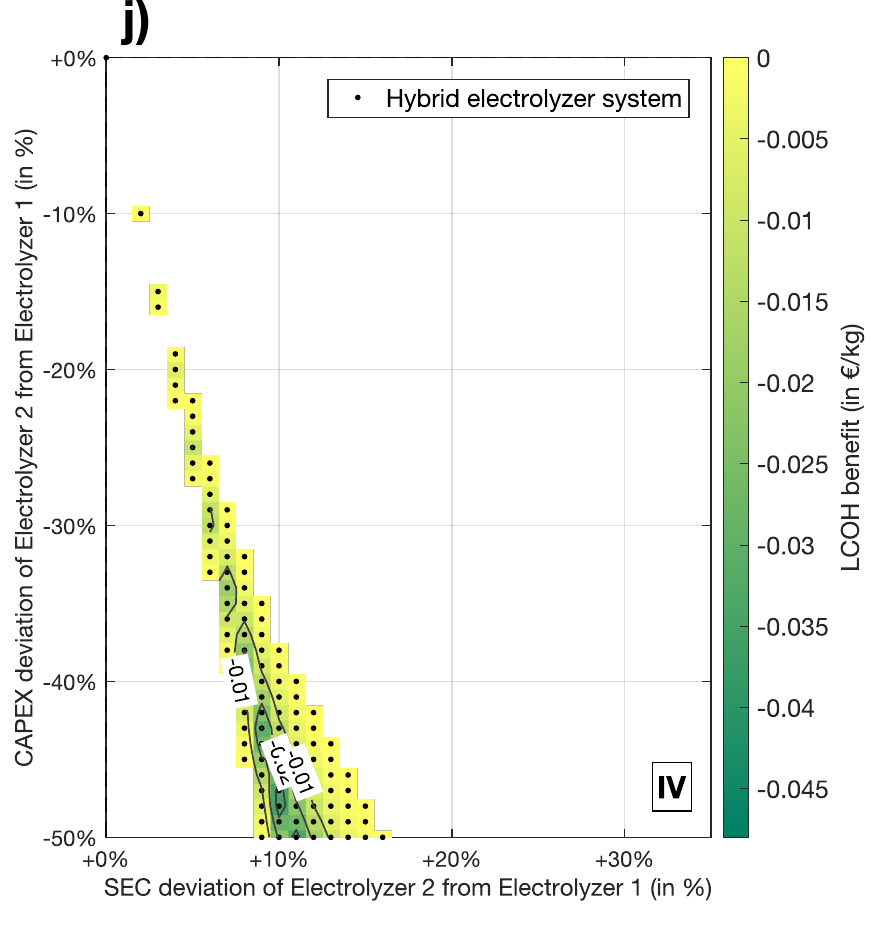}
\end{subfigure}
\caption{$LCOH$ benefit of HES depending on PPA price variation. CAPEX deviation of electrolyzer 2 from electrolyzer 1 in \% is shown on the y-axis. SEC deviation of electrolyzer 2 from electrolyzer 1 in \% is shown on the x-axis. The dashed lines mark where electrolyzer 2 has the same CAPEX value respectively SEC value as electrolyzer 1. CAPEX and SEC parameter combinations leading to HES are marked by black dots. The exemplary HES result is marked by a red cross.  For HES results, $LCOH$ benefit in €/kgH2 is shown. Labeled contour lines clarify $LCOH$ benefit values. a) Quadrant II is shown for a PPA deviation of -60\%. b) Quadrant II is shown for a PPA deviation of -30\%. c) Quadrant II is shown for no PPA deviation. d) Quadrant II is shown for a PPA deviation of +30\%. e) Quadrant II is shown for a PPA deviation of +60\%. f) Quadrant IV is shown for a PPA deviation of -60\%. g) Quadrant IV is shown for a PPA deviation of -30\%. h) Quadrant IV is shown for no PPA deviation. i) Quadrant IV is shown for a PPA deviation of +30\%. j) Quadrant IV is shown for a PPA deviation of +60\%.}
\label{Fig: LCOH benefit PPA cost variation}
\end{figure}

\noindent To get a comprehensive picture of these results, Figure \ref{Fig: LCOH benefit PPA cost variation} presents the $LCOH$ benefit for varying PPA prices in quadrant II and quadrant IV. Subfigures a)–e) correspond to quadrant II and subfigures f)–j) to quadrant IV, each showing PPA price variations of -60\%, -30\%, 0\%, +30\%, and +60\%. Accordingly, Figure \ref{Fig: LCOH benefit PPA cost variation} c) and h) show the same as Figure \ref{Fig: LCOH} b) and c). Regarding the $LCOH$ benefits, the maximum of 0.048 €/kgH2 is observed in Figure \ref{Fig: LCOH benefit PPA cost variation} h) without PPA price variation, showing the results for the base PPA price assumption already presented in Figure \ref{Fig: LCOH} c). Regarding the appearance of the HES regions in quadrant II, a tilt to the left is notable for lower PPA values in a) and b), as well as a tilt to the right for higher PPA values in d) and e). The opposite behavior is visible for the HES region in quadrant IV. These tilting effects result from changes in the relative techno-economic relevance of SEC and CAPEX with varying PPA price. For lower PPA values, the optimization increasingly favors lower CAPEX over lower SEC. Consequently, in quadrant II, electrolyzer 1 with the lower CAPEX value compared to electrolyzer 2 becomes more favorable, shifting the HES region to the left and reducing the single electrolyzer 2 region. In quadrant IV, the opposite effect occurs, shifting the HES region to the right and reducing the single electrolyzer 1 region. For higher PPA values, however, SEC becomes more techno-economically relevant and CAPEX become less relevant, resulting in the opposite shifting and tilting behavior. \newline

\begin{figure}[H]
\centering
\begin{subfigure}[t]{0.28\textwidth}
    \includegraphics[width=\linewidth]{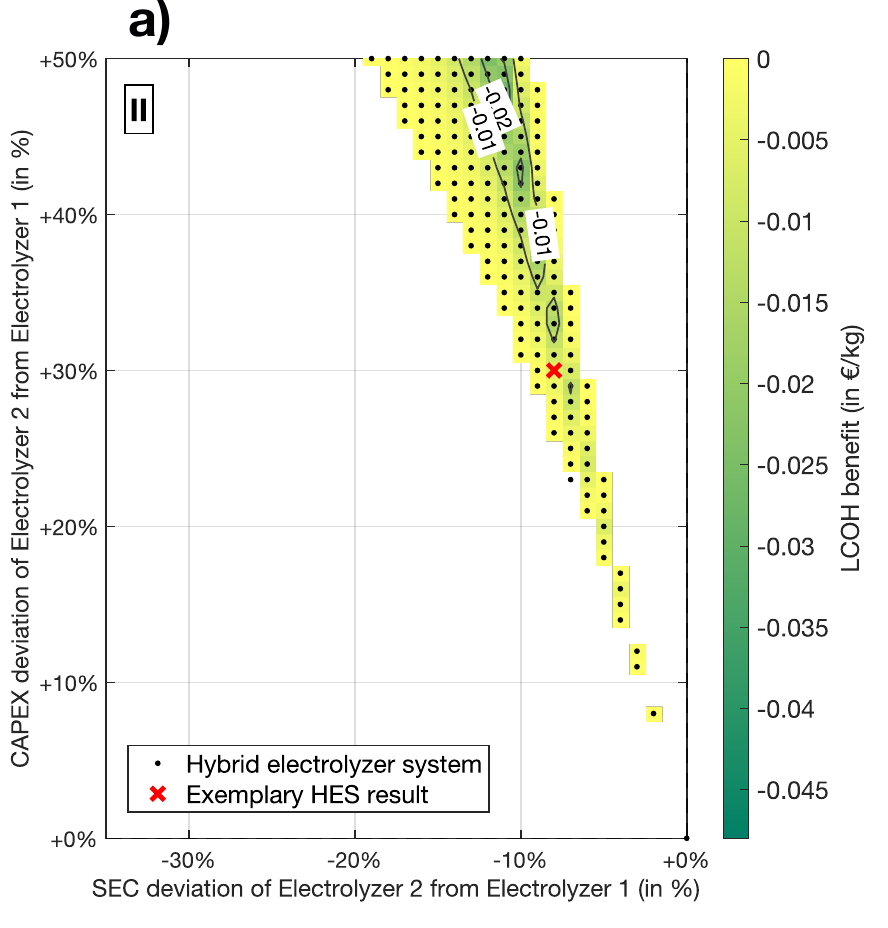}
\end{subfigure}
\begin{subfigure}[t]{0.28\textwidth}
    \includegraphics[width=\linewidth]{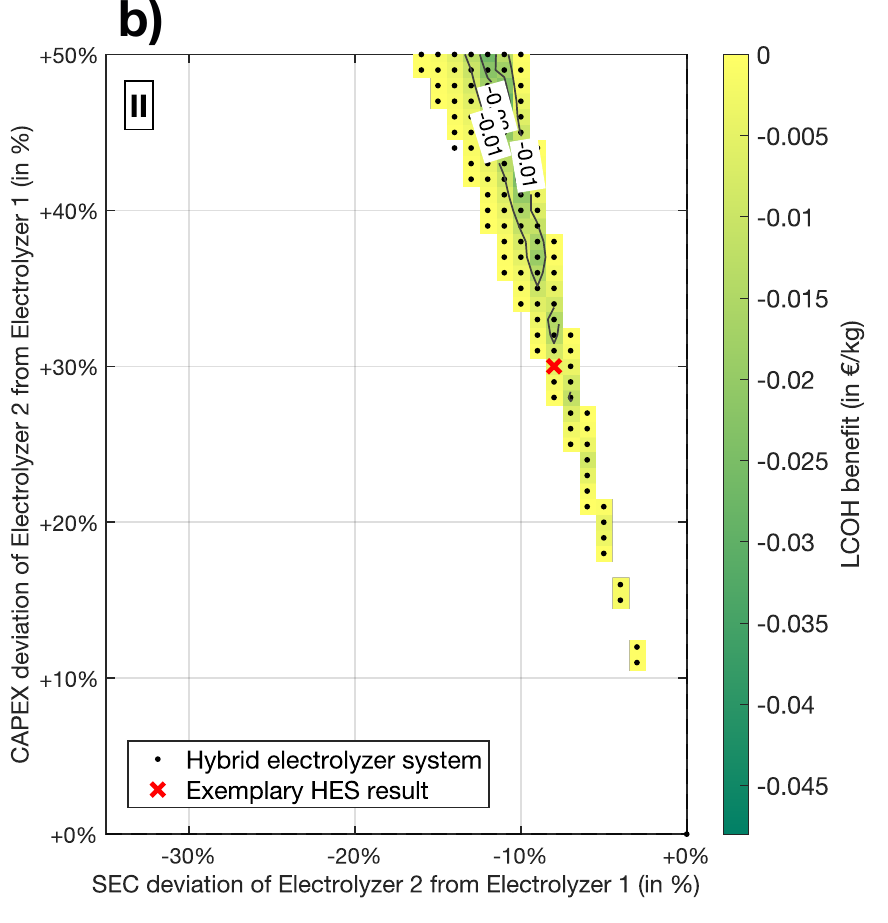}
\end{subfigure}
\begin{subfigure}[t]{0.28\textwidth}
    \includegraphics[width=\linewidth]{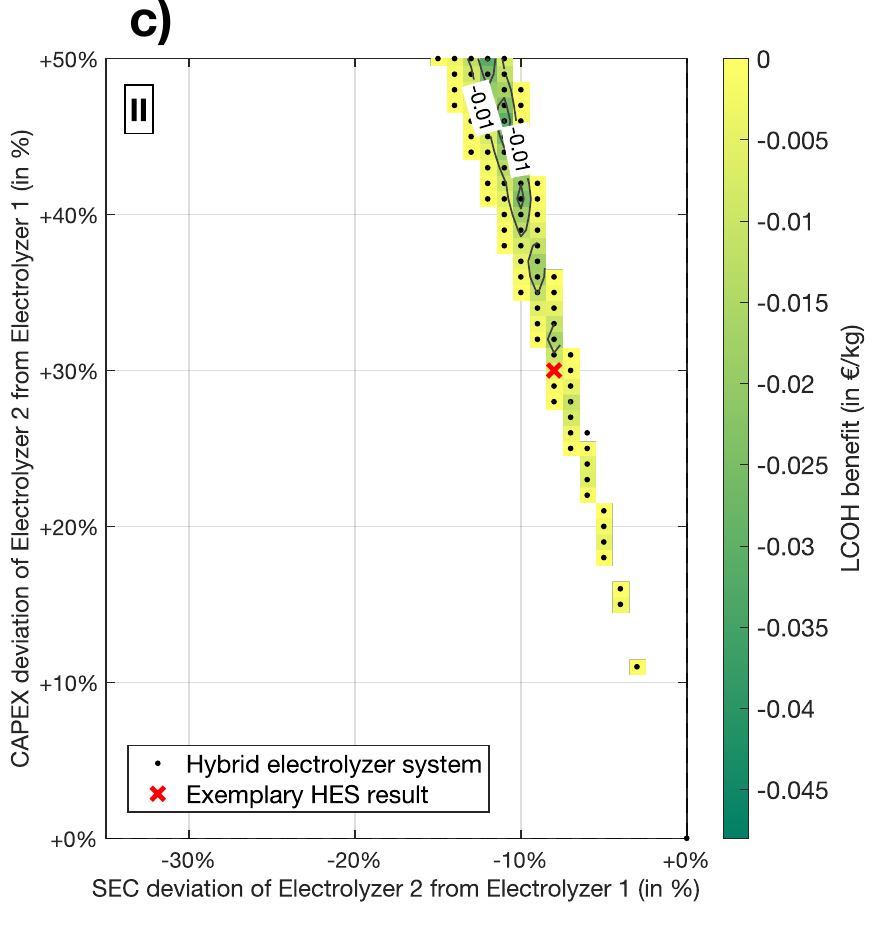}
\end{subfigure}
\vspace{0.5em}
\begin{subfigure}[t]{0.28\textwidth}
    \includegraphics[width=\linewidth]{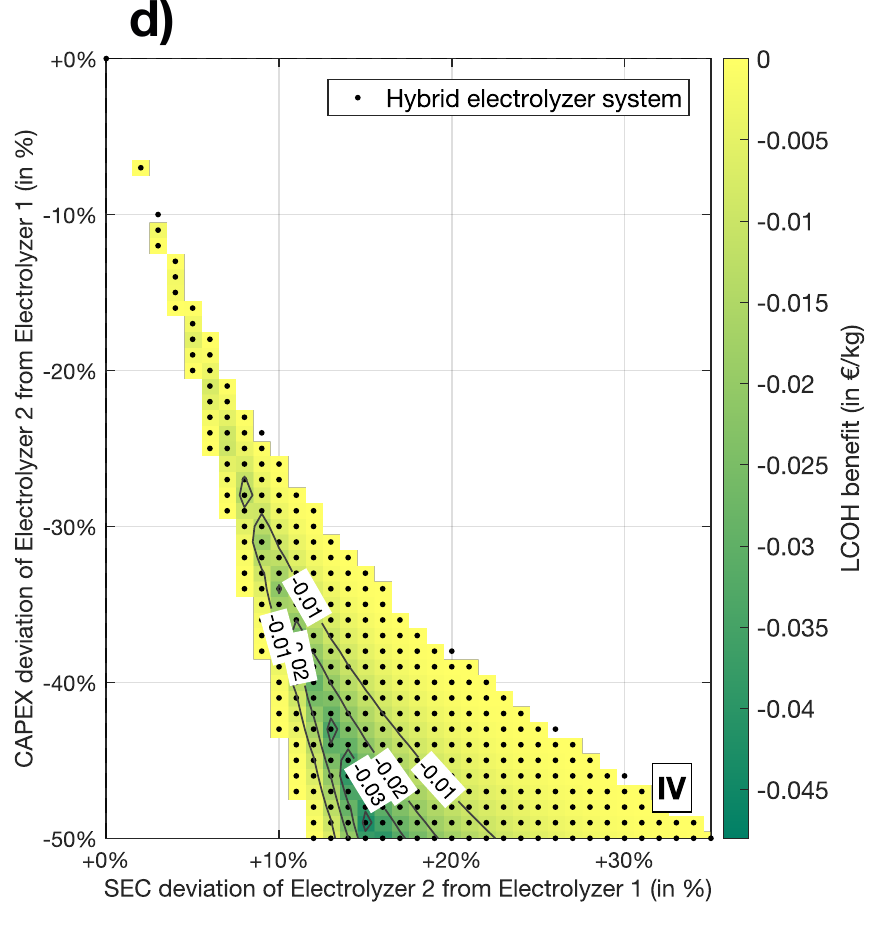}
\end{subfigure}
\begin{subfigure}[t]{0.28\textwidth}
    \includegraphics[width=\linewidth]{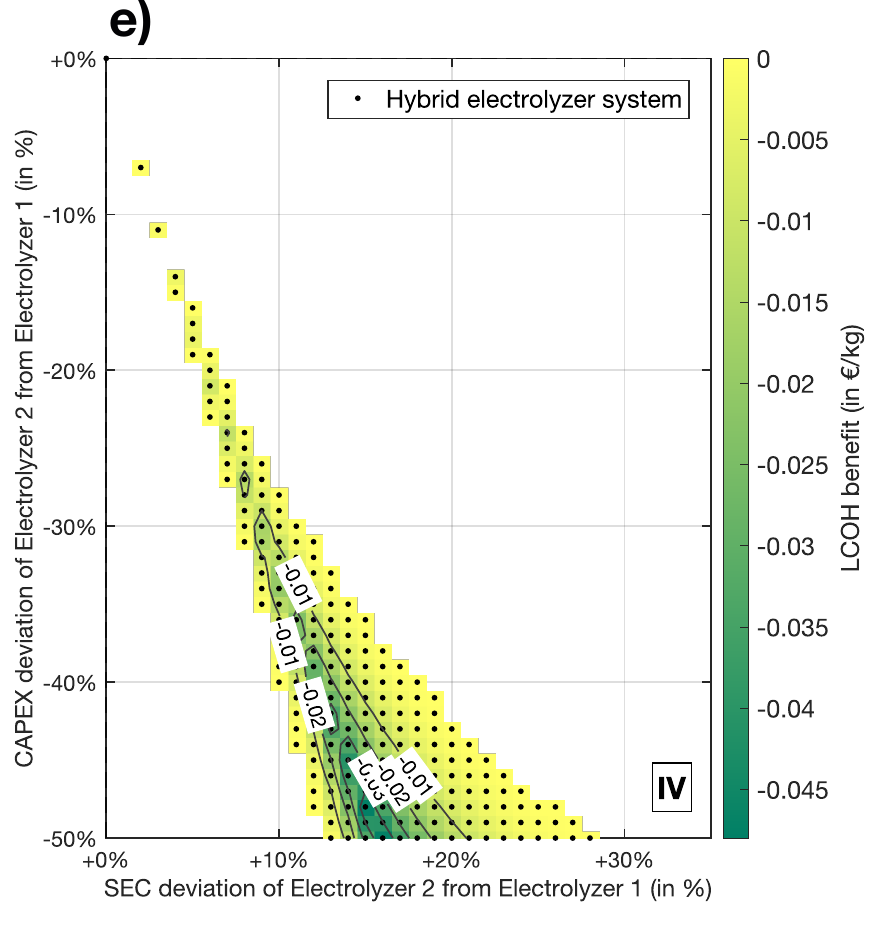}
\end{subfigure}
\begin{subfigure}[t]{0.28\textwidth}
    \includegraphics[width=\linewidth]{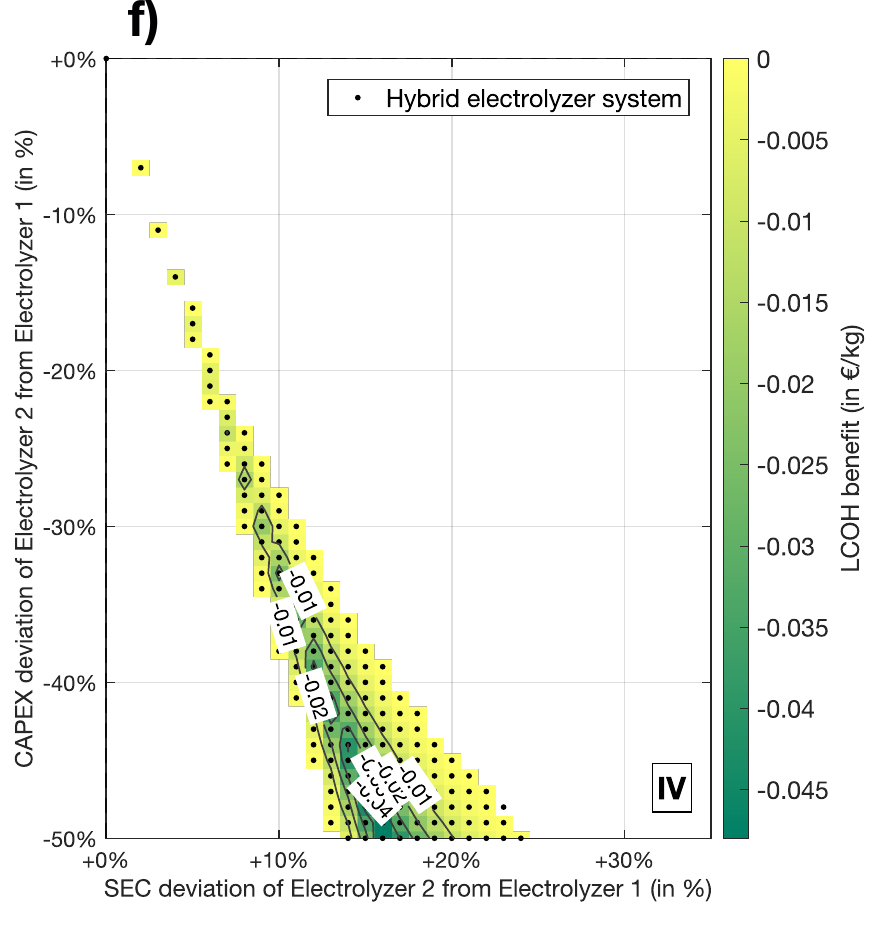}
\end{subfigure}
\caption{$LCOH$ benefit of HES depending on baseline SEC variation. CAPEX deviation of electrolyzer 2 from electrolyzer 1 in \% is shown on the y-axis. SEC deviation of electrolyzer 2 from electrolyzer 1 in \% is shown on the x-axis. The dashed lines mark where electrolyzer 2 has the same CAPEX value respectively SEC value as electrolyzer 1. CAPEX and SEC parameter combinations leading to HES are marked by black dots. The exemplary HES result is marked by a red cross.  For HES results, $LCOH$ benefit in €/kgH2 is shown. Labeled contour lines clarify $LCOH$ benefit values. a) Quadrant II is shown for a baseline SEC deviation of -30\%. b) Quadrant II is shown for no baseline SEC deviation. c) Quadrant II is shown for a baseline SEC deviation of +30\%.   d) Quadrant IV is shown for a baseline SEC deviation of -30\%. e) Quadrant IV is shown for no baseline SEC deviation. f) Quadrant IV is shown for a baseline SEC deviation of +30\%.}
\label{Fig: LCOH benefit baseline SEC variation}
\end{figure}

\noindent In addition to the discussed PPA price variation, Figure \ref{Fig: LCOH benefit baseline SEC variation} presents the $LCOH$ benefit of varying the baseline SEC values following the same presentation as in Figure \ref{Fig: LCOH benefit PPA cost variation}. However, the baseline SEC fixed for electrolyzer 1 is only varied within a range of $\pm$30\% relative to the baseline introduced in Section \ref{Sec:Study_design}, since larger deviations would not be technically feasible. The maximum $LCOH$ benefit of 0.057 €/kgH2 is observed in Figure \ref{Fig: LCOH benefit baseline SEC variation} f) for a baseline SEC increase of 30\%. As the baseline SEC increases from -30\% to +30\%, the HES regions in both quadrants decrease considerably, from approximately 5.0\% to 2.5\%. Concurrently, the shares of both single electrolyzer systems increase steadily, with electrolyzer 1 increasing from 46.3\% to 47.8\% and electrolyzer 2 from 48.7\% to 49.7\% across the same baseline SEC range. This behavior can be explained by the parameterization of the SEC variation. Since the SEC of electrolyzer 2 is defined as a relative deviation from the baseline SEC fixed to electrolyzer 1, a lower baseline SEC results in smaller SEC differences between both technologies, whereas a higher baseline SEC leads to larger SEC differences. Consequently, for lower baseline SEC values and smaller SEC differences, HES become economically more favorable leading to an increasing HES share. In contrast, for higher baseline SEC values and larger SEC differences, single electrolyzer systems become economically more favorable leading to a decreasing HES share. In contrast to the effects discussed for the PPA price variation, neither a tilting nor a shifting effect is present here. \newline

\begin{figure}[H]
\centering
\vspace{0.8em}
\begin{minipage}{\textwidth}
\centering
\begin{subfigure}[t]{0.48\textwidth}
    \centering
    \includegraphics[width=\linewidth]{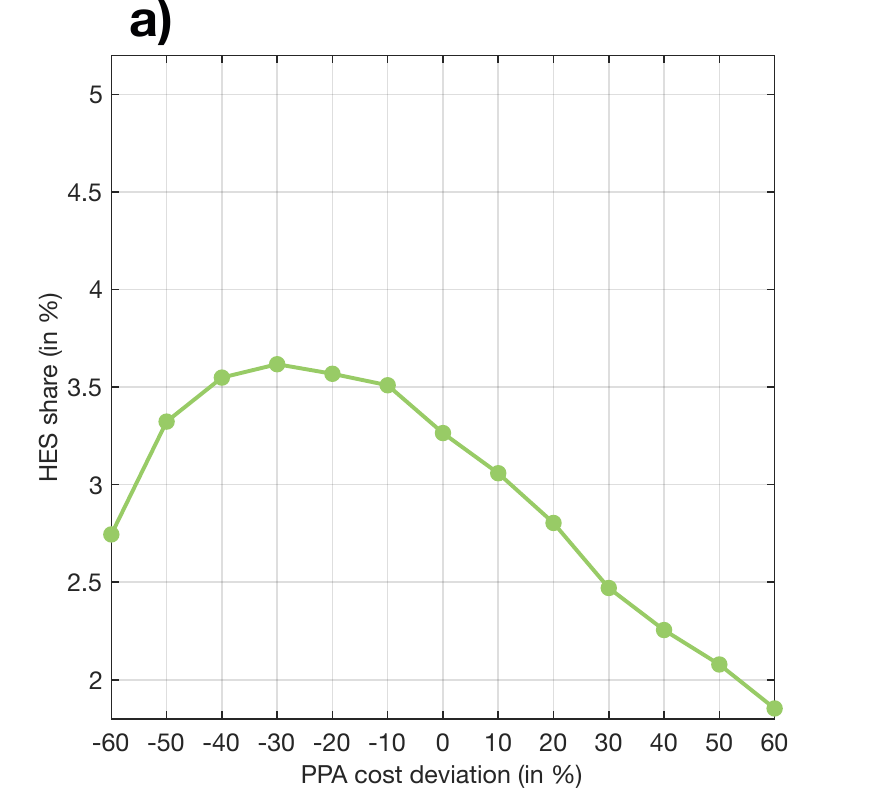}
\end{subfigure}
\begin{subfigure}[t]{0.48\textwidth}
    \centering
    \includegraphics[width=\linewidth]{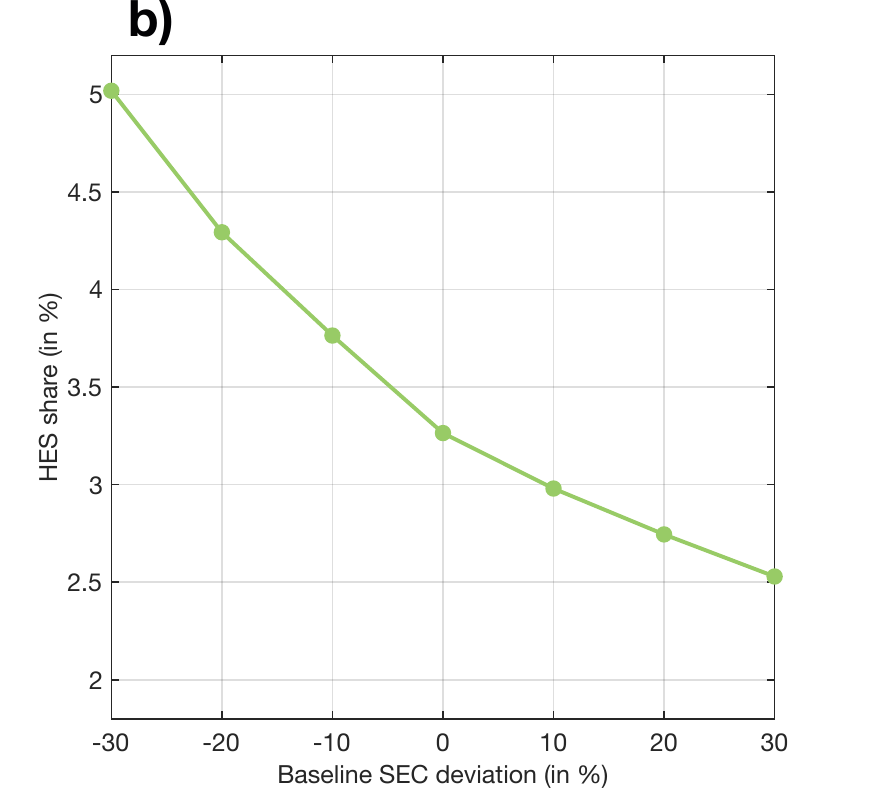}
\end{subfigure}
\end{minipage}
\caption{HES share depending on PPA price and baseline SEC deviation. The y-axis shows the HES share in \%. The x-axis shows the parameter deviation in \%. a) PPA price deviation ranging from -60\% to +60\%. b) Baseline SEC deviation ranging from -30\% to +30\%.}
\label{Fig: HES share}
\end{figure}

\noindent Figure \ref{Fig: HES share} presents the HES share as a function of PPA price variation within a range of $\pm$60\% in a) and as a function of baseline SEC variation within a range of $\pm$30\% in b). For the PPA price sensitivity, the HES share reaches a maximum of 3.6\% at a PPA price reduction of 30\%. Both higher and lower PPA prices relative to this maximum result in a decreasing HES share. For the baseline SEC sensitivity, the HES share reaches a maximum of 5\% at a baseline SEC reduction of 30\%. With increasing baseline SEC, the HES share continuously decreases. Overall, these results indicate that HES remain  minor in comparison to single electrolyzer systems, even when the ratio between SEC and CAPEX is changed through variations in PPA price or baseline SEC covering technically feasible and economically imaginable ranges. \newline

\noindent All in all, HES are cost optimal compared to single electrolyzer systems for a maximum share of only 5.0\% of the investigated cases. Moreover, the maximum cost benefit of 0.057 €/kgH2 only corresponds to approximately 1\% of the total production cost ($LCOH$). Therefore, the techno-economic benefit of HES could be interpreted as limited. Although the investigated sensitivities PPA price and baseline SEC do affect the share and cost benefit of HES, they do not substantially alter this overall conclusion. In particular, variations in PPA price lead to a shifting and tilting of the HES regions, whereas variations in baseline SEC influence the HES share, which decreases with increasing baseline SEC values. In contrast, variations in storage fees and availability of RES have negligible impact and are therefore presented in Section S1 of the Supplementary material.

\subsubsection*{Deep dive: CAPEX/SEC trade-off}\label{Sec: Deep dive: CAPEX/SEC trade-off}

\noindent Independent of HES, this section focuses exclusively on electrolyzer 2 operated as a single system in order to analyze the trade-off between CAPEX and SEC. For this purpose, the same parameter variation space methodology as used in Section \ref{Sec: System design and cost} is applied. However, only quadrant III of the $LCOH$ heat map shown in Figure \ref{Fig: LCOH} a) is considered, where electrolyzer 2 is solely chosen. Within this quadrant, the slope of the $LCOH$ contour lines enables a cost-based comparison of CAPEX and SEC defining the ratio between both quantities. The resulting cost-based ratio between SEC and CAPEX provides an indication of whether prioritizing investment cost reductions over efficiency improvements is economically justified, and if so, to what extent. This question is particularly relevant, since the current development of PEMWE technology primarily focuses on investment cost reduction rather than efficiency improvements \cite{Smolinka2018}. \newline

\begin{figure}[H]
    \centering \includegraphics[width=0.6\textwidth]{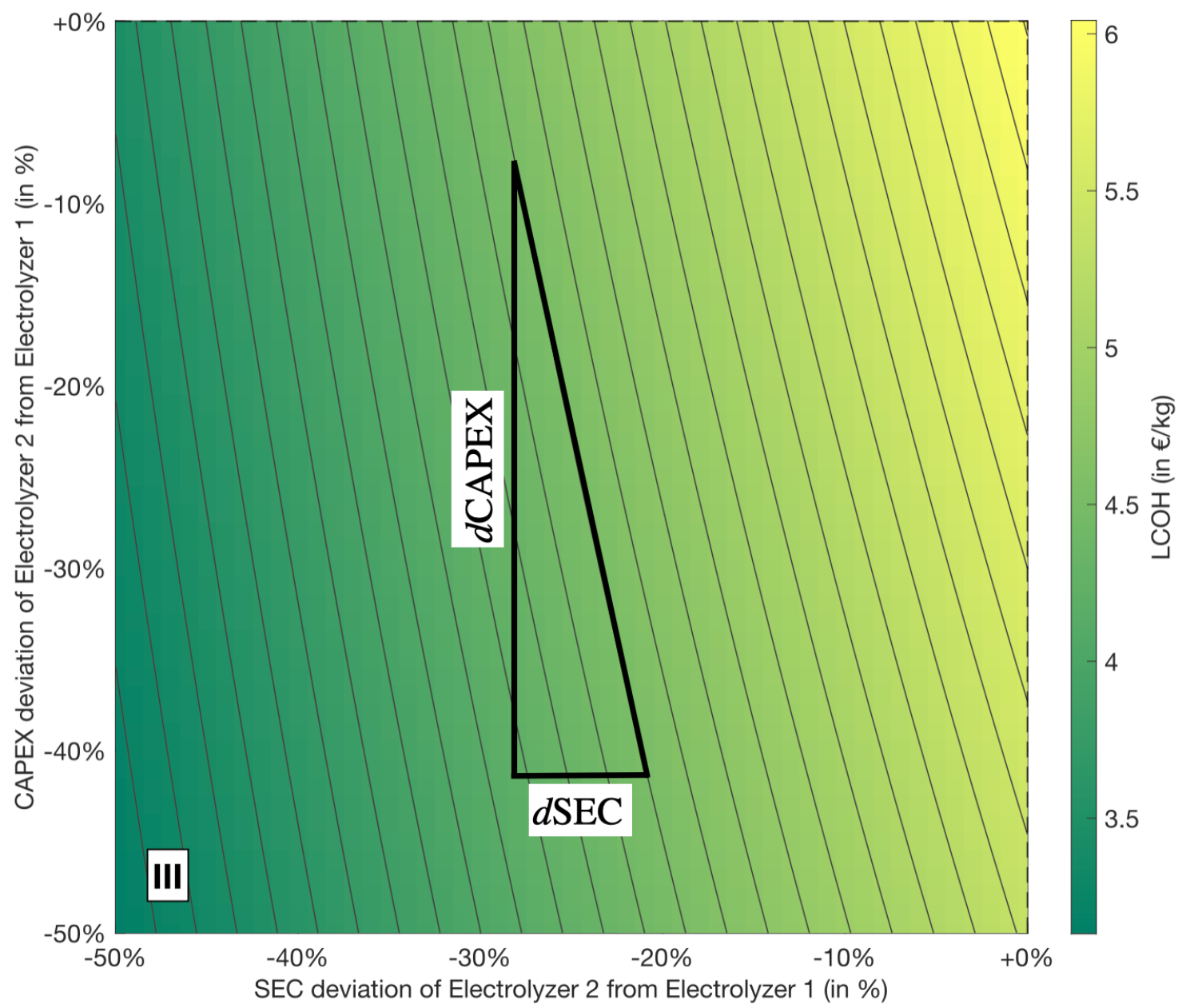}
    \caption{Illustration of the calculation of the CAPEX/SEC trade-off. CAPEX deviation of electrolyzer 2 from electrolyzer 1 in \% is shown on the y-axis. SEC deviation of electrolyzer 2 from electrolyzer 1 in \% is shown on the x-axis. The $LCOH$ heat map in €/kgH2 for quadrant III is presented including contour lines of equal values. A gradient triangle is marked in black with the labels $d\text{CAPEX}$ and $d\text{SEC}$.}
    \label{Fig: Rho analysis visualization}
\end{figure}

\noindent The basis for this analysis, quadrant III of the $LCOH$ heat map, is presented in Figure \ref{Fig: Rho analysis visualization}. Contour lines of equal $LCOH$ values are plotted and a gradient triangle is marked in black for one exemplary contour line. For ten evenly distributed contour lines, $\rho$ is calculated with the respective gradient triangles according to Equation \eqref{Eq: rho}. Afterwards, the average is determined.

\begin{align}\label{Eq: rho}
   \rho = \frac{d\text{CAPEX}}{d\text{SEC}} \left[\frac{\text{€/kW}}{\text{kWh/kg}}\right]
\end{align}

\noindent The resulting average value for $\rho$ is 135,38 €/kW per kWh/kg. This implies, that an improvement of 1 kWh/kg in SEC allows for a CAPEX improvement of up to 135,38 €/kW, without increased $LCOH$. Generally, a higher $\rho$ value indicates, that CAPEX improvements have a smaller impact on $LCOH$ compared to SEC improvements, and a lower $\rho$ value indicates a higher impact of CAPEX improvements. \newline

\begin{figure}[H]
    \centering \includegraphics[width=0.6\textwidth]{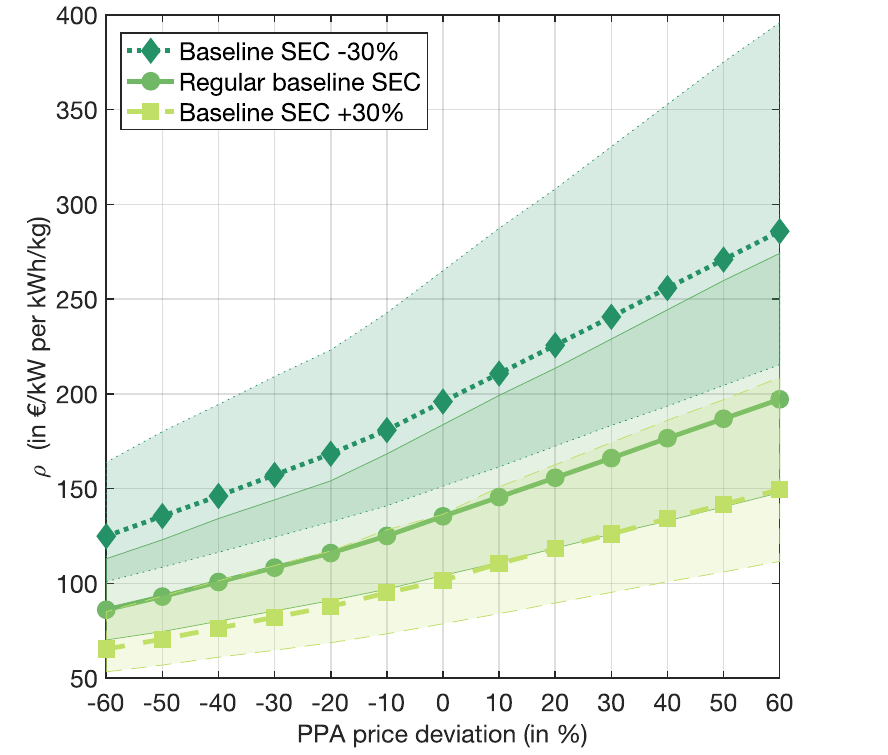}
    \caption{$\rho$ values resulting for variation of PPA price and baseline SEC value. The y-axis shows $\rho$ in €/kW per kWh/kg. The x-axis shows the relative PPA price deviation from -60\% to 60\%. The average $\rho$ values resulting for regular baseline SEC are presented by the green line and dots. The average $\rho$ values resulting for 30\% higher baseline SEC  are presented by the light green dashed line and squares. The average $\rho$ values resulting for 30\% lower baseline SEC are presented by the dark green dotted line and diamonds. For each average $\rho$ values, the transparent shaded area in the corresponding color indicates the range between the minimum and maximum $\rho$ values, which are represented by boundary lines in the respective line style.}
    \label{Fig: rho sensitivity}
\end{figure}

\noindent For a comprehensive picture of $\rho$, relative PPA price variations in the interval of [-60,60]\% are performed. In addition, these PPA price variations are evaluated for a baseline SEC variation within a range of $\pm$30\% relative to the baseline introduced in Section \ref{Sec:Study_design}, following the procedure in Section \ref{Sec: System design and cost}. The results are shown in Figure \ref{Fig: rho sensitivity}. The average $\rho$ values that result from regular baseline SEC are presented by a green line and dots. The average $\rho$ values that result from a 30\% higher baseline SEC are presented by a light green dashed line and squares. The average $\rho$ values that result from a 30\% lower baseline SEC are presented by a dark green dotted line and diamonds. For each average $\rho$ value, the transparent shaded area in the corresponding color indicates the range between the minimum and maximum $\rho$ values, which are represented by boundary lines in the respective line style. The results show that for increasing PPA price, $\rho$ generally increases. This implies, that the impact of CAPEX improvements on $LCOH$ decreases and the impact of SEC improvements increases. This underlines the finding of Section \ref{Sec: System design and cost}, that the techno-economic relevance of SEC increases in comparison to CAPEX with increasing PPA price. For the 30\% higher baseline SEC value (presented by the light green dashed line and squares), $\rho$ is generally shifted to lower values. For the 30\% lower baseline SEC value (presented by the dark green dotted line and diamonds), $\rho$ is generally shifted to higher values. Thus, the effect of increased baseline SEC values is comparable to the effect of decreased PPA price, because both variations lead to lower $\rho$ values and a higher techno-economic relevance of CAPEX in comparison to SEC. The variation in storage fees and RES availability has negligible impact on the resulting $\rho$ value. The corresponding results are presented in Section S2 of the Supplementary material. \newline

\noindent Overall, the calculation of $\rho$ provides a direct cost-based ratio between the technical parameter SEC and the economic parameter CAPEX. Thus, the cost trade-off of both parameters is quantifiable. Nevertheless, the exact ratio of SEC and CAPEX depends on the PPA price and the selected baseline SEC value, which therefore should be deliberately chosen. 

\subsection{Electrolyzer operation}\label{Sec: Dispatch results}

\begin{figure}[H]
  \centering
  \begin{minipage}[t]{0.48\textwidth}
    \centering
    \begin{subfigure}[t]{\textwidth}
    \includegraphics[width=\linewidth]{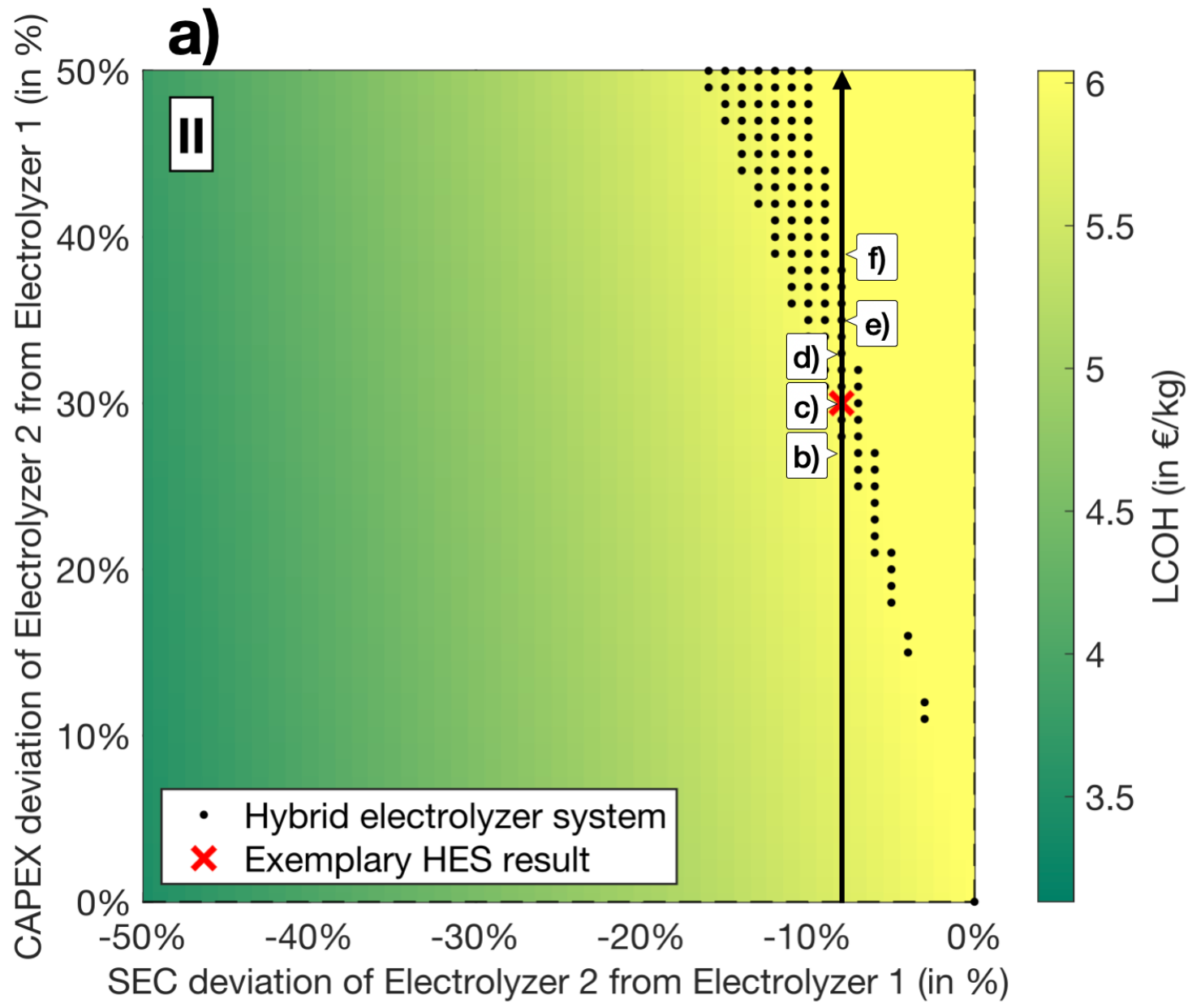}
    \end{subfigure}
    \begin{subfigure}[t]{0.95\textwidth}
      \includegraphics[width=\linewidth]{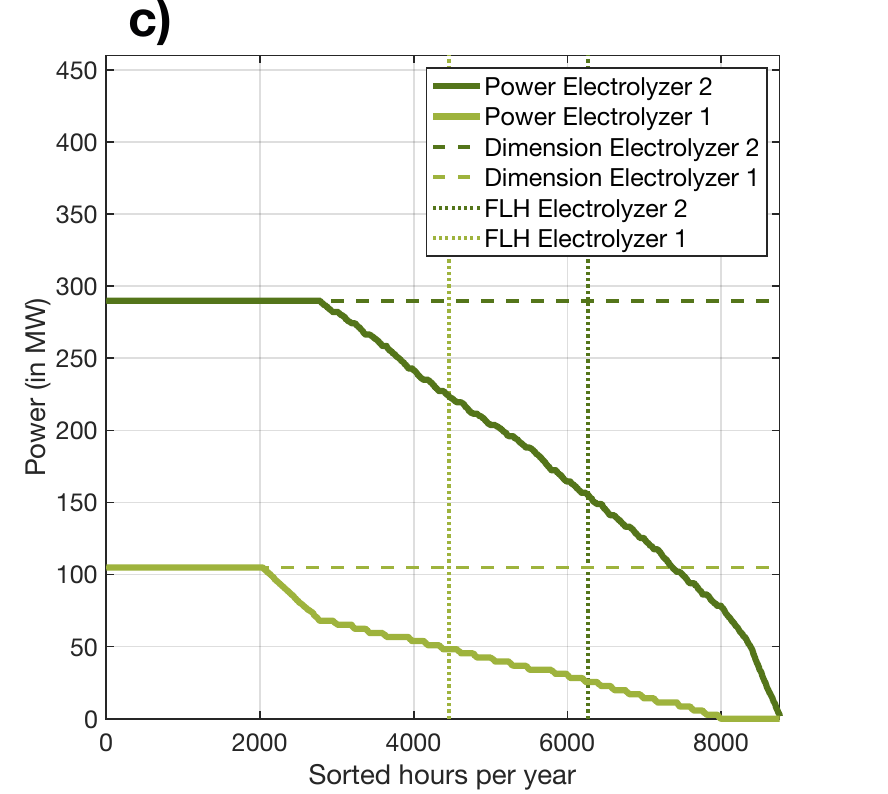}
    \end{subfigure}
    \begin{subfigure}[t]{0.95\textwidth}
      \includegraphics[width=\linewidth]{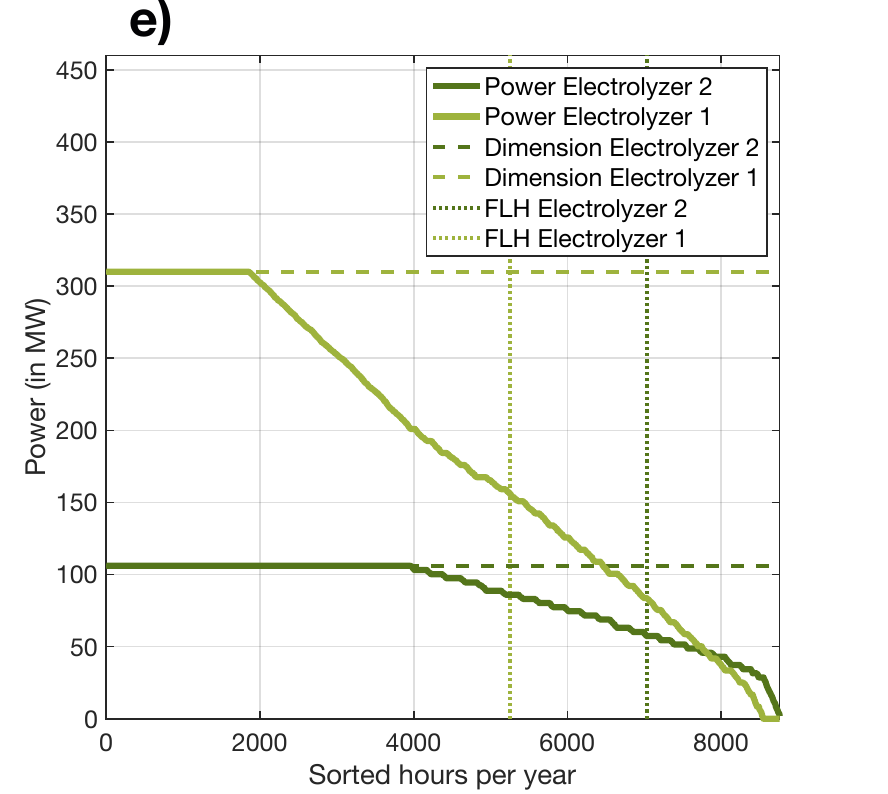}
    \end{subfigure}
  \end{minipage}
  \hfill
  \begin{minipage}[t]{0.48\textwidth}
    \centering
    \begin{subfigure}[t]{0.95\textwidth}
      \includegraphics[width=\linewidth]{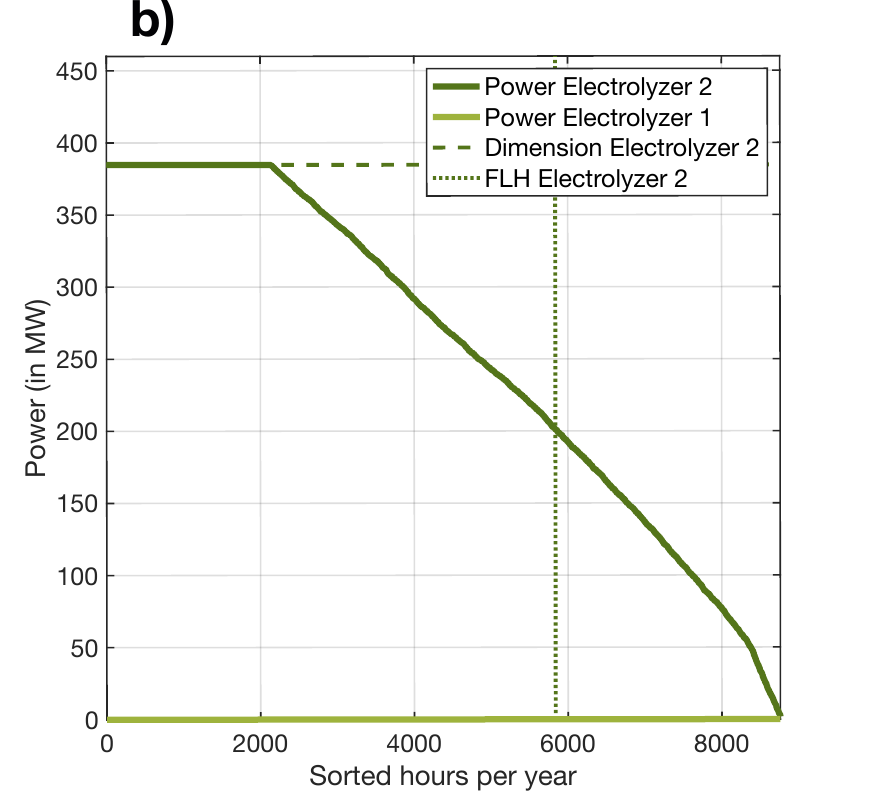}
    \end{subfigure}
    \begin{subfigure}[t]{0.95\textwidth}
      \includegraphics[width=\linewidth]{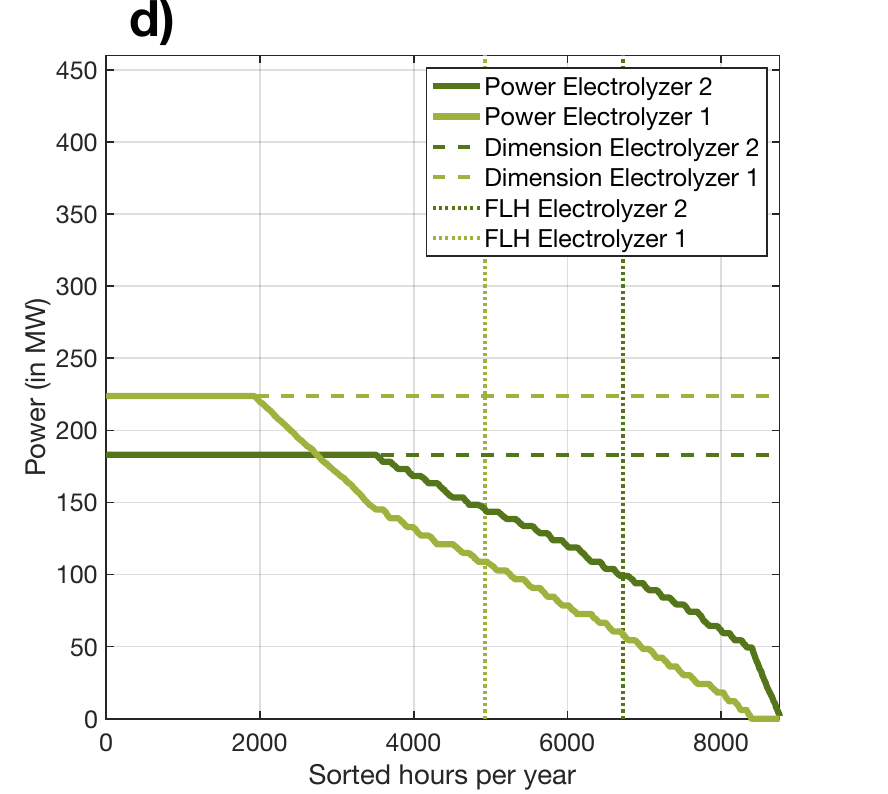}
    \end{subfigure}
    \begin{subfigure}[t]{0.95\textwidth}
      \includegraphics[width=\linewidth]{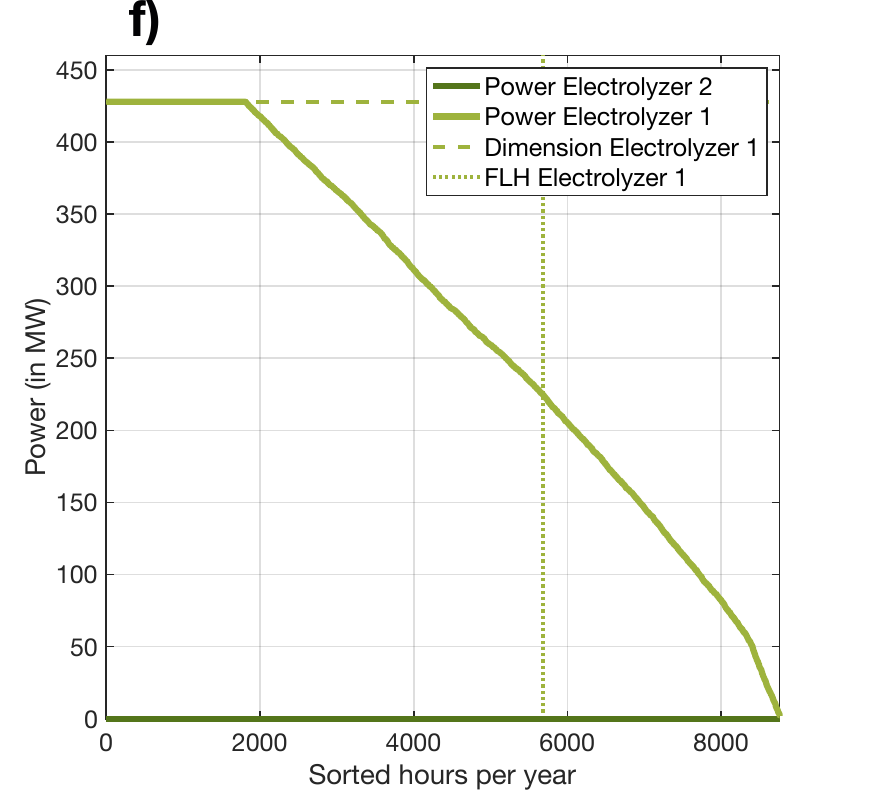}
    \end{subfigure}
  \end{minipage}
  \caption{APDCs resulting from CAPEX variation and set SEC value. a) Quadrant II of the $LCOH$ heat map is shown. A black arrow marks the set SEC value and the direction of the CAPEX variation. Additionally, the data points considered for the APDCs in subfigures b)-f) are labeled accordingly. b)-f) contain APDC as continuous lines of electrolyzer 1 in light green and of electrolyzer 2 in dark green. The dashed lines in the respective colors mark each electrolyzer dimension, the dotted lines mark each electrolyzers FLH.}
  \label{Fig: Dispatch SEC level}
\end{figure}

\begin{figure}[H]
  \centering
  \begin{minipage}[t]{0.48\textwidth}
    \centering
    \begin{subfigure}[t]{\textwidth}
    \includegraphics[width=\linewidth]{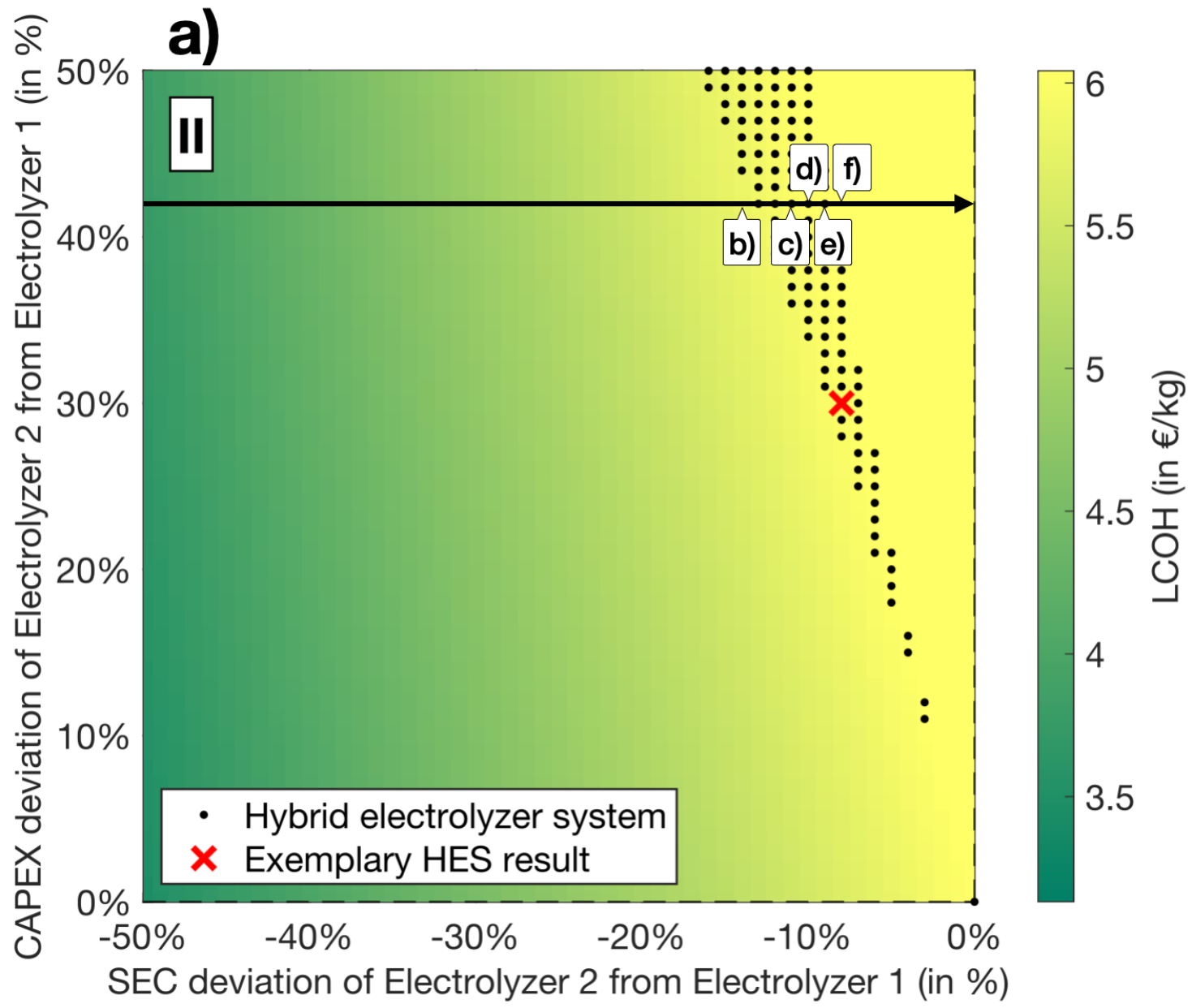}
    \end{subfigure}
    \begin{subfigure}[t]{0.95\textwidth}
      \includegraphics[width=\linewidth]{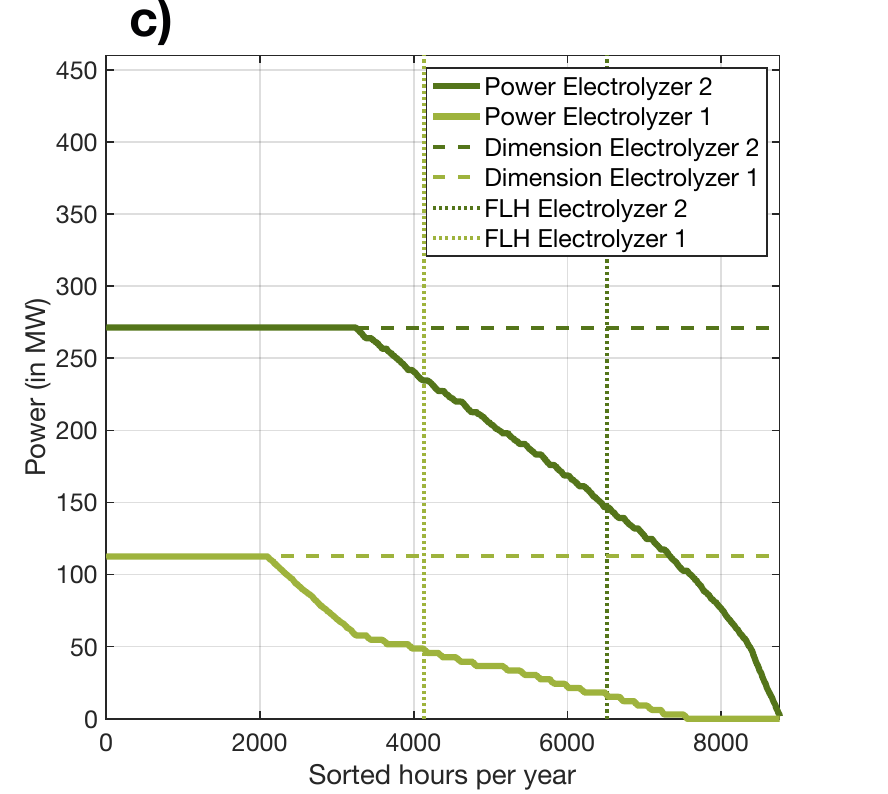}
    \end{subfigure}
    \begin{subfigure}[t]{0.95\textwidth}
      \includegraphics[width=\linewidth]{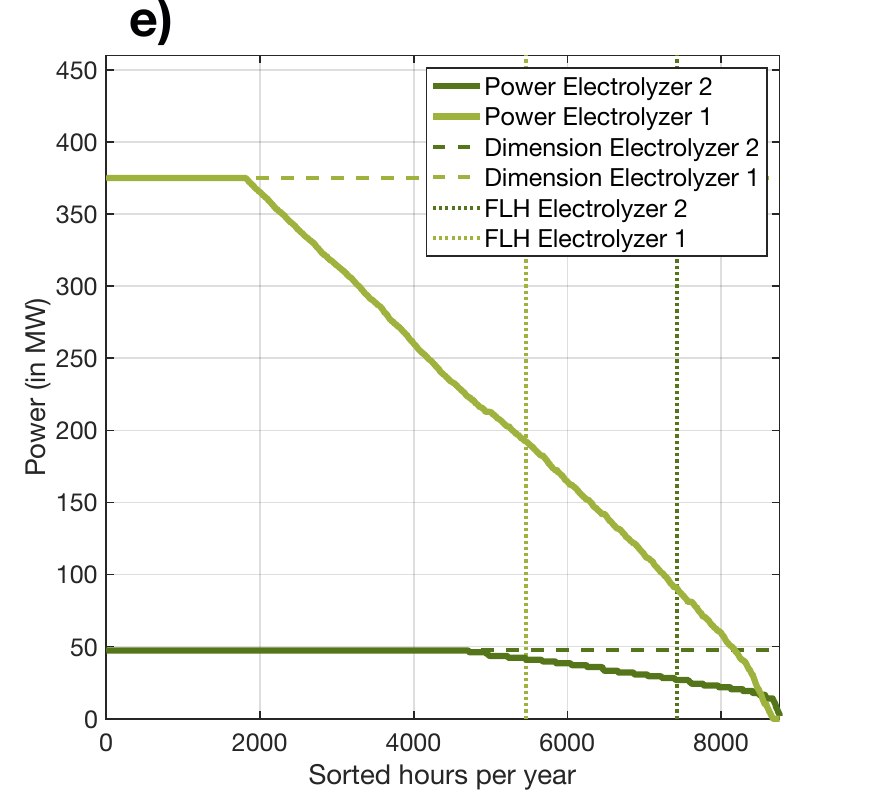}
    \end{subfigure}
  \end{minipage}
  \hfill
  \begin{minipage}[t]{0.48\textwidth}
    \centering
    \begin{subfigure}[t]{0.95\textwidth}
      \includegraphics[width=\linewidth]{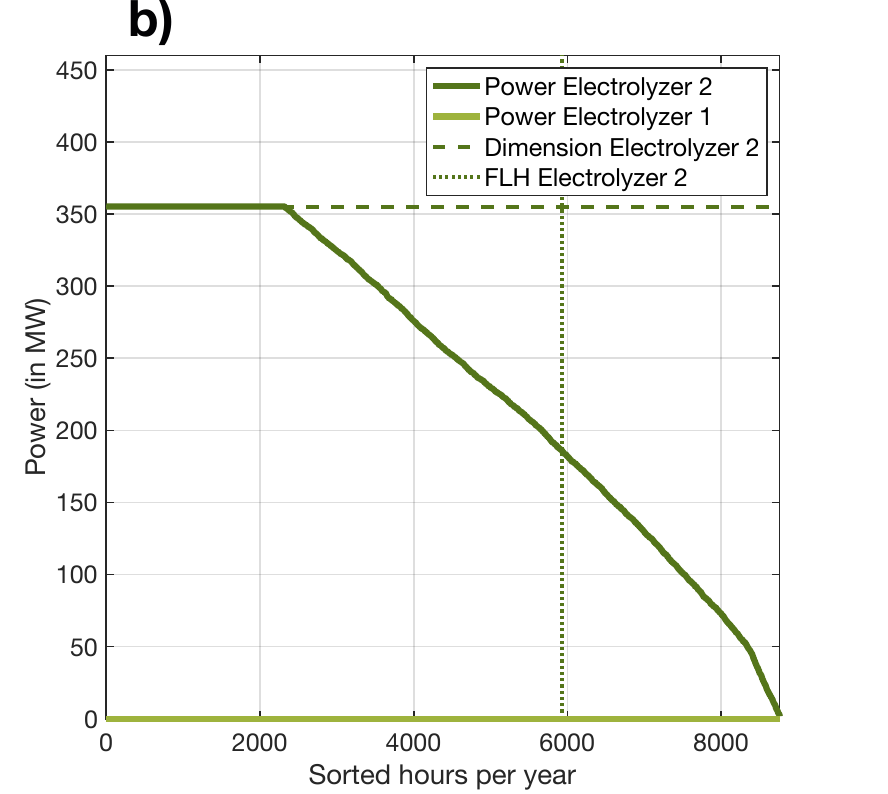}
    \end{subfigure}
    \begin{subfigure}[t]{0.95\textwidth}
      \includegraphics[width=\linewidth]{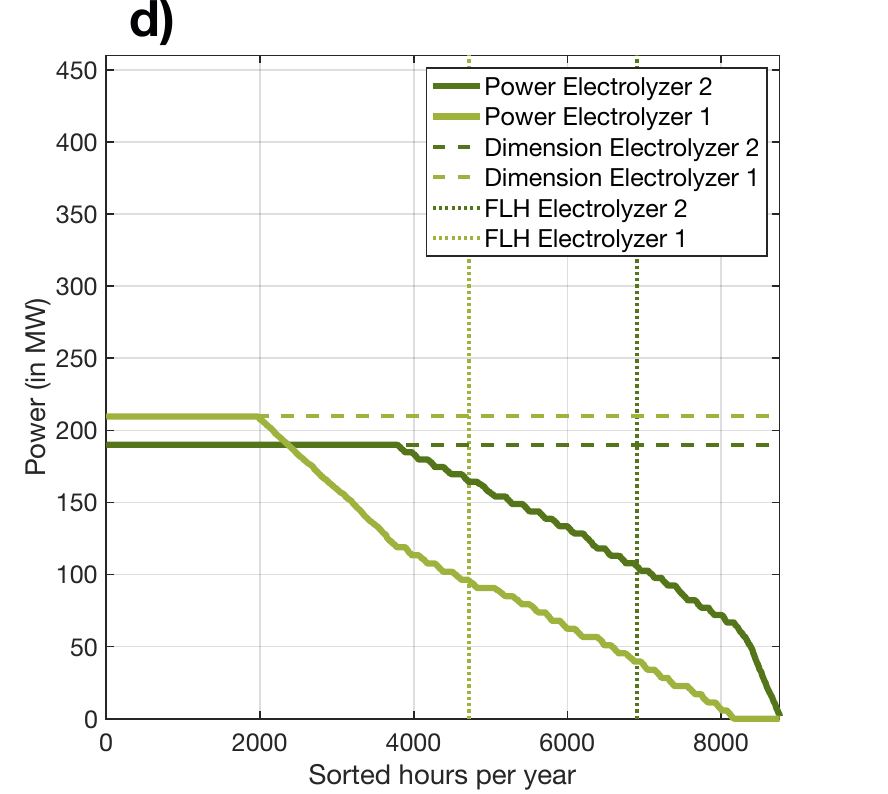}
    \end{subfigure}
    \begin{subfigure}[t]{0.95\textwidth}
      \includegraphics[width=\linewidth]{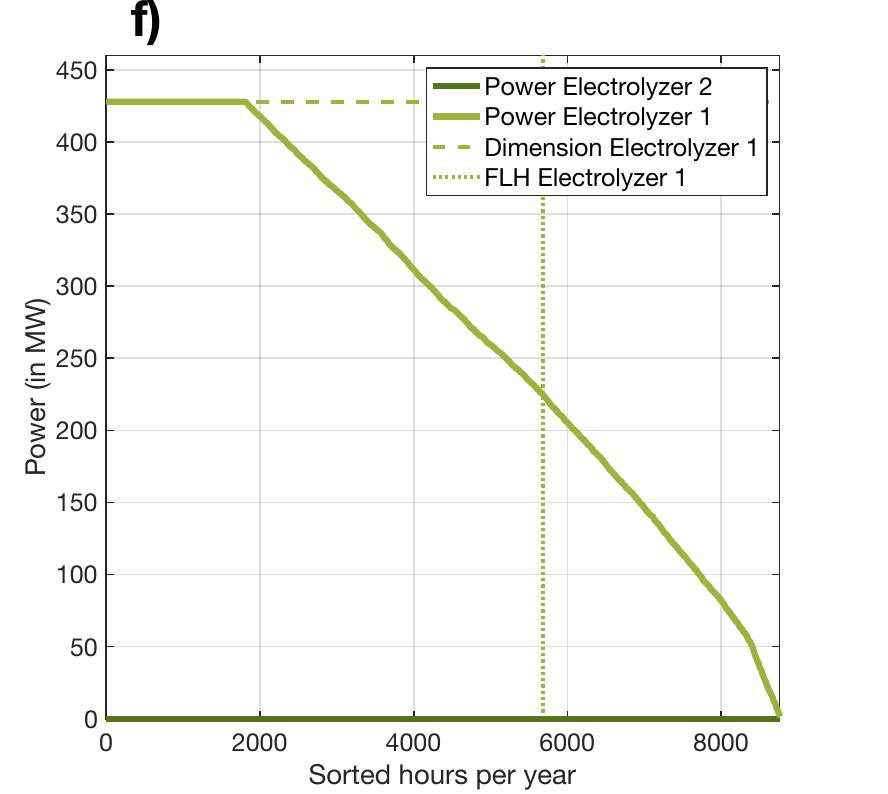}
    \end{subfigure}
  \end{minipage}
  \caption{APDCs resulting from SEC variation and set CAPEX value. a) Quadrant II of the $LCOH$ heat map is shown. A black arrow marks the set CAPEX value and the direction of the SEC variation. Additionally, the data points considered for the APDCs in subfigures b)-f) are labeled accordingly. b)-f) contain APDC as continuous lines of electrolyzer 1 in light green and of electrolyzer 2 in dark green. The dashed lines in the respective colors mark each electrolyzer dimension, the dotted lines mark each electrolyzers FLH.}
  \label{Fig: Dispatch CAPEX level}
\end{figure}

\noindent Having analyzed the design and cost of the HES in detail, this section focuses on its operational dispatch. For comparison, the dispatch behavior of single electrolyzer systems is also examined. For this purpose, Figure \ref{Fig: Dispatch SEC level} and Figure \ref{Fig: Dispatch CAPEX level} present annual power duration curves (APDCs) for both single electrolyzer systems and HES. APDCs of electrolyzer 1 are presented in light green, APDCs of electrolyzer 2 are presented in dark green. The exemplary HES APDC, which was already discussed in Section \ref{Sec:Exemplary HES result} and shown in Figure \ref{Fig: Exemplary HES} d), is presented in \ref{Fig: Dispatch SEC level} c) as a reference. Figure \ref{Fig: Dispatch SEC level} shows different aspects of electrolyzer operation for a CAPEX variation and a set SEC value in quadrant II of the $LCOH$ heat map known from Figure \ref{Fig: LCOH} a). Figure \ref{Fig: Dispatch SEC level} a) shows this quadrant for visualization. Here, a black arrow marks the set SEC value of -8\% and the direction of the CAPEX variation. Additionally, the data points considered for the APDCs in subfigures b)-f) are labeled accordingly. First, subfigures  b)-f) present the operation of electrolyzer 2 as a single system, followed by the operational behavior in the HES region, and finally the operation of electrolyzer 1 as a single system. In addition to the APDCs, which are presented as continuous lines, the dimensions of both electrolyzers are shown as dashed lines and the FLH are shown as dotted lines in the respective color. \newline

\noindent Figure \ref{Fig: Dispatch SEC level} b) presents the APDC of electrolyzer 2 as a single system. The dimension and FLH at this data point results in 385 MW and 5837 h, respectively.
Figure \ref{Fig: Dispatch SEC level} c) shows the exemplary HES operation, which was already discussed in detail in Section \ref{Sec:Exemplary HES result}. Electrolyzer 2 has a larger dimension of 290 MW than electrolyzer 1 with 105 MW. Additionally, electrolyzer 2 tends to operate in base load behavior with 6272 FLH compared to electrolyzer 1, that tends to operate in peak load behavior with about 1800 fewer FLH.
\noindent Figure \ref{Fig: Dispatch SEC level} d) shows a change in the dimension ratio of the electrolyzers. 
Electrolyzer 1 has a larger dimension of 224 MW than electrolyzer 2 with 183 MW. However, electrolyzer 2 still continues to operate in base load behavior with approximately 1800 more FLH compared to electrolyzer 1.
\noindent This trend continues in Figure \ref{Fig: Dispatch SEC level} e) with the nominal power of electrolyzer 1 increasing to 310 MW and the nominal power of electrolyzer 2 decreasing to 106 MW. Still, the FLH ratio remains the same, so does the operational behavior. 
\noindent Finally, Figure \ref{Fig: Dispatch SEC level} f) shows the APDC of electrolyzer 1 as a single system with a nominal power of 428 MW and 5680 FLH. Compared to electrolyzer 2 as a single system shown in Figure \ref{Fig: Dispatch SEC level}  b), the dimension of electrolyzer 1 as a single system is larger, which is due to its higher SEC. Concurrently, because of the larger dimension of electrolyzer 1, it needs less FLH for meeting the defined hydrogen demand. Figure \ref{Fig: Dispatch CAPEX level} presents the same analysis, but for a set CAPEX value of +42\% and a SEC variation. Comparing the APDCs in Figure \ref{Fig: Dispatch CAPEX level} b)-f) with those shown in Figure \ref{Fig: Dispatch SEC level} b)-f), the overall operational behavior remains similar. An only noticeable difference is only observed in Figure \ref{Fig: Dispatch CAPEX level} e), where the  dimension of electrolyzer 2, at 47 MW, corresponds to just about one-eighth of the dimension of electrolyzer 1. Nevertheless, electrolyzer 2 still operates in base load behavior with 7420 FLH, comparable to the operation shown in Figure \ref{Fig: Dispatch SEC level} e). \newline

\noindent It basically applies, that in case of a HES, the more efficient electrolyzer with the lower SEC operates in base load behavior and at least 1800 more FLH, regardless of its dimension. Concurrently, the less efficient electrolyzer with the higher SEC rather operates in peak load behavior. The same applies for the HES region in quadrant IV. The related figures, as well as the negligible impact of PPA price variation on the operational behavior, are presented in Section S3 of the Supplementary material.

\section{Conclusion}\label{Sec: Conclusion}

\noindent Hybrid electrolyzer systems (HES) are assumed to efficiently use the techno-economic advantages of the mature technologies alkaline water electrolysis (AWE) and proton exchange membrane water electrolysis (PEMWE) while potentially compensating for their respective disadvantages. Thus, green hydrogen production costs are expected to decrease, which would help the green hydrogen market ramp-up to proceed. In this paper, we investigated HES independently of technology-specific characteristics typically associated with AWE and PEMWE. Instead, we focused on a comprehensive analysis of the key techno-economic parameters investment cost (respectively capital expenditures, CAPEX) and efficiency (respectively specific energy consumption, SEC) of two electrolyzers integrated into a green hydrogen supply chain. By cost optimizing the design and dispatch of this hydrogen infrastructure, a HES is selected only if the combination of CAPEX and SEC of both electrolyzers favorably complement each other, resulting in a cost benefit over single electrolyzer systems. \newline

\noindent As a first result, HES configurations occur only in 5.0\% of the cases investigated in this study. Second, the maximum cost benefit for any HES configuration is 0.057 €/kgH2 in peripheral areas of the parameter space considered, which is approximately only 1\% of total green hydrogen production costs. Additional analyses considering variations in energy purchase prices, storage fees, availability of renewable energy, and baseline SEC value yield negligible changes to these results. Regarding the operation of a HES, this study found that the more efficient electrolyzer generally operates in base load behavior and has at least 1800 more full load hours compared to the less efficient electrolyzer. The increased operational flexibility of a HES is found to potentially result in reduced hydrogen storage capacity requirements. Nevertheless, this reduction in storage requirement is minor at 0.1\% and should therefore not be considered a key factor in planning decisions for green hydrogen production projects. Overall, the analysis conducted in this study indicates a very limited economic feasibility of HES, even when the ratios of the varied parameters are pushed to their technological and economic limits. Additionally, it provides only a negligible cost benefit compared to the next best single electrolyzer technology. Thus, HES do not provide sufficient techno-economic advantages to justify their implementation. This answers the research question by underlining that HES are more an economic fallacy than a smart strategy for electrolyzer operators. Furthermore, this paper explored the trade-off between CAPEX and SEC by defining a cost-based ratio based on the comprehensive sensitivity analysis conducted in this study. With regard to base assumptions, the results show that an improvement of 1 kWh/kg in SEC allows for a CAPEX improvement of up to 135,38 €/kW, without increased total production cost. \newline

\noindent It should be noted, that the conclusions of this study are based on the assumptions made. As the focus of our research lies on comprehensively analyzing the impact of CAPEX and SEC on HES, other technical parameters were outside of the scope. For example, degradation, ramping rate and operation range are uncertain technical parameters as well, and they are also assumed different for AWE and PEMWE. These technical uncertainties could therefore be subject of further investigations. Additionally, this study is based on a constant demand profile, reflecting large-scale industrial applications such as the chemical industry, which is expected to play a key role in early green hydrogen demand. Nevertheless, variations in the demand configuration could influence the HES results, as it would increase flexibility needs. Thus, a more detailed investigation could further improve the understanding of the applicability of HES. Due to these limitations and the complexity of the subject, this study alone cannot be expected to account for all existing technical and economic uncertainties related to HES. Nevertheless, the findings highlight key cost-driving parameters and evaluate the techno-economic viability of HES accordingly, providing further insights into the profitability of green hydrogen production.

\section*{CRediT authorship contribution statement}

\noindent \textbf{Marie Arnold:} Writing - original draft, Writing - review \& editing, Conceptualization, Data curation, Formal analysis, Investigation, Methodology, Software, Visualization. \textbf{Jonathan Brandt:} Writing - review \& editing, Supervision. \textbf{Geert Tjarks:} Writing - review \& editing, Conceptualization, Supervision. \textbf{Richard Hanke-Rauschenbach:} Conceptualization, Methodology, Supervision. 

\section*{Declaration of competing interest}

\noindent The authors declare that they have no known competing financial interest or personal relationships that could have appeared to influence the work reported in this paper.

\section*{Data and code availability}

\noindent The model code, input data and key results of this study are publicly available at Zenodo \cite{Zenodo}.

\section*{Acknowledgements}

\noindent The results presented were achieved by computations carried out on the cluster system at Leibniz Universität Hannover, Germany.

\appendix
\section{Mathematical description of optimization problem}\label{app1}

\noindent The objective function in Equation \eqref{Eq.: obj} aims to minimize the total annual expenditures of the system under consideration presented in Section \ref{Fig: System under consideration} (compare with Equation \eqref{Eq.: objective function}).

\begin{align}\label{Eq.: obj}
    \min_X \quad C^{\text{PPA}} + C^{\text{Storage}} + \sum_{i=1}^N C_i^{\text{Electrolyzer}} - R^{\text{Surplus}}
\end{align}

\noindent The total PPA expenses $C^{\text{PPA}}$ are calculated as a double summation over all time steps $t$ and PPA options, shown in Equation \eqref{Eq.: PPA costs}. Thereby, the individual PPA costs are calculated as the product of the rated nominal power of the respective PPA option $P^{\text{PPA}}$, the specific power purchase price $p^{\text{PPA}}$, the capacity factor $f_t^{\text{PPA}}$, and the length of time step $\Delta t$.

\begin{align}\label{Eq.: PPA costs}
    C^{\text{PPA}} = \sum_{t=1}^{\text{T}} \sum_{\text{PPA}} p^{\text{PPA}} \cdot P^{\text{PPA}} \cdot f_t^{\text{PPA}} \cdot \Delta t
\end{align}

\noindent The costs of both electrolyzer systems ${C_i}^{\text{Electrolyzer}}$ are equally calculated as shown in Equation \eqref{Eq.:Elycostfirst}. The electrolyzer cost consist of the peripherals cost share $C_i^\text{Peri}$, the stack cost share $C_i^\text{Stacks}$ and the costs for the electrolyzer operation $C_i^\text{Operation}$. 
The peripherals cost calculation in Equation \eqref{Eq.:Elycostsecond} is defined as the multiplication of the nominal power of the respective electrolyzer $P_i^{\text{Ely,Nom}}$ with the allocated CAPEX $c^{\text{Ely,CAPEX}}$, the peripherals cost share $S^{\text{Ely,Peri}}$ and the annuity factor of the peripherals $A^{\text{Ely,Peri}}$.
The stack cost calculation in Equation \eqref{Eq.:Elycostthird} is defined as the multiplication of the nominal power of the respective electrolyzer $P_i^{\text{Ely,Nom}}$ with the allocated CAPEX $c^{\text{Ely,CAPEX}}$, the stack cost share  $S^{\text{Ely,Stacks}}$ and the annuity factor of the stacks $A^{\text{Ely,Stacks}}$.
The operation cost add up from the nominal power of the respective electrolyzer $P_i^{\text{Ely, Nom}}$ multiplied with the maintenance costs $c^{\text{Ely,OPEX}}$ as well as the water costs of the electrolyzer $c^{\text{Water}}$ multiplied with the specific water consumption $w^{\text{Ely}}$, the sum of the hydrogen produced by the respective electrolyzer $\dot{m}^{\text{Ely},i}_{t}$ and the length of time step $\Delta t$. The respective calculation is shown in Equation \eqref{Eq.:Elycostlast}.

\begin{align}
   \label{Eq.:Elycostfirst} C_i^{\text{Electrolyzer}} &= C_i^\text{Peri} + C_i^\text{Stacks} + C_i^\text{Operation} && \forall i \in \{1,2\} \\ \label{Eq.:Elycostsecond}
    C_i^\text{Peri} &= P_i^{\text{Ely,Nom}} \cdot c_i^{\text{Ely,CAPEX}} \cdot s^{\text{Ely,Peri}} \cdot A^\text{Ely,Peri}  &&  
    \forall i \in \{1,2\} \\ \label{Eq.:Elycostthird}
    C_i^\text{Stacks} &= P_i^{\text{Ely,Nom}} \cdot c_i^{\text{Ely,CAPEX}} \cdot s^{\text{Ely,Stacks}} \cdot A^\text{Ely,Stacks}  && \forall i \in \{1,2\} \\ \label{Eq.:Elycostlast}
    C_i^\text{Operation} &= P_i^{\text{Ely,Nom}} \cdot c^{\text{Ely,OPEX}} +  c^{\text{Water}} \cdot w^{\text{Ely}} \cdot \sum_{t=1}^T \dot{m}^{\text{Ely},i}_{t} \cdot \Delta t  &&  \forall i \in \{1,2\} 
\end{align}

\noindent The calculation of $A^{\text{Ely,Peri}}$ is presented in Equation \eqref{Eq.:Peri_Annuity} with the interest rate representing the real weighted average cost of capital $r_{\text{in}}^{\text{Ely}}$ and the depreciation time of the electrolyzers' peripherals $t^{\text{Ely,Peri}}_{\text{dep}}$.

\begin{align}\label{Eq.:Peri_Annuity}
    A^{\text{Ely,Peri}} = \frac{r_{\text{in}}^{\text{Ely}} \cdot \left(1 + r_{\text{in}}^{\text{Ely}}\right)^{t^{\text{Ely,Peri}}_{\text{dep}}}}{\left(1 + r_{\text{in}}^{\text{Ely}}\right)^{t^{\text{Ely,Peri}}_{\text{dep}}} - 1}
\end{align}

\noindent The calculation of $A^{\text{Ely,Stacks}}$ shown in Equation \eqref{Eq.:Stacks_Annuity} is performed accordingly with the exception of a differing depreciation time of the stacks $t^{\text{Ely,Stacks}}_{\text{dep}}$.

\begin{align}\label{Eq.:Stacks_Annuity}
    A^{\text{Ely,Stacks}} = \frac{r_{\text{in}}^{\text{Ely}} \cdot \left(1 + r_{\text{in}}^{\text{Ely}}\right)^{t^{\text{Ely,Stacks}}_{\text{dep}}}}{\left(1 + r_{\text{in}}^{\text{Ely}}\right)^{t^{\text{Ely,Stacks}}_{\text{dep}}} - 1}
\end{align}

\noindent The costs of the hydrogen storage $C^{\text{Storage}}$ are calculated as follows by adding the booking costs resulting from multiplying the capacity fee $p^{\text{Storage,cap}}$ with the booked storage capacity $m^{\text{Storage,max}}$ to the operational costs resulting from multiplying the usage fee $p^{\text{Storage,turn}}$ and the sum of the time-dependent mass flow rate $\dot{m}^{\text{Storage,in}}_{t}$ multiplied by the length of time step $\Delta t$.

\begin{align}
    C^{\text{Storage}} = p^{\text{Storage,cap}} \cdot m^{\text{Storage,max}} + p^{\text{Storage,turn}} \cdot \sum_{t=1}^T \dot{m}^{\text{Storage,in}}_{t} \cdot \Delta t
\end{align}

\noindent The surplus revenues $R^{\text{Surplus}}$ are calculated by multiplying the sum of the surplus power $P^{\text{Surplus}}_{t}$ by the grid electricity price $p^{\text{Grid}}$ and the length of time step $\Delta t$, shown in Equation \eqref{Eq: Surplus}.

\begin{align}\label{Eq: Surplus}
    R^{\text{Surplus}} = \sum_{t=1}^T P^{\text{Surplus}}_{t} \cdot p^{\text{Grid}} \cdot \Delta t
\end{align}

\noindent The following equality constraints define the technical operation of the system under consideration shown in Figure \ref{Fig: System under consideration}. The respective optimization parameters are listed in Table \ref{Tab:Optimization parameters} and the variables in Table \ref{Tab:Optimization variables}.
\begin{align}
  0 &= \sum_{i=1}^N \dot{m}^{\text{Ely},i}_{t} - \left( \dot{m}_{t}^{\text{Storage,in}} - \dot{m}_{t}^{\text{Storage,out}} \right) - \dot{m}_{t}^{\text{Demand}} && \forall t \in \{1,2,3,\dots,T\}, N\in\{1,2\} \\
  0 &= P_t^{\text{Onshore}} + P_t^{\text{Offshore}} + P_t^{\text{Solar}} - \sum_{i=1}^N P_t^{\text{Ely},i} - P_t^{\text{Grid}} && \forall t \in \{1,2,3,\dots,T\}, N\in\{1,2\} \\
  m_t^{\text{Storage}} &= m_{t-1}^{\text{Storage}} + \left( \dot{m}_t^{\text{Storage,in}} - \dot{m}_t^{\text{Storage,out}} \right) \cdot \Delta t && \forall t \in \{2,3,4,\dots,T\} \\
  m_1^{\text{Storage}} &= m_T^{\text{Storage}} + \left( \dot{m}_1^{\text{Storage,in}} - \dot{m}_1^{\text{Storage,out}} \right) \cdot \Delta t && \forall t \in \{1,2,3,\dots,T\}
\end{align}

\noindent In the following, the inequality constraints of the optimization problem are shown.

\begin{align}
  P_t^{\text{Onshore}} &\leq P^{\text{PPA,Onshore}} \cdot f_t^{\text{PPA,Onshore}} && \forall t \in \{1,2,3,\dots,T\} \\
  P_t^{\text{Offshore}} &\leq P^{\text{PPA,Offshore}} \cdot f_t^{\text{PPA,Offshore}} && \forall t \in \{1,2,3,\dots,T\} \\
  P_t^{\text{Solar}} &\leq P^{\text{PPA,Solar}} \cdot f_t^{\text{PPA,Solar}} && \forall t \in \{1,2,3,\dots,T\} \\
  P_t^{\text{Grid}} &\geq 0 && \forall t \in \{1,2,3,\dots,T\} \\
  0 &\leq m_t^{\text{Storage}} \leq m^{\text{Storage,max}} && \forall t \in \{1,2,3,\dots,T\} \\
   R^\text{Storage,in} &\geq  \frac{\dot{m}^\text{Storage,in}_t}{m^\text{Storage,max}} && \forall t \in \{1,2,3,\dots,T\}  \\
  R^\text{Storage,out} &\geq \frac{\dot{m}^\text{Storage,out}_t}{m^\text{Storage,max}} && \forall t \in \{1,2,3,\dots,T\}  \\
\end{align}

\noindent The load-dependency of the energy demand of each electrolyzer $i \in \{1,2\}$ is integrated into the optimization problem by the linearization method used in \cite{Brandt2024}. The respective constraints constructing a convex search space are shown in the following, with the y-axis intersect $b^{\text{lin}}_j$ of the respective linear constraint $j$.

\begin{align}
 \dot{m}^{\text{Ely},i}_{t} - a^{\text{lin}}_j \cdot P_t^{\text{Ely},i} - b^{\text{lin}}_j \leq 0, \ &\forall t \in \{1,2,3,\dots,T\},\ \forall j \in \{1,2,3,\dots,J{-}1\}, \ \forall i \in \{1,2\} \\
    a^{\text{lin}}_j = \frac{j+1}{\epsilon^{\text{Ely},i}_{j+1}} - \frac{j}{\epsilon^{\text{Ely},i}_{j}} ,\ &\forall j \in \{1,2,3,\dots,J{-}1\}, \ \forall i \in \{1,2\} \\
       b^{\text{lin}}_j = \frac{\frac{j}{J-1} \cdot P_i^{\text{Ely,Nom}}}{\epsilon^{\text{Ely},i}_j} -
    a^{\text{lin}}_j \cdot \frac{j}{J-1} \cdot P_i^{\text{Ely,Nom}} ,\ &\forall j \in \{1,2,3,\dots,J{-}1\}, \ \forall i \in \{1,2\}
\end{align}

\noindent The number of linearization steps used in this study is 37. To reduce the search space and thereby the optimization time, a lower bound is defined and shown in Equation \eqref{Eq:lower_bound}.

\begin{align}\label{Eq:lower_bound}
    P_i^{\text{Ely,Nom}} - \dot{m}^{\text{Ely,i}}_{t} \cdot \epsilon_i^{\text{Ely,Nom}} &\leq 0 && \forall t \in \{1,2,3,\dots,T\}, \ \forall i \in \{1,2\}
\end{align}

\begin{table}[H]
\centering
\caption{Optimization parameters}
\renewcommand{\arraystretch}{1.5} 
\begin{tabular}{ |p{2.5cm}|p{10cm}|} 
\hline
Parameter & Description \\
\hline
$\Delta t$                & Length of time step in hours\\ 
$f_t^{\text{Onshore}}$    & Onshore capacity factor at time step $t$\\ 
$f_t^{\text{Offshore}}$   & Offshore capacity factor at time step $t$ \\ 
$f_t^{\text{Solar}}$      & Solar capacity factor at time step $t$ \\ 
$\epsilon_i^{\text{Ely,Nom}}$    & Specific energy demand of electrolyzer $i$ at nominal power\\ 
$a^{\text{lin}}_j$        & Gradient of linearized characteristic curve of the electrolyzer between linearization steps $j$ and $j+1$ \\ 
$b^{\text{lin}}_j$        & Y-axis intersect of the linearized characteristic curve of the electrolyzer between linearization steps $j$ and $j+1$ \\ 
$\epsilon^{\text{Ely}}_{i,j}$     & Specific energy demand of electrolyzer $i$ at $\frac{j}{J-1} \cdot 100\%$ of nominal power\\ 
$\dot{m}^{\text{Demand}}_{t}$  & Predefined hydrogen demand at time step $t$\\
\hline
\end{tabular}
\label{Tab:Optimization parameters}
\end{table}

\begin{table}[H]
\centering
\caption{Optimization variables}
\renewcommand{\arraystretch}{1.5} 
\begin{tabular}{ |p{2.5cm}|p{10cm}| } 
\hline
Variable & Description \\
\hline
$P_t^{\text{Onshore}}$         & Onshore wind power at time step $t$ \\
$P_t^{\text{Offshore}}$        & Offshore wind power at time step $t$ \\ 
$P_t^{\text{Solar}}$           & Solar power at time step $t$\\ 
$P^{\text{PPA,Onshore}}$       & Booked onshore PPA for one year \\ 
$P^{\text{PPA,Offshore}}$      & Booked offshore PPA for one year \\ 
$P^{\text{PPA,Solar}}$         & Booked solar PPA for one year\\ 
$P_t^{\text{Grid}}$            & Surplus power at time step $t$ \\ 
$P_{t}^{\text{Ely,i}}$         & Power consumption of electrolyzer $i$ at time step $t$ \\ 
$P_i^{\text{Ely,Nom}}$      & Nominal power of electrolyzer $i$ \\
$\dot{m}_t^{\text{Ely,i}}$             & Hydrogen produced by electrolyzer $i$ at time step $t$  \\ 
$\dot{m}_t^{\text{Storage,in}}$     & Hydrogen stored at time step $t$ \\ 
$\dot{m}_t^{\text{Storage,out}}$    & Hydrogen provided by the storage at time step $t$ \\ 
$m_t^{\text{Storage}}$         & Stored hydrogen mass at time step $t$ \\ 
$m^{\text{Storage,max}}$       & Booked storage capacity for one year \\ 
\hline
\end{tabular}
\label{Tab:Optimization variables}
\end{table}

\section{Technical and economic parameter assumptions}\label{app2}

\noindent The data for the capacity factors of the renewable energy sources used in Equation \eqref{Eq.: PPA costs} was taken from \cite{RenewablesNinja}, whereby the solar data is based on \cite{Pfenninger2016} and the wind data is based on \cite{Staffell2016}. Weather year 2023 was chosen in this study as well as three renewable power production sites located in Northern Germany. The exact configuration is given in Table \ref{renewableninjatable}.

\begin{table}[H]
\centering
\caption{Renewables configuration}
\renewcommand{\arraystretch}{1.5} 
\begin{tabular}{ |p{3cm}|p{3cm}|p{3.5cm}|p{2.5cm}|} 
\hline
Parameter & Wind onshore & Wind offshore & Solar \\
\hline
Local time  & Europe/Berlin  & Europe/Berlin & Europe/Berlin  \\ 

Location & Dietrichsfeld  & Riffgat & Oldenburg  \\ 

Electricity & kW & kW & kW \\ 

Latitude & 53.5288° & 53.6903° & 53.1756°  \\ 

Longitude & 7.4704° & 6.4811° & 8.1719°  \\ 

Dataset & Merra2 & Merra2 & Merra2  \\ 

Capacity & 1 & 1 & 1  \\ 

Tilt/Azimuth & / & / & 35°/180°  \\ 

Height & 135 m & 90 m & / \\ 

Turbine & Enercon E126 3500 & Siemens SWT 3.6 120 & /  \\ 
\hline
\end{tabular}
\label{renewableninjatable}
\end{table}

\begin{table}[H]
\caption{Technical and economic parameter assumptions}
\centering
\renewcommand{\arraystretch}{1.5} 
\begin{tabular}{ |p{2cm}|p{1.5cm}|p{3.3cm}|p{1.3cm}|p{1.8cm}|p{1.3cm}|} 
\hline
Component & Parameter & Description & Value & Unit & Reference \\
\hline
Wind onshore & $p^{\text{PPA,Onshore}}$& Pay-as-produced price & 0.0729 & €$_{2024}$/kWh & \cite{Brandt2026}\\ 
\hline
Wind offshore & $p^{\text{PPA,Offshore}}$& Pay-as-produced price & 0.0883 & €$_{2024}$/kWh & \cite{Brandt2026}\\ 
\hline
Solar & $p^{\text{PPA,Solar}}$ & Pay-as-produced price & 0.0555 & €$_{2024}$/kWh & \cite{Brandt2026} \\ 
\hline
Grid & $p^{\text{Grid}}$ & Electricity price & 0.01976 & €$_{2024}$/kWh &  \cite{Brandt2026}\\ 
\hline
 Electrolyzer (applies equally to electrolyzer 1 and electrolyzer 2)& $c^{\text{Ely,OPEX}}$ & Maintenance OPEX fix  & 23.45 & €$_{2024}$/(kW$\cdot$a) & \cite{Holst2021} \\ 
 & $s^{\text{Ely,Peri}}$ &  Cost share peripherals & 75 & \% & \cite{Holst2021} \\ 
 & $s^{\text{Ely,Stacks}}$ &  Cost share stacks & 25 & \% & \cite{Holst2021} \\ 
 & $t^{\text{Ely,Peri}}_{\text{dep}}$ & Depreciation time peripherals & 20 & a & \cite{UdoLubenau2022} \\ 
 & $t^{\text{Ely,Stacks}}_{\text{dep}}$ & Depreciation time peripherals & 10 & a &  \cite{SiemensEnergy} \\ 
 & $r_{\text{in}}^{\text{Ely}}$ & Interest rate  & 7 & \% & \cite{Brandt2026} \\ 
 &  & Decrease of specific energy demand  & 1 & \% per 10\% load reduction & \cite{Brandt2026} \\
 & $w^{\text{Ely}}$ & Specific water consumption  & 14 & kgH$_2$O/kgH$_2$ & \cite{Brandt2026}\\ 
 & $c^{\text{Water}}$ & Water costs  & 3.725 & €$_{2024}$/m$^3$H$_2$O & \cite{Brandt2026}\\ 
\hline 
Hydrogen cavern storage & $p^{\text{Storage,cap}}$ & Capacity fee  & 12.75 & $\text{€}_{2024}$/$(\text{kg} \cdot \text{a})$ & \cite{Brandt2026} \\
 & $p^{\text{Storage,turn}}$ & Usage fee  & 0.36 & $\text{€}_{2024}$/$\text{kg}$ & \cite{Brandt2026}\\ 
 & $R^{\text{Storage,in}}$ & Specific storage injection rate  & 0.001 & 1/h & \cite{Frischmuth2024}\\ 
 & $R^{\text{Storage,out}}$ & Specific storage withdrawal rate  & 0.002 & 1/h & \cite{Frischmuth2024}\\ 
 \hline
 Hydrogen demand & $\dot{m}^{\text{Demand}}_{t}$ &  Amount  & 5500 & kg/h & Own assumption\\ 
\hline
\end{tabular}
\end{table}

\section{Software}\label{app3}
\noindent The optimization problem was implemented in Matlab \cite{MATLAB}. Gurobi \cite{gurobi} was used as the mathematical solver for all optimizations run in this study.\newpage

\bibliographystyle{elsarticle-num}
\bibliography{HES}

@article{Sezer2025,
   author = {Nurettin Sezer and Sertac Bayhan and Ugur Fesli and Antonio Sanfilippo},
   doi = {10.1016/j.mset.2024.07.006},
   issn = {25892991},
   journal = {Materials Science for Energy Technologies},
   month = {1},
   pages = {44-65},
   publisher = {KeAi Communications Co.},
   title = {A comprehensive review of the state-of-the-art of proton exchange membrane water electrolysis},
   volume = {8},
   year = {2025}
}

@techreport{Smolinka2018,
   author = {T. Smolinka and N. Wiebe and P. Sterchele and A. Palzer and F. Lehner and M. Jansen and S. Kiemel and R. Miehe and S. Wahren and F. Zimmermann},
   title = {{Studie IndWEDe Industrialisierung der Wasserelektrolyse in Deutschland: Chancen und Herausforderungen für nachhaltigen Wasserstoff für Verkehr, Strom und Wärme}},
   year = {2018},
   url = {https://www.now-gmbh.de/wp-content/uploads/2020/09/indwede-studie_v04.1.pdf}
}

@article{Sayed-Ahmed2024,
   author = {H. Sayed-Ahmed and I. Toldy and A. Santasalo-Aarnio},
   doi = {10.1016/j.rser.2023.113883},
   issn = {18790690},
   journal = {Renewable and Sustainable Energy Reviews},
   month = {1},
   publisher = {Elsevier Ltd},
   title = {Dynamic operation of proton exchange membrane electrolyzers—Critical review},
   volume = {189},
   year = {2024}
}

@article{ShivaKumar2022,
   author = {S. Shiva Kumar and Hankwon Lim},
   doi = {10.1016/j.egyr.2022.10.127},
   issn = {23524847},
   journal = {Energy Reports},
   month = {11},
   pages = {13793-13813},
   publisher = {Elsevier Ltd},
   title = {An overview of water electrolysis technologies for green hydrogen production},
   volume = {8},
   year = {2022}
}

@misc{Accelera,
   author = {Accelera by Cummins},
   title = {HyLYZER 1000-30},
   url = {https://www.accelerazero.com/sites/default/files/2025-09/HyLYZER-1000-spec-sheet-2025.pdf}
}

@misc{NelPEMWE,
   author = {Nel},
   title = {PEM 100},
   url = {https://nelhydrogen.com/wp-content/uploads/2025/03/PEM-100-Standard-Plant-Solution_PD-0600-0144-Rev-B-1.pdf}
}

@misc{NelAWE,
   author = {Nel},
   title = {AWE 100},
   url = {https://nelhydrogen.com/wp-content/uploads/2025/08/AWE-100-Standard-Plant-Solution_DOC006451_01-1.pdf}
}

@misc{HydrogenPro,
   author = {Hydrogen Pro},
   title = {Technical electrolyzer data},
   url = {https://hydrogenpro.com/wp-content/uploads/2025/06/Technical-details_1.3.pdf}
}

@misc{QuestOne,
   author = {QuestOne},
   title = {Modular Hydrogen Platform},
   url = {https://www.questone.com/fileadmin/user_upload/Quest_One_Data-Sheet_MHP_DE_0825.pdf}
}

@book{IRENA2020,
   author = {IRENA},
   city = {Abu Dhabi},
   isbn = {9789292602956},
   publisher = {International Renewable Energy Agency},
   title = {Green Hydrogen Cost Reduction Scaling up Electrolysers to Meet the 1.5°C Climate Goal},
   url = {https://www.irena.org/-/media/Files/IRENA/Agency/Publication/2020/Dec/IRENA_Green_hydrogen_cost_2020.pdf},
   year = {2020}
}

@misc{RWEAG2025,
   author = {RWE AG},
   month = {3},
   title = {RWE and TotalEnergies agree groundbreaking long-term offtake agreement for green hydrogen},
   url = {https://www.rwe.com/en/press/rwe-ag/2025-03-12-rwe-and-totalenergies-agree-long-term-offtake-agreement-for-green-h2/},
   year = {2025}
}

@misc{GreenHydrogenSystems,
   author = {Green Hydrogen Systems},
   title = {Hyprovide X-Series},
   url = {https://www.greenhydrogensystems.com/electrolysers/hyprovide-x-series-6mw-modular-electrolyser}
}

@article{Ma2025,
   author = {Suliang Ma and Zeqing Meng and Yang Mei and Mingxuan Chen and Yuan Jiang},
   doi = {10.3390/app15063135},
   issn = {20763417},
   issue = {6},
   journal = {Applied Sciences (Switzerland)},
   month = {3},
   publisher = {Multidisciplinary Digital Publishing Institute (MDPI)},
   title = {A Multi-Optimization Method for Capacity Configuration of Hybrid Electrolyzer in a Stand-Alone Wind-Photovoltaic-Battery System},
   volume = {15},
   year = {2025}
}

@article{Shin2026,
   author = {Haeseong Shin and Dohyung Jang and Hee Sun Shin and Sungtae Park and Sanggyu Kang},
   doi = {10.1016/j.apenergy.2026.127451},
   issn = {03062619},
   journal = {Applied Energy},
   month = {4},
   publisher = {Elsevier Ltd},
   title = {High-resolution dynamic modeling and techno-economic optimization of off-grid PV–electrolysis–BESS systems for green hydrogen production},
   volume = {409},
   year = {2026}
}

@article{Mingxuan2024,
   author = {Chen Mingxuan and He Bin and Zhang Baoping and Ren Yongfeng and Pan Yu and Yun Pingping and Mi Yue},
   doi = {10.1080/14786451.2024.2403494},
   issn = {1478646X},
   issue = {1},
   journal = {International Journal of Sustainable Energy},
   publisher = {Taylor and Francis Ltd.},
   title = {Intraday energy management strategy for wind-hydrogen coupled systems based on hybrid electrolysers},
   volume = {43},
   year = {2024}
}

@article{Liang2024,
   author = {Yuan Liang and Haoyuan Ma and Zhonghao Liang and Hongqing Wang and Jianlin Li},
   doi = {10.1016/j.segan.2024.101539},
   issn = {23524677},
   journal = {Sustainable Energy, Grids and Networks},
   month = {12},
   publisher = {Elsevier Ltd},
   title = {A method for configuring hybrid electrolyzers based on joint wind and photovoltaic power generation modeling using copula functions},
   volume = {40},
   year = {2024}
}

@article{Tang2025,
   author = {Yuzhen Tang and Zhuoqun Zheng and Fanqi Min and Jingying Xie and Hengzhao Yang},
   doi = {10.1016/j.renene.2025.122555},
   issn = {18790682},
   journal = {Renewable Energy},
   month = {4},
   publisher = {Elsevier Ltd},
   title = {An optimization framework for component sizing and energy management of hybrid electrolyzer systems considering physical characteristics of alkaline electrolyzers and proton exchange membrane electrolyzers},
   volume = {243},
   year = {2025}
}

@article{Shi2026,
   author = {Shujing Shi and Yuzhe Pan and Yihang Li and Hao Wang and Youjun Lu},
   doi = {10.1016/j.renene.2025.124814},
   issn = {18790682},
   journal = {Renewable Energy},
   month = {2},
   publisher = {Elsevier Ltd},
   title = {Capacity configuration and optimization of an off-grid wind-solar-hydrogen integrated system considering hybrid hydrogen production with alkaline electrolyzers and proton exchange membrane electrolyzers},
   volume = {258},
   year = {2026}
}

@article{Xu2024,
   author = {Guanxin Xu and Yan Wu and Shuo Tang and Yufei Wang and Xinhai Yu and Mingyan Ma},
   doi = {10.1016/j.energy.2024.131827},
   issn = {18736785},
   journal = {Energy},
   month = {9},
   publisher = {Elsevier Ltd},
   title = {Optimal design of hydrogen production processing coupling alkaline and proton exchange membrane electrolyzers},
   volume = {302},
   year = {2024}
}

@article{Yu2024,
   author = {Binbin Yu and Guangyao Fan and Kai Sun and Jing Chen and Bo Sun and Peigen Tian},
   doi = {10.1016/j.energy.2024.131508},
   issn = {18736785},
   journal = {Energy},
   month = {8},
   publisher = {Elsevier Ltd},
   title = {Adaptive energy optimization strategy of island renewable power-to-hydrogen system with hybrid electrolyzers structure},
   volume = {301},
   year = {2024}
}

@article{Yang2025,
   author = {Wenlong Yang and Yafeng Hu and Bingxin Guo and Wenchao Zhu and Changjun Xie and Li You and Liangli Xiong and Leiqi Zhang},
   doi = {10.1016/j.energy.2025.138607},
   issn = {18736785},
   journal = {Energy},
   month = {11},
   publisher = {Elsevier Ltd},
   title = {Adaptive operation strategy for wind-hydrogen systems integrating alkaline and proton exchange membrane electrolyzers},
   volume = {337},
   year = {2025}
}

@article{Zhang2024,
   author = {Qinjin Zhang and Di Xie and Yuji Zeng and Yancheng Liu and Heyang Yu and Siyuan Liu},
   doi = {10.1016/j.renene.2024.121116},
   issn = {18790682},
   journal = {Renewable Energy},
   month = {10},
   publisher = {Elsevier Ltd},
   title = {Optimizing wind-solar hydrogen production through collaborative strategy with ALK/PEM multi-electrolyzer arrays},
   volume = {232},
   year = {2024}
}

@article{Munther2025,
   author = {Hassan Munther and Qusay Hassan and Anees A. Khadom and Hameed B. Mahood},
   doi = {10.1016/j.uncres.2024.100122},
   issn = {26665190},
   journal = {Unconventional Resources},
   month = {1},
   publisher = {KeAi Communications Co.},
   title = {Evaluating the techno-economic potential of large-scale green hydrogen production via solar, wind, and hybrid energy systems utilizing PEM and alkaline electrolyzers},
   volume = {5},
   year = {2025}
}

@article{Wang2023,
   author = {Xiaohua Wang and Andrew G. Star and Rajesh K. Ahluwalia},
   doi = {10.3390/en16134964},
   issn = {19961073},
   issue = {13},
   journal = {Energies},
   month = {7},
   publisher = {Multidisciplinary Digital Publishing Institute (MDPI)},
   title = {Performance of Polymer Electrolyte Membrane Water Electrolysis Systems: Configuration, Stack Materials, Turndown and Efficiency},
   volume = {16},
   year = {2023}
}

@inproceedings{Guo2019,
   author = {Yujing Guo and Gendi Li and Junbo Zhou and Yong Liu},
   doi = {10.1088/1755-1315/371/4/042022},
   issn = {17551315},
   issue = {4},
   booktitle = {IOP Conference Series: Earth and Environmental Science},
   month = {12},
   publisher = {Institute of Physics Publishing},
   title = {Comparison between hydrogen production by alkaline water electrolysis and hydrogen production by PEM electrolysis},
   volume = {371},
   year = {2019}
}

@article{Tan2025,
   author = {Qingshan Tan and Ke Li and Longquan Zeng and Lu Xie and Man Cheng and Wei He},
   doi = {10.3390/en18082008},
   issn = {19961073},
   issue = {8},
   journal = {Energies},
   month = {4},
   publisher = {Multidisciplinary Digital Publishing Institute (MDPI)},
   title = {A Multi-State Rotational Control Strategy for Hydrogen Production Systems Based on Hybrid Electrolyzers},
   volume = {18},
   year = {2025}
}

@article{Hu2025,
   author = {Haowen Hu and Fengxiang Chen and Yejing Xu and Huan Ye and Zhipeng Hou and Bo Zhang and Xiuxiang Chen},
   doi = {10.1016/j.energy.2025.137647},
   issn = {18736785},
   journal = {Energy},
   month = {10},
   publisher = {Elsevier Ltd},
   title = {Triple-level cost-effective sizing optimization for solar-powered hybrid hydrogen system with PEM and alkaline electrolyzers},
   volume = {334},
   year = {2025}
}

@misc{IEA2025,
   author = {IEA},
   title = {Tracking electrolysers},
   url = {https://www.iea.org/energy-system/low-emission-fuels/electrolysers},
   year = {2025}
}

@techReport{Ffe2025,
   author = {Ffe},
   title = {{Von der Theorie zur Praxis: Warum grüner Wasserstoff teurer ist als gedacht}},
   url = {https://www.ffe.de/wp-content/uploads/2025/07/Discussion_Paper-Investitionskosten_Elektrolyse-2.pdf},
   year = {2025}
}

@article{Wei2019,
   author = {A. Weiß and A. Siebel and M. Bernt and T.-H. Shen and V. Tileli and H. A. Gasteiger},
   doi = {10.1149/2.0421908jes},
   issn = {0013-4651},
   issue = {8},
   journal = {Journal of The Electrochemical Society},
   pages = {F487-F497},
   publisher = {The Electrochemical Society},
   title = {Impact of Intermittent Operation on Lifetime and Performance of a PEM Water Electrolyzer},
   volume = {166},
   year = {2019}
}

@article{Grigoriev2020,
   author = {S. A. Grigoriev and V. N. Fateev and D. G. Bessarabov and P. Millet},
   doi = {10.1016/j.ijhydene.2020.03.109},
   issn = {03603199},
   issue = {49},
   journal = {International Journal of Hydrogen Energy},
   month = {10},
   pages = {26036-26058},
   publisher = {Elsevier Ltd},
   title = {Current status, research trends, and challenges in water electrolysis science and technology},
   volume = {45},
   year = {2020}
}

@article{Arnold2025,
   author = {Marie Arnold and Jonathan Brandt and Geert Tjarks and Anna Vanselow and Richard Hanke-Rauschenbach},
   doi = {10.1016/j.ecmx.2025.101261},
   title = {Cost-optimized replacement strategies for water electrolysis systems affected by degradation},
   year = {2025}
}

@article{Brandt2024,
   author = {Jonathan Brandt and Thore Iversen and Christoph Eckert and Florian Peterssen and Boris Bensmann and Astrid Bensmann and Michael Beer and Hartmut Weyer and Richard Hanke-Rauschenbach},
   doi = {10.1038/s41560-024-01511-z},
   issn = {20587546},
   issue = {6},
   journal = {Nature Energy},
   month = {6},
   pages = {703-713},
   publisher = {Nature Research},
   title = {Cost and competitiveness of green hydrogen and the effects of the European Union regulatory framework},
   volume = {9},
   year = {2024}
}

@article{Omar2025,
   author = {Mas Omar and Fitri Yakub and Rahayu Tasnim and Fikri Zhafran and Ahmad Aiman Azmi and Takafumi Morisaki and Sathiabama T. Thirugnana and A. Bakar Jaafar and Yasuyuki Ikegami},
   doi = {10.1016/j.ijhydene.2025.05.080},
   issn = {03603199},
   journal = {International Journal of Hydrogen Energy},
   month = {6},
   pages = {339-357},
   publisher = {Elsevier Ltd},
   title = {Techno-economic analysis of sustainable hydrogen production via 10 MW hybrid ocean thermal energy conversion offshore plant integrated alkaline and proton exchange membrane electrolyzers},
   volume = {136},
   year = {2025}
}

@article{Knoebl2026,
   author = {Melanie Knoebl and Christoph Mueller and Darja Markova},
   doi = {10.1007/s00267-025-02331-x},
   issn = {14321009},
   issue = {2},
   journal = {Environmental Management},
   month = {2},
   pmid = {41436602},
   publisher = {Springer},
   title = {Implementation of a Green Hydrogen Ecosystem in Central Europe: A Stakeholder Analysis},
   volume = {76},
   year = {2026}
}

@article{Odenweller2025,
   author = {Adrian Odenweller and Falko Ueckerdt},
   doi = {10.1038/s41560-024-01684-7},
   issn = {2058-7546},
   issue = {1},
   journal = {Nature Energy},
   month = {1},
   pages = {110-123},
   title = {The green hydrogen ambition and implementation gap},
   volume = {10},
   year = {2025}
}

@misc{CEPCI,
   author = {{Chemical Engineering}},
   month = {6},
   title = {The Chemical Engineering Plant Cost Index},
   year = {2024},
    URL = {https://www.chemengonline.com/pci-home}
}

@misc{EuropeanCommission2023,
   author = {European Commission},
   month = {2},
   title = {Delegated regulation (EU) 2023/1184 on Union methodology for RFNBOs },
   url = {http://data.europa.eu/eli/reg_del/2023/1184},
   year = {2023}
}

@article{Brandt2026,
	author={Brandt, Jonathan and Bensmann, Astrid and Hanke-Rauschenbach, Richard},
	title={Negative redispatch power for green hydrogen production: Game changer or lame duck? A German perspective},
	journal={Progress in Energy},
	url={http://iopscience.iop.org/article/10.1088/2516-1083/ae6011},
	year={2026}
}

@techReport{Holst2021,
   author = {Marius Holst and Stefan Aschbrenner and Tom Smolinka and Christopher Voglstätter and Gunter Grimm},
   title = {Cost forecast for low tepmerature electrolysis-technology driven bottom-up prognosis for PEM and alkaline water electrolysis systems},
    year = {2021},
    url = {https://publica-rest.fraunhofer.de/server/api/core/bitstreams/459031bc-d874-4a24-b2af-5300add8a045/content}
}

@techReport{UdoLubenau2022,
   author = {Hagen Bültemeier and Cruz Marrune and Jens Hüttenrauch and Udo Lubenau and Matthias Janssen Maximiliane Reger},
   title = {{H2-Kurzstudie: Wasserstoffqualität in einem gesamtdeutschen Wasserstoffnetz}},
   year = {2022},
   url = {https://www.dvgw.de/medien/dvgw/forschung/berichte/g202140-abschlussbericht-h2-qualitaet.pdf}
}

@article{Frischmuth2024,
   author = {Felix Frischmuth and Mattis Berghoff and Martin Braun and Philipp Härtel},
   doi = {10.1016/j.apenergy.2024.123991},
   issn = {03062619},
   journal = {Applied Energy},
   month = {12},
   publisher = {Elsevier Ltd},
   title = {Quantifying seasonal hydrogen storage demands under cost and market uptake uncertainties in energy system transformation pathways},
   volume = {375},
   year = {2024}
}

@misc{SiemensEnergy,
   author = {Siemens Energy},
   title = {Hydrogen and Power-to-X solutions},
   url = {https://assets.siemens-energy.com/dam/973beb1b-97e0-4150-9c2e-b29800761a8c/Electrolyzer_Brochure_Hydrogen_PowertoX-pdf_Original%20file.pdf}
}

@misc{RenewablesNinja,
  title        = {Renewables Ninja},
  note         = {\url{https://www.renewables.ninja} [Accessed: 2025-06-30]}
}

@article{Pfenninger2016,
   author = {Stefan Pfenninger and Iain Staffell},
   doi = {10.1016/j.energy.2016.08.060},
   issn = {03605442},
   journal = {Energy},
   month = {11},
   pages = {1251-1265},
   publisher = {Elsevier Ltd},
   title = {Long-term patterns of European PV output using 30 years of validated hourly reanalysis and satellite data},
   volume = {114},
   year = {2016}
}

@article{Staffell2016,
   author = {Iain Staffell and Stefan Pfenninger},
   doi = {10.1016/j.energy.2016.08.068},
   issn = {03605442},
   journal = {Energy},
   month = {11},
   pages = {1224-1239},
   publisher = {Elsevier Ltd},
   title = {Using bias-corrected reanalysis to simulate current and future wind power output},
   volume = {114},
   year = {2016}
}

@misc{MATLAB,
year = {2024},
author = {The MathWorks Inc.},
title = {MATLAB version: 23.2.0.2515942 (R2023b) Update 7},
publisher = {The MathWorks Inc.},
address = {Natick, Massachusetts, United States},
url = {https://www.mathworks.com}
}

@misc{gurobi,
  author = {{Gurobi Optimization, LLC}},
  title = {{Gurobi Optimizer Reference Manual}},
  year = 2024,
  url = "https://www.gurobi.com"
}

@article{Wang2026,
   author = {Bowen Wang and Zhaoqing Liang and Kai Yang and Lei Xing and Heng Shao and Zhuorui Wu and Yixin Liu and Li Guo and Ning Yang and Bing Hu and Chengshan Wang and Kui Jiao},
   doi = {10.1016/j.eng.2026.02.020},
   issn = {20958099},
   journal = {Engineering},
   month = {5},
   publisher = {Elsevier Ltd},
   title = {Multi-Timescale Scheduling Optimization of ALK/PEM Hybrid Electrolyzers System Considering Flexible Hydrogen Demand},
   year = {2026}
}

@misc{BMJ_EnWG,
   author = {{Bundesministerium der Justiz}},
   title = {{"Gesetz über die Elektrizitäts- und Gasversorgung (Energiewirtschaftsgesetz EnWG) § 118 Übergangsregelungen (6)}},
   url = {https://www.gesetze-im-internet.de/enwg_2005/__118.html}
}

@misc{BMJ_StromStG,
   author = {{Bundesministerium der Justiz}},
   title = {{"Stromsteuergesetz (StromStG) § 9a Erlass, Erstattung oder Vergütung der Steuer für bestimmte Prozesse und Verfahren (1)}},
   url = {https://www.gesetze-im-internet.de/stromstg/__9a.html}
}

@misc{Zenodo,
  doi = {https://doi.org/10.5281/zenodo.21026655},
  author = {Marie Arnold},
  language = {en},
  title = {Hybrid electrolyzer systems: Smart strategy or economic fallacy? - Model and results, {Zenodo}},
  publisher = {Zenodo},
  year = {2026}
}

\end{document}